\documentclass[a4paper,11pt]{article}
\pdfoutput=1 

\usepackage{jheppub} 
\usepackage{amsmath,latexsym}
\usepackage{longtable}
\usepackage{color}
\usepackage[small]{subfigure}

\usepackage[T1]{fontenc}

\title{\boldmath Charm Mass Determination from QCD Charmonium
Sum Rules at Order $\alpha_s^3$}

\preprint{\begin{flushright} MIT--CTP 4208, MPP--2011--4, IFIC/11-04\\UWThPh-2011-1,
LPN11-05\end{flushright}\vspace*{-1cm}}

\author[a,b]{Bahman Dehnadi}

\author[b,c]{Andre H. Hoang}

\author[c,d,e,1]{Vicent Mateu,\note{Corresponding author.}}

\author[a]{S. Mohammad Zebarjad}

\affiliation[a]{Shiraz University, Physics Department,
        Shiraz 71454, Iran.}
\affiliation[b]{University of Vienna, Faculty of Physics,
        Boltzmanngasse 5,  A-1090 Vienna, Austria.}
\affiliation[c]{Max-Planck-Institut f\"ur Physik (Werner-Heisenberg-Institut),
        F\"ohringer Ring 6, D-80805 M\"unchen, Germany.}
\affiliation[d]{Center for Theoretical Physics, Massachusetts Institute of 
        Technology, Cambridge, MA 02139.}
\affiliation[e]{Instituto de F\'\i sica Corpuscular, Universitat de Val\`encia -- Consejo Superior de 
        Investigaciones Cient\'\i ficas, Parc Cient\'\i fic, E-46980 Paterna (Valencia), Spain.}

\emailAdd{bahman.dehnadi@univie.ac.at}
\emailAdd{andre.hoang@univie.ac.at}
\emailAdd{mateu@ific.uv.es}
\emailAdd{zebarjad@susc.ac.ir}

\abstract{We determine the $\overline{{\rm MS}}$ charm quark mass
from a charmonium QCD sum rules analysis. 
On the theoretical side we use input from perturbation theory at 
${\cal O}(\alpha_s^3)$. Improvements with respect to previous ${\cal
  O}(\alpha_s^3)$ analyses include (1) an account of all available $e^+e^-$
hadronic cross section data and (2) a thorough analysis of perturbative
uncertainties. Using a data clustering method to combine hadronic
cross section data sets from different measurements we demonstrate that using
all available experimental data up to c.m.\ energies of $10.538$~GeV 
allows for determinations of experimental moments and their correlations with
small errors and that there is no need to rely on theoretical input
above the charmonium resonances. We also show that good convergence
properties of the perturbative series for the theoretical sum rule moments need
to be considered with some care when extracting the charm mass and demonstrate
how to set up a suitable set of scale variations to obtain a proper estimate of
the perturbative uncertainty. As the final outcome of our analysis we obtain
$\overline m_c(\overline m_c) = 1.282
\, \pm \, (0.006)_{\rm stat} 
\, \pm \, (0.009)_{\rm syst} 
\, \pm \, (0.019)_{\rm pert}
\, \pm \, (0.010)_{\alpha_s} 
\, \pm \, (0.002)_{\langle GG\rangle}$~GeV.
The perturbative error is an order of magnitude larger than the one obtained 
in previous ${\cal O}(\alpha_s^3)$ sum rule analyses.
} 

\keywords{QCD, perturbative corrections}
\begin{document}
\maketitle
\flushbottom

\section{Introduction}
\label{sectionintroduction}

Accurate determinations of the charm quark mass are an important ingredient
in the prediction of inclusive and radiative $B$ decays or exclusive
kaon decays such as $K\to\pi\nu\bar{\nu}$. Since these decays are
instruments to either measure CKM matrix elements or to search
for new physics effects, appropriate and realistic estimates of the
uncertainties are also an important element of these analyses
\cite{Antonelli:2009ws}.

One of the most powerful methods to determine the charm quark mass
is based on sum rules for the charm-anticharm production rate in $e^{+}e^{-}$
annihilation \cite{Novikov:1977dq}. Here, moments of the correlation function
of two charm vector currents at zero momentum transfer
\begin{eqnarray}
\label{momentdef1}
M_{n}^{{\rm th}} & = & \dfrac{12\pi^2 Q_c^2}{n!}\,\dfrac{{\rm d}}{{\rm
    d}q^{2n}}\left.\Pi(q^{2})\right|_{q^{2}=0}\,,\\
\left(g_{\mu\nu}q^{2}-q_{\mu}q_{\nu}\right)\,\Pi(q^{2}) 
& = & 
-\, i\int\mathrm{d}x\, e^{iqx}\left\langle \,0\left|T\,
    j_{\mu}(x)j_{\nu}(0)\right|0\,\right\rangle 
\,,\nonumber \\[2mm]
j^{\mu}(x) 
& = & 
\bar{\psi}(x)\gamma^{\mu}\psi(x)
\,,\nonumber 
\end{eqnarray}
$Q_c$ being the charm quark electric charge, can be related to weighted
integrals of the normalized charm cross section
\begin{eqnarray}
\label{momentdef2}
M_{n} & = & 
\int\dfrac{{\rm d}s}{s^{n+1}}R_{e^{+}e^{-}\to\, c\bar{c}\,+X}(s)\,,\\
R_{e^{+}e^{-}\to\, c\bar{c}\,+X}(s) & = & \dfrac{\sigma_{e^{+}e^{-}\to\,
c\bar{c}\,+X}(s)}{\sigma_{e^{+}e^{-}\to\,\mu^{+}\mu^{-}}(s)}\,,\nonumber 
\end{eqnarray}
which can be obtained from experiments. For small values of $n$ such
that $m_{c}/n\gtrsim\Lambda_{{\rm QCD}}$ the theoretical
moments $M_{n}^{{\rm th}}$ can be computed in an operator product expansion
(OPE) where the dominant 
part is provided by perturbative QCD supplemented by small vacuum
condensates that parametrize nonperturbative
effects~\cite{Shifman:1978bx,Shifman:1978by}. The leading gluon condensate power
correction term has a surprisingly small numerical effect and is essentially 
negligible for the numerical analysis as long as $n$ is small.

This allows to determine the charm mass in a short distance scheme such as 
$\overline{{\rm MS}}$ to high precision \cite{Kuhn:2001dm}. This method to 
determine the $\overline{{\rm MS}}$ charm mass is frequently called charmonium
sum rules. For the theoretical
moments the perturbative part of the OPE is known at ${\mathcal O}(\alpha_{s}^{0})$
and ${\mathcal O}(\alpha_{s})$ for any value of $n$ \cite{Kallen:1955fb}.
At ${\mathcal O}(\alpha_{s}^{2})$ the moments are known to high values of $n$
\cite{Chetyrkin:1995ii,Chetyrkin:1996cf, Boughezal:2006uu,
Czakon:2007qi,Maier:2007yn}, and to ${\mathcal O}(\alpha_{s}^{3})$
for $n=1$ \cite{Chetyrkin:2006xg,Boughezal:2006px},
$n=2$ \cite{Maier:2008he}, and $n=3$ \cite{Maier:2009fz}. Higher
moments at ${\cal O}(\alpha_s^3)$ have been determined by a semianalytical
procedure \cite{Hoang:2008qy,Kiyo:2009gb} (see also \cite{Greynat:2010kx}).
The Wilson coefficient of the gluon condensate contribution is known
to ${\mathcal O}(\alpha_{s})$ \cite{Broadhurst:1994qj}. On the experimental
side the total hadronic cross section in $e^{+}e^{-}$ annihilation
is known from various experimental measurements for c.m.\ energies up to
$10.538\,$GeV. None of the experimental analyses actually ranges over the
entire energy region between the charmonium region and $10.538$~GeV, but
different analyses overlapping in energy exist such that 
energies up to $10.538\,$GeV are completely covered
\cite{Bai:1999pk,Bai:2001ct,Ablikim:2004ck,Ablikim:2006aj,Ablikim:2006mb,:2009jsa,Osterheld:1986hw,Edwards:1990pc,Ammar:1997sk,Besson:1984bd,:2007qwa,CroninHennessy:2008yi,Blinov:1993fw,Criegee:1981qx,Siegrist:1976br,Rapidis:1977cv,Abrams:1979cx,Siegrist:1981zp}.\footnote{As a word of caution we mention that, except for the contributions of
the $J/\psi$ and $\psi'$ resonances, no experimental separation of the
charm and non-charm contributions in the hadronic cross section has been
provided in available data, although charm-tagged measurements are possible, see
e.g.\ Ref.~\cite{CroninHennessy:2008yi} (CLEO collaboration). So the charm pair
production rate from above the $J/\psi$ and $\psi^\prime$ that enters the
charmonium sum rules  
in Eq.~(\ref{momentdef2}) is usually obtained partly from the measured total
\mbox{$R$-ratio} with theory motivated subtractions of the non-charm rate,
and partly by using theory predictions for the charm production rate.
}
Interestingly, to the best of our knowledge, the complete set of all available
experimental data on the hadronic cross section has never been used in previous
charmonium sum rule analyses to determine the experimental moments. Rather, sum
rule analyses have relied heavily on theoretical input using different
approaches to determine the corresponding ``experimental error'' and
intrinsically leading to a sizable modeling uncertainty for energy regions
below $10.538\,$GeV for low values of $n$~\cite{Hoang:2004xm}.

The most recent charmonium sum rule analysis based on Eqs.~(\ref{momentdef1}) and
(\ref{momentdef2}), carried out by
K\"uhn~et~al.~\cite{Chetyrkin:2009fv,Kuhn:2007vp}\footnote{Ref.~
\cite{Chetyrkin:2009fv} is actually Chetyrkin~et~al.} using input from perturbative
QCD (pQCD) at ${\mathcal O}(\alpha_{s}^{3})$ for the perturbative contribution,
obtained $\overline{m}_{c}(\overline{m}_{c})=
1279\pm(2)_{{\rm pert}}\pm(9)_{{\rm exp}}
\pm(9)_{\alpha_{s}}\pm(1)_{{\rm \left\langle GG\right\rangle}}\,$~MeV
where the first quoted error is the perturbative uncertainty and the second is
the experimental one. The third and the
fourth quoted uncertainties come from $\alpha_s$ and the gluon condensate
correction, respectively.\footnote{In Refs.~\cite{Chetyrkin:2009fv,
Kuhn:2007vp} the main quoted result is for $\overline{m}_{c}(3{\rm GeV})$.
For $\overline{m}_{c}(\overline{m}_{c})$ the central value and total errors are
quoted. We have extracted the individual errors from results in their paper.}
To our knowledge this result, the outcome of 
similar analyses in Ref.~\cite{Chetyrkin:2006xg} and by Boughezal, Czakon and 
Schutzmeier~\cite{Boughezal:2006px}\footnote{Since
the analyses of Refs.~\cite{Chetyrkin:2006xg,Boughezal:2006px} were based on
outdated and 
less precise data for the $J/\psi$ and $\psi^\prime$ electronic partial widths 
\cite{Yao:2006px}, we frequently only compare our numerical results with those
of Refs.~\cite{Chetyrkin:2009fv,Kuhn:2007vp}. However the perturbative input of
Refs.~\cite{Chetyrkin:2006xg,Boughezal:2006px,Chetyrkin:2009fv,Kuhn:2007vp} and
ours is identical.}, a closely related analysis based on lattice results instead
of data for pseudoscalar moments~\cite{Allison:2008xk, McNeile:2010ji}, and the
finite-energy sum rules analysis by Bodenstein~et~al. \cite{Bodenstein:2011ma} 
represent the analyses with the
highest precision achieved so far in the literature. If confirmed, any further 
investigations and attempts concerning a more precise charm quark
$\overline{{\rm MS}}$ mass would likely be irrelevant for any foreseeable future.

We therefore find it warranted to reexamine the charmonium sum rule
analysis with special attention on the way how perturbative and experimental
uncertainties have been treated in Refs.~\cite{Chetyrkin:2009fv, Kuhn:2007vp}. A
closer look into their analysis reveals that the quoted
perturbative uncertainty results from a specific way to arrange
the $\alpha_s$ expansion for the charm mass extractions and, in addition,
by setting the $\overline{{\rm MS}}$ renormalization scales in $\alpha_{s}$
and in the charm mass (which we call $\mu_{\alpha}$ and $\mu_{m}$,
respectively) equal to each other (i.e., they use $\mu_{\alpha}=\mu_{m}$).
Moreover, concerning the experimental moments, only data 
up to $\sqrt{s}=4.8\,$GeV from the BES experiments \cite{Bai:2001ct,Ablikim:2006mb}
were used, while for $\sqrt{s}>4.8$~GeV perturbative QCD predictions
were employed. Conceptually this approach is somewhat related to the method
of finite energy sum rules (see e.g.\ Ref.~\cite{Penarrocha:2001ig}), which we,
however, do not discuss
in this work. While this approach might be justified to estimate the overall
nominal contribution for the experimental moments from  $\sqrt{s}>4.8$~GeV,
since perturbative QCD predictions describe quite well the measured total
hadronic cross section outside the resonance regions, it is not obvious
how this method can provide an experimental uncertainty. Since the region
$\sqrt{s}>4.8$~GeV constitutes about $30\%$ of the first moment
$M_{1}$, which is theoretically most reliable, this approach contains a
significant intrinsic model dependence that cannot be quantified unambiguously.

In this work we reexamine the charmonium sum rules analysis for low values of
$n$ using the latest ${\mathcal O}(\alpha_{s}^{3})$ perturbative results,
and we implement
improvements which concern the two issues just mentioned:
\begin{enumerate}
\item We analyze several different types of perturbative expansions
and examine in detail how the result for the $\overline{{\rm MS}}$
charm mass depends on independent choices of $\mu_{\alpha}$ and $\mu_{m}$. 
We show in particular that the interplay of certain choices for
the perturbative expansion
and the scale setting  $\mu_{\alpha}=\mu_{m}$ used in previous
${\mathcal O}(\alpha_s^3)$ analyses
leads to sizable cancellations of the dependence on $\mu_{\alpha}$ and $\mu_{m}$
that in the light of our new analysis has to be considered as accidental.
As the outcome of our analysis we quantify the current ${\mathcal O}(\alpha_{s}^{3})$
perturbative error as around $20$~MeV, which is an order of magnitude
larger than that of Refs.~\cite{Chetyrkin:2009fv, Kuhn:2007vp, Chetyrkin:2006xg, Boughezal:2006px}.
\item Using a clustering method 
\cite{Agostini:1993uj,Takeuchi:1995xe,Hagiwara:2003da}
to combine correlated data from many different experimental measurements
we show that the $e^{+}e^{-}$ total hadronic cross section relevant
for the charmonium sum rules can be determined with a
complete coverage of center of mass energies above the $J/\psi$ and
$\psi'$ resonances up to $10.538\,$GeV.
Conservatively estimated modeling uncertainties
coming from the energy range above $10.538\,$GeV then only lead to an
insignificant contribution to the total uncertainty of the experimental
moments. We also take the opportunity to include recent 
updates concerning the data on $\psi^\prime$ charmonium resonance.
\end{enumerate}

This paper is organized as follows: In Sec.~\ref{sectiontheory} we introduce the
theoretical framework and review the current status of perturbative computations. We
also show various equivalent ways of arranging the perturbative  
series in $\alpha_s$ for the charm mass. Finally we discuss how to properly estimate theoretical
uncertainties due to the truncation of  
the perturbative series. In Sec.~\ref{sectiondata} we present all the
experimental information that goes into our  
analysis. We discuss a clustering fit procedure that allows to combine data from
different experiments accounting for their correlation and show the results.  
In Sec.~\ref{sectionanalysis} we carry out the numerical charm mass analysis
concentrating on the first moment
$M_1$ using arbitrary values of $\alpha_s$, and we present our
final charm mass result.  magenta In Sec.~\ref{sectionanalysishigher}
we discuss the charm mass results obtained from higher moments $M_{2,3,4}$ 
and find agreement with the outcome of the first moment analysis.
In Appendix~A we present more details on the outcome of
our clustering fit procedure for the charm \mbox{$R$-ratio}, and in Appendix~B we prove
the equivalence of different versions of $\chi^2$ functions when auxiliary fit
parameters are employed. Appendix~C shows the dependence of the higher moment
charm mass results on the strong coupling.

\section{Theoretical Input}
\label{sectiontheory}

\subsection{Perturbative Contribution}
\label{subsectionperturbative}

The moments of the vector current correlator are defined in
Eq.~(\ref{momentdef1}). Their perturbative contribution in the framework of the 
OPE has a non-linear dependence on the charm quark mass. Thus in principle no
conceptual preference can be imposed on any of the possible perturbative 
series that arises when solving for the charm mass. As a consequence, different
versions of the expansion should be considered to obtain reliable estimates of
the perturbative uncertainty. As indicated
in Sec.~\ref{sectionintroduction} we use in the following $\mu_\alpha$ 
as the renormalization scale in $\alpha_s$ and $\mu_m$ as the renormalization
scale in the $\overline{\rm MS}$ charm quark mass $\overline m_c$. 

\vskip 5mm
\noindent
{\bf (a) Standard fixed-order expansion}\\
Writing the perturbative vacuum polarization function as 
\begin{eqnarray}
\label{Mnpertfixedorder1}
\Pi^{\rm pert}(q^2,\alpha_s(\mu_\alpha), \overline m_c(\mu_m), 
\mu_\alpha, \mu_m) 
\, = \, \dfrac{1}{12\pi^2 Q_c^2}\sum_{n=0}^\infty q^{2n} M_n^{\rm pert}
\,,
\end{eqnarray}
we have for the perturbative moments $M_n^{\rm pert}$
\begin{eqnarray}
\label{Mnpertfixedorder2}
M_n^{\rm pert} & = &
\frac{1}{(4\overline m^2_c(\mu_m))^{n}}
\sum_{i,a,b} \left(\frac{\alpha_s(\mu_\alpha)}{\pi}\right)^i
C^{a,b}_{n,i}\,\ln^a\left(\frac{\overline{m}^2_c(\mu_m)}{\mu^2_m}\right)   
\ln^b\left(\frac{\overline{m}^2_c(\mu_m)}{\mu^2_\alpha}\right).
\end{eqnarray}
This is the standard fixed-order expression for the perturbative moments. At
${\cal O}(\alpha_s^3)$ the coefficients $C^{0,0}_{n,3}$ were recently determined
for $n=1$~\cite{Chetyrkin:2006xg, Boughezal:2006px}, $n=2$ \cite{Maier:2008he},
$n=3$ \cite{Maier:2009fz} and higher \cite{Hoang:2008qy, Kiyo:2009gb,  
Greynat:2010kx}. We refer to Ref.~\cite{Boughezal:2006uu, Maier:2007yn} for the 
coefficients at ${\cal O}(\alpha_s^2)$. For convenience we have summarized the numerical 
expressions for the $C^{a,b}_{n,i}$ coefficients of the first four moments in 
Tab.~\ref{tabcfixedorder}.

The standard fixed-order expansion in Eq.~(\ref{Mnpertfixedorder2}) is the common
way to represent the perturbative moments. However, written in this form the
non-linear dependence on $\overline m_c$ does for some values of the experimental 
moments and the renormalization scales not yield numerical
solutions\footnote{This tends to happen frequently
at any order in $\alpha_s$ for $n>1$ and for $\mu_m\sim 3$~GeV.}
for $\overline m_c$.

\renewcommand{\arraystretch}{1.2}\setlength{\LTcapwidth}{\textwidth}

\begin{table}[t!]\begin{center}
\scriptsize
\begin{tabular}{|c|ccccccccc|}
\hline 
 & $C_{n,i}^{0,0}$ & $C_{n,i}^{1,0}$ & $C_{n,i}^{2,0}$ & $C_{n,i}^{3,0}$ & $C_{n,i}^{0,1}$ & $C_{n,i}^{1,1}$ & $C_{n,i}^{2,1}$ & $C_{n,i}^{0,2}$ & $C_{n,i}^{1,2}$\tabularnewline\hline
\multicolumn{10}{|c|}{$n=1$} \\\hline
$i=0$ & $1.06667$ & $0$ & $0$ & $0$ & $0$ & $0$ & $0$ & $0$ & $0$\tabularnewline
$i=1$ & $2.55473$ & $2.13333$ & $0$ & $0$ & $0$ & $0$ & $0$ & $0$ & $0$\tabularnewline
$i=2$ & $2.49671$ & $8.63539$ & $4.35556$ & $0$ & $-5.32236$ & $-4.44444$ & $0$ & $0$ & $0$\tabularnewline
$i=3$ & $-5.64043$ & $22.6663$ & $32.696$ & $8.95309$ & $-18.5994$ & $-42.8252$ & $-18.1481$ & $11.0882$ & $9.25926$\tabularnewline
\hline
\multicolumn{10}{|c|}{$n=2$} \\\hline
$i=0$ & $0.457143$ & $0$ & $0$ & $0$ & $0$ & $0$ & $0$ & $0$ & $0$\tabularnewline
$i=1$ & $1.10956$ & $1.82857$ & $0$ & $0$ & $0$ & $0$ & $0$ & $0$ & $0$\tabularnewline
$i=2$ & $2.77702$ & $7.46046$ & $5.5619$ & $0$ & $-2.31158$ & $-3.80952$ & $0$ & $0$ & $0$\tabularnewline
$i=3$ & $-3.49373$ & $21.8523$ & $38.6277$ & $15.1407$ & $-15.1307$ & $-36.9519$ & $-23.1746$ & $4.81579$ & $7.93651$\tabularnewline
\hline
\multicolumn{10}{|c|}{$n=3$} \\\hline
$i=0$ & $0.270899$ & $0$ & $0$ & $0$ & $0$ & $0$ & $0$ & $0$ & $0$\tabularnewline
$i=1$ & $0.519396$ & $1.6254$ & $0$ & $0$ & $0$ & $0$ & $0$ & $0$ & $0$\tabularnewline
$i=2$ & $1.63882$ & $5.8028$ & $6.56931$ & $0$ & $-1.08207$ & $-3.38624$ & $0$ & $0$ & $0$\tabularnewline
$i=3$ & $-2.83951$ & $16.0684$ & $40.3042$ & $22.2627$ & $-8.4948$ & $-29.3931$ & $-27.3721$ & $2.25432$ & $7.05467$\tabularnewline
\hline
\multicolumn{10}{|c|}{$n=4$} \\\hline
$i=0$ & $0.184704$ & $0$ & $0$ & $0$ & $0$ & $0$ & $0$ & $0$ & $0$\tabularnewline
$i=1$ & $0.203121$ & $1.47763$ & $0$ & $0$ & $0$ & $0$ & $0$ & $0$ & $0$\tabularnewline
$i=2$ & $0.795555$ & $4.06717$ & $7.44974$ & $0$ & $-0.42317$ & $-3.0784$ & $0$ & $0$ & $0$\tabularnewline
$i=3$ & $-3.349$ & $8.91524$ & $38.2669$ & $30.2128$ & $-3.96649$ & $-21.6873$ & $-31.0406$ & $0.881603$ & $6.41334$\tabularnewline
\hline
\end{tabular}\end{center}
\caption{Numerical values of the coefficients for Eq.~(\ref{Mnpertfixedorder2}).
(Standard fixed-order expansion).\label{tabcfixedorder}}
\end{table}

\vskip 5mm
\noindent
{\bf (b) Linearized expansion}\\
Concerning the charm mass dependence, a more linear way to organize the
perturbative expansion is to take the \mbox{$2n$-th} root of
Eq.~(\ref{Mnpertfixedorder2}):
\footnote{A similar expansion was employed in Ref.~\cite{Allison:2008xk}.}
\begin{eqnarray}
\label{Mnpertlinearized1}
\Big(M_n^{\rm th, pert}\Big)^{1/2n} & = &
\frac{1}{2\overline{m}_c(\mu_m)} \,\sum_{i,a,b}\left(\frac{\alpha_s(\mu_\alpha)}{\pi}\right)^i \tilde C_{n,i}^{a,b}
\ln^a\left(\frac{\overline{m}^2_c(\mu_m)}{\mu^2_m}\right)   \ln^b\left(\frac{\overline{m}^2_c(\mu_m)}{\mu^2_\alpha}\right)
,
\end{eqnarray} 
or equivalently
\begin{eqnarray}
\label{Mnpertlinearized2}
\overline m_c(\mu_m) & = & \frac{1}{2\Big(M_n^{\rm th, pert}\Big)^{1/2n}} \sum_{i,a,b}\,\left(\frac{\alpha_s(\mu_\alpha)}{\pi}\right)^i\tilde C_{n,i}^{a,b}
\ln^a\left(\frac{\overline{m}^2_c(\mu_m)}{\mu^2_m}\right)   \ln^b\left(\frac{\overline{m}^2_c(\mu_m)}{\mu^2_\alpha}\right)
.
\end{eqnarray}
The coefficients $\tilde C_{n,i}^{a,b}$ using again $\mu_\alpha$ for the
renormalization scale in $\alpha_s$ and $\mu_m$ for the renormalization scale in
the $\overline{\rm MS}$ charm mass are given in
Tab.~\ref{tabctildefixedorder}. Although relation (\ref{Mnpertlinearized2}) involves a non-linear
dependence on ${\overline m}_c$, we find that it always has a numerical solution.
\begin{table}[t!]\begin{center} {\scriptsize
\begin{tabular}{|c|ccccccccc|}
\hline 
 & $\tilde{C}_{n,i}^{0,0}$ & $\tilde{C}_{n,i}^{1,0}$ & $\tilde{C}_{n,i}^{2,0}$ & $\tilde{C}_{n,i}^{3,0}$ & $\tilde{C}_{n,i}^{0,1}$ & $\tilde{C}_{n,i}^{1,1}$ & $\tilde{C}_{n,i}^{2,1}$ & $\tilde{C}_{n,i}^{0,2}$ & $\tilde{C}_{n,i}^{1,2}$\tabularnewline\hline
\multicolumn{10}{|c|}{$n=1$} \\
\hline
$i=0$ & $1.0328$ & $0$ & $0$ & $0$ & $0$ & $0$ & $0$ & $0$ & $0$\tabularnewline
$i=1$ & $1.2368$ & $1.0328$ & $0$ & $0$ & $0$ & $0$ & $0$ & $0$ & $0$\tabularnewline
$i=2$ & $0.46816$ & $2.94379$ & $1.59223$ & $0$ & $-2.57668$ & $-2.15166$ & $0$ & $0$ & $0$\tabularnewline
$i=3$ & $-3.2913$ & $6.97983$ & $10.9784$ & $2.74217$ & $-5.91875$ & $-15.5793$ & $-6.63428$ & $5.36808$ & $4.48262$\tabularnewline\hline
\multicolumn{10}{|c|}{$n=2$} \\\hline
$i=0$ & $0.822267$ & $0$ & $0$ & $0$ & $0$ & $0$ & $0$ & $0$ & $0$\tabularnewline
$i=1$ & $0.498944$ & $0.822267$ & $0$ & $0$ & $0$ & $0$ & $0$ & $0$ & $0$\tabularnewline
$i=2$ & $0.79463$ & $1.85797$ & $1.26766$ & $0$ &$-1.03947$ & $-1.71306$ & $0$ & $0$ & $0$\tabularnewline
$i=3$ & $-3.20128$ & $3.15216$ & $7.99164$ & $2.1832$ & $-4.91174$ & $-10.3796$ & $-5.28192$ & $2.16555$ & $3.56887$\tabularnewline\hline
\multicolumn{10}{|c|}{$n=3$} \\\hline
$i=0$ & $0.804393$ & $0$ & $0$ & $0$ & $0$ & $0$ & $0$ & $0$ & $0$ \tabularnewline
$i=1$ & $0.257044$ & $0.804393$ & $0$ & $0$ & $0$ & $0$ & $0$ & $0$ & $0$\tabularnewline
$i=2$ & $0.605688$ & $1.58653$ & $1.24011$ & $0$ & $-0.535508$ & $-1.67582$ & $0$ & $0$ & $0$\tabularnewline
$i=3$ & $-2.46047$ & $1.56737$ & $7.46171$ & $2.13574$ & $-3.34838$ & $-9.19128$ & $-5.1671$ & $1.11564$ & $3.49129$\tabularnewline\hline
\multicolumn{10}{|c|}{$n=4$} \\\hline
$i=0$ & $0.809673$ & $0$ & $0$ & $0$ & $0$ & $0$ & $0$ & $0$ & $0$ \tabularnewline
$i=1$ & $ 0.111301$ & $0.809673$ & $0$ & $0$ & $0$ & $0$ & $0$ & $0$ & $0$\tabularnewline
$i=2$ & $0.382377$ & $1.44951$ & $1.24825$ & $0$ & $-0.231877$ & $-1.68682$ & $0$ & $0$ & $0$\tabularnewline
$i=3$ & $-2.21776$ & $0.492406$ & $7.2834$ & $2.14976$ & $-1.95033$ & $-8.63733$ & $-5.20102$ & $0.483077$ & $3.51421$\tabularnewline
\hline
\end{tabular}
\caption{Numerical values of the coefficients for Eq.~(\ref{Mnpertlinearized1}).
  (Linearized expansion). \label{tabctildefixedorder}}
}\end{center}\end{table}

\vskip 5mm
\noindent
{\bf (c) Iterative linearized expansion}\\
For the standard and the linearized expansions in Eqs.~(\ref{Mnpertfixedorder2})
and (\ref{Mnpertlinearized1}) one searches for numerical
solutions of the charm mass $\overline m_c(\mu_m)$ keeping the exact mass
dependence on the respective RHS at each order in $\alpha_s$. An alternative way
that is consistent within perturbation theory is to solve for $\overline
m_c(\mu_m)$ iteratively order-by-order supplementing appropriate lower order
values for $\overline m_c(\mu_m)$ in higher order perturbative
coefficients. To be more explicit, we describe the method in the following.
As the basis for the iterative expansion carried out in our analysis we use the
linearized expansion of Eq.~(\ref{Mnpertlinearized2}).

In the first step we determine $\overline m_c(\mu_m)$ employing the tree-level
relation 
\begin{eqnarray}
\label{Mnpertiterative1}
\overline m_c^{(0)} =
\frac{1}{2\Big(M_n^{\rm th, pert}\Big)^{1/2n}} \,\tilde C_{n,0}^{0,0}
\,,
\end{eqnarray}
giving the tree-level charm mass $\overline m_c^{(0)}$.
In the next step one employs the relation 
\begin{eqnarray}
\label{Mnpertiterative2}
\overline m_c^{(1)}(\mu_m) &=&
\frac{1}{2\Big(M_n^{\rm th, pert}\Big)^{1/2n}}\,\Biggr\{\,
\tilde C_{n,0}^{0,0} \, + \,
\frac{\alpha_s(\mu_\alpha)}{\pi}\bigg[\tilde C_{n,1}^{0,0}+\tilde C_{n,1}^{1,0}
\ln\left(\frac{\overline{m}_c^{(0)\,2}}{\mu^2_m}\right)\Bigg]\Biggr\}
\,,\quad
\end{eqnarray}
to determine the ${\cal O}(\alpha_s)$ charm mass $\overline m_c^{(1)}(\mu_m)$.
In the ${\cal O}(\alpha_s)$ terms on the RHS of Eq.~(\ref{Mnpertiterative2}) the
tree-level charm mass $\overline m_c^{(0)}$ is used, which is consistent to
${\cal O}(\alpha_s)$. At ${\cal O}(\alpha_s^2)$ for the determination of
$\overline m_c^{(2)}(\mu_m)$ one uses $\overline m_c^{(0)}$ for the ${\cal
  O}(\alpha_s^2)$ coefficient and $\overline m_c^{(1)}(\mu_m)$ for the ${\cal
  O}(\alpha_s)$ correction, which in the strict $\alpha_s$ expansion yields 
\begin{eqnarray}
\label{Mnpertiterative3}
\overline m_c^{(2)} (\mu_m)& = &
\frac{1}{2\Big(M_n^{\rm th, pert}\Big)^{1/2n}}\,\Biggr\{\,
\tilde C_{n,0}^{0,0} \, +\,
\frac{\alpha_s(\mu_\alpha)}{\pi}\Bigg[\tilde C_{n,1}^{0,0}+\tilde C_{n,1}^{1,0}
\ln\left(\frac{\overline{m}_c^{(0)\,2}}{\mu^2_m}\right)\Bigg]+\nonumber\\
&&\left(\frac{\alpha_s(\mu_\alpha)}{\pi}\right)^2\Bigg[2\,\dfrac{\tilde C_{n,1}^{1,0}\,\tilde C_{n,1}^{0,0}}{\tilde C_{n,0}^{0,0}}+
2\,\dfrac{(\tilde C_{n,1}^{1,0})^2}{\tilde C_{n,0}^{0,0}}
\ln\left(\frac{\overline{m}_c^{(0)\,2}}{\mu^2_m}\right)+\nonumber\\
&&\sum_{a,b}\,\tilde C_{n,2}^{a,b}
\ln^a\left(\frac{\overline{m}_c^{(0)\,2}}{\mu^2_m}\right)
\ln^b\left(\frac{\overline{m}_c^{(0)\,2}}{\mu^2_\alpha}\right)\Bigg]\Biggr\}
\,.
\end{eqnarray}
Here the second line contains the derivative of the ${\mathcal O}(\alpha_s)$ terms with
respect to the charm mass. The determination of the ${\cal O}(\alpha_s^3)$ charm
mass $\overline 
m_c^{(3)}(\mu_m)$ is then carried out in an analogous way involving the second (first)
derivative with respect to the mass in the ${\cal O}(\alpha_s)$ (${\cal
  O}(\alpha_s^2)$) correction and using again $\overline m_c^{(0)}$ for the  
${\cal O}(\alpha_s^3)$ coefficient.

In general we can write the iterative expansion as follows:
\begin{equation}
{\overline m}_c(\mu_m)\,=\,{\overline m}_c^{(0)}
\sum_{i,a,b}\left(\frac{\alpha_s(\mu_\alpha)}{\pi}\right)^i \hat C_{n,i}^{a,b}\,
\ln^a\left(\frac{{\overline m}_c^{(0)\,2}}{\mu^2_m}\right)   
\ln^b\left(\frac{{\overline m}_c^{(0)\,2}}{\mu^2_\alpha}\right),
\label{eq:iterative-general}
\end{equation}
where the numerical value of the coefficients $\hat C_{n,i}^{a,b}$ are collected in Tab.~\ref{tab:chat}.

The iterative way to treat the perturbative series for the charm mass has the
advantage that solving for the charm mass involves
equations that are strictly linear in the charm mass at any order of the
$\alpha_s$ expansion and thus by construction always have
solutions. In this way any possible influence on the analysis arising from a
non-linear dependence is eliminated. 

\begin{table}[t!]\begin{center}{\scriptsize
\begin{tabular}{|c|ccccccccc|}
\hline 
 & $\hat{C}_{n,i}^{0,0}$ & $\hat{C}_{n,i}^{1,0}$ & $\hat{C}_{n,i}^{2,0}$ & $\hat{C}_{n,i}^{3,0}$ & $\hat{C}_{n,i}^{0,1}$ & $\hat{C}_{n,i}^{1,1}$ & $\hat{C}_{n,i}^{2,1}$ & $\hat{C}_{n,i}^{0,2}$ & $\hat{C}_{n,i}^{1,2}$\tabularnewline\hline
\multicolumn{10}{|c|}{$n=1$} \\
\hline
$i=0$ & $1$ & $0$ & $0$ & $0$ & $0$ & $0$ & $0$ & $0$ & $0$\tabularnewline
$i=1$ & $1.19753$ & $1$ & $0$ & $0$ & $0$ & $0$ & $0$ & $0$ & $0$\tabularnewline
$i=2$ & $2.84836$ & $4.85031$ & $1.54167$ & $0$ & $-2.49486$ & $-2.08333$ & $0$ & $0$ & $0$\tabularnewline
$i=3$ & $1.92718$ & $17.1697$ & $14.7131$ & $2.65509$ & $-15.7102$ & $-23.418$ & $-6.42361$ & $5.19762$ & $4.34028$\tabularnewline\hline
\multicolumn{10}{|c|}{$n=2$} \\\hline
$i=0$ & $1$ & $0$ & $0$ & $0$ & $0$ & $0$ & $0$ & $0$ & $0$\tabularnewline
$i=1$ & $0.60679$ & $1$ & $0$ & $0$ & $0$ & $0$ & $0$ & $0$ & $0$\tabularnewline
$i=2$ & $2.17997$ & $4.25957$ & $1.54167$ & $0$ &$-1.26415$ & $-2.08333$ & $0$ & $0$ & $0$\tabularnewline
$i=3$ & $1.30653$ & $14.3435$ & $13.8024$ & $2.65509$ & $-11.03$ & $-20.9565$ & $-6.42361$ & $2.63364$ & $4.34028$\tabularnewline\hline
\multicolumn{10}{|c|}{$n=3$} \\\hline
$i=0$ & $1$ & $0$ & $0$ & $0$ & $0$ & $0$ & $0$ & $0$ & $0$ \tabularnewline
$i=1$ & $0.31955$ & $1$ & $0$ & $0$ & $0$ & $0$ & $0$ & $0$ & $0$\tabularnewline
$i=2$ & $1.39208$ & $3.97233$ & $1.54167$ & $0$ & $-0.66573$ & $-2.08333$ & $0$ & $0$ & $0$\tabularnewline
$i=3$ & $0.458292$ & $12.5064$ & $13.3595$ & $2.65509$ & $-6.82554$ & $-19.7597$ & $-6.42361$ & $1.38694$ & $4.34028$\tabularnewline\hline
\multicolumn{10}{|c|}{$n=4$} \\\hline
$i=0$ & $1$ & $0$ & $0$ & $0$ & $0$ & $0$ & $0$ & $0$ & $0$ \tabularnewline
$i=1$ & $ 0.137464$ & $1$ & $0$ & $0$ & $0$ & $0$ & $0$ & $0$ & $0$\tabularnewline
$i=2$ & $0.747189$ & $3.79024$ & $1.54167$ & $0$ & $-0.286383$ & $-2.08333$ & $0$ & $0$ & $0$\tabularnewline
$i=3$ & $-0.850143$ & $11.1964$ & $13.0788$ & $2.65509$ & $-3.55432$ & $-19.001$ & $-6.42361$ & $0.596632$ & $4.34028$\tabularnewline
\hline
\end{tabular}
\caption{Numerical values of the coefficients for
  Eq.~(\ref{eq:iterative-general}).
(Iterative linearized expansion).\label{tab:chat}}
}\end{center}\end{table}

\begin{figure}[t]
\center
 \includegraphics[width=0.3\textwidth]{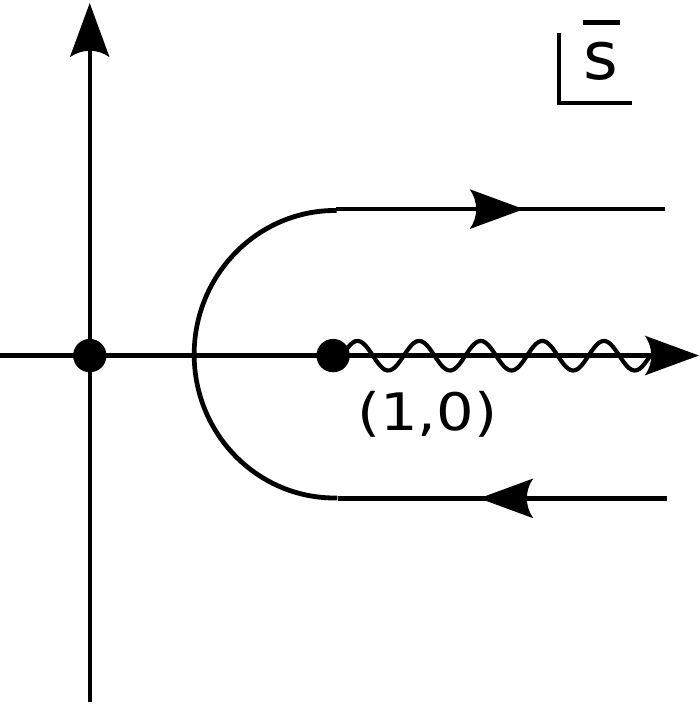}
 \caption{
Path of integration in the complex \mbox{$\bar s$-plane} for the computation
of the moments.
 \label{fig:contour} }
\end{figure}

\vskip 5mm
\noindent
{\bf (d) Contour improved expansion}\\
For the expansion methods (a)-(c) the moments and the charm quark mass are
computed for a fixed choice of the renormalization scale $\mu_\alpha$ in the
strong coupling $\alpha_s$. In analogy to the contour improved methods used for
\mbox{$\tau$-decays} (see e.g.\ Refs.~\cite{LeDiberder:1992te, Pivovarov:1991rh, 
Braaten:1991qm, Narison:1988ni, Braaten:1988ea, Braaten:1988hc})
one can employ a path-dependent $\mu_\alpha$ in
the contour integration that defines the perturbative moments~\cite{Hoang:2004xm}, 
see Fig.~\ref{fig:contour},
\begin{eqnarray}
\label{Mnpertcontour1}
M_n^{\rm c, pert} & = &
\frac{6\pi Q_c^2}{i}\,\int_c\,\frac{{\rm d}s}{s^{n+1}}
\Pi(q^2, \alpha_s(\mu_\alpha^c(s,\overline m_c^2)), \overline{m}_c(\mu_m),
\mu_\alpha^c(s,\overline m_c^2), \mu_m)
\,.
\end{eqnarray}
Due to the independence of the moments on $\mu_\alpha$ and since no large
logarithms are being generated anywhere for a path with distance of order
$\overline m_c$ from the cut on the real axis, this method is a viable
alternative to
carry out the perturbative expansion. The different orders in the expansion of
the contour improved moments $M_n^{\rm c, pert}$ are generated from the fixed-order
$\alpha_s$ expansion of the vacuum polarization function $\Pi$ in
Eq.~(\ref{Mnpertcontour1}). 

A useful path-dependent choice for $\mu_\alpha^c$ is given by~\cite{Hoang:2004xm}
\begin{eqnarray}
\label{mualphacontour}
(\mu_\alpha^c)^2(s,\overline m_c^2) & = &
\mu_\alpha^2\,\bigg(\,1-\frac{s}{4\overline m_c^2(\mu_m)}\,\bigg) 
\,,
\end{eqnarray}
which implements a modified weighting of threshold versus high energy
contributions. 
It is straightforward to prove that the resulting moments $M_n^{\rm c,
  pert}$ can be obtained from the small-$q^2$ expansion of the perturbative
vacuum polarization function using $\mu_\alpha^c$ as the renormalization scale
of $\alpha_s$,
\begin{eqnarray}
\label{Mnpertcontour2}
\Pi^{\overline {\rm MS}}\Big(q^2,
\alpha_s(\mu_\alpha^c(q^2,\overline m_c^2)),
\overline m_c(\mu_m), 
\mu_\alpha^c(q^2,\overline m_c^2),
\mu_m\Big) & = &
\sum\limits_{n=0}^\infty \, q^{2n}\,M_n^{\rm c, pert}
\,.
\end{eqnarray}
From Eq.~(\ref{Mnpertcontour2}) we can see that the $M_n^{\rm c, pert}$ can be
derived from the expressions for the fixed-order moments $M_m^{\rm pert}$ with
$m\le n$ given in Eq.~(\ref{Mnpertfixedorder2}).\footnote{
This works in general as long as the
path dependent $\mu_\alpha^c(q^2,\overline m_c^2)$ does not produce a spurious
cut in $\alpha_s$ starting at $q^2=0$ and running towards $-\infty$. This
condition is implemented into Eq.~(\ref{mualphacontour}) by $(\mu_\alpha^c)^2$
being negative along the physical cut of the vacuum polarization function above the
charm pair threshold.   
} They also depend on the QCD
\mbox{$\beta$-function} and its derivatives, which arise in the small-$q^2$ expansion
of $\alpha_s(\mu^c_\alpha(q^2,\overline m_c^2))$. 
Note that the \mbox{$\beta$-function} that has to be employed must be exactly the
same that is used for the contour integration in Eq.~(\ref{Mnpertcontour1}). 
Expanding the dependence of
the \mbox{$\beta$-function} on $\alpha_s$ strictly in fixed-order one recovers the
fixed-order moments $M_n^{\rm pert}$. So the dependence of the contour improved
moments $M_n^{\rm c, pert}$ on the fixed-order moments $M_m^{\rm pert}$ with
$m<n$ is only residual due to the truncation of the $\alpha_s$ series
and vanishes in the large order limit. The contour improved
expansion thus represents yet another alternative parametrization of higher order 
perturbative corrections. Due to their residual dependence on
lower moments the contour improved moments have a sensitivity to the
UV-subtraction scheme for the vacuum polarization
function, i.e.\ on $\Pi(0)=M_0^{\rm pert}$. Using the ``on-shell'' scheme with
$\Pi(0)=0$ one finds that $M_1^{\rm c,pert} = M_1^{\rm pert}$. For our analysis
we employ the $\overline{\rm MS}$ scheme for $\Pi(0)$ defined for $\mu={\overline m}_c({\overline m}_c)$.
Expressed in terms of $\alpha_s(\mu_\alpha)$ and ${\overline m}_c(\mu_m)$ it has the form \cite{Chetyrkin:2006xg}
\begin{eqnarray}
\label{Pi0msbar}
\Pi^{\overline {\rm MS}}(0) & = &
\sum_{i,a,b} \left(\frac{\alpha_s(\mu_\alpha)}{\pi}\right)^i
C^{a,b}_{0,i}\,\ln^a\left(\frac{\overline{m}^2_c(\mu_m)}{\mu^2_m}\right)   \ln^b\left(\frac{\overline{m}^2_c(\mu_m)}{\mu^2_\alpha}\right)
\,.
\end{eqnarray}  
The numerical values for the coefficients 
$C^{a,b}_{0,i}$ can be found in Tab.~\ref{tabcPi0}. Using contour improved moments to determine $\overline m_c$ also
involves non-linear relations, which implies that in some cases there is no
solution. Again this can happen for $n \geq 1$ and for $\mu_m\sim 3$~GeV at any order.

\begin{table}[t!]\begin{center}
\begin{tabular}{|c|cccc|}
\hline 
 & $C_{0,i}^{0,0}$ & $C_{0,i}^{1,0}$ & $C_{0,i}^{0,1}$ & $C_{0,i}^{0,2}$\tabularnewline
\hline
$i=0$ & $0$ & $0$ & $0$ & $0$\tabularnewline
$i=1$ & $1.44444$ & $0$ & $0$ & $0$\tabularnewline
$i=2$ & $2.83912$ & $0$ & $-3.00926$ & $0$\tabularnewline
$i=3$ & $-5.28158$ & $6.01852$ & $-16.4639$ & $6.26929$\tabularnewline
\hline
\end{tabular}\end{center}
\caption{Numerical values of the coefficients for Eq.~(\ref{Pi0msbar}).
($\Pi(0)$ in the $\overline{\mbox{MS}}$ scheme.)\label{tabcPi0}}
\end{table}

\subsection{Gluon Condensate Contribution}
\label{subsectioncondensate}
The dominant subleading contribution in the OPE for the theory moments $M_n^{\rm
  th}$ is arising from the gluon condensate giving~\cite{Novikov:1977dq,Baikov:1993kc}
\begin{eqnarray}
\label{MnOPE1} 
M_n^{\rm th} & = & M_n^{\rm pert} + \Delta M_n^{\langle G^2\rangle}\,+\,\ldots
\,,
\end{eqnarray}
where the ellipses represent higher order power-suppressed condensate
contributions of the OPE. 
The ${\cal O}(\alpha_s)$ corrections to the Wilson
coefficient of the gluon condensate corrections have been determined in
Ref.~\cite{Broadhurst:1994qj}. In Ref.~\cite{Chetyrkin:2010ic} it is suggested that 
the Wilson coefficient of the gluon condensate should be expressed in terms of
the pole rather than the $\overline{\mbox{MS}}$ mass based on the observation
that the pole mass leads to a condensate correction that is numerically quite
stable for higher moments. We confirm this behavior, and mention that it
results from the strong inverse power-dependence on the charm mass that
generates a large $n$-dependence in the ${\cal O}(\alpha_s)$ corrections. Since
we parameterize all mass dependence in terms of the
$\overline{\mbox{MS}}$ charm mass, we adopt an analytic expression for the gluon
condensate correction, where the corrections associated to the pole mass are
grouped together with the $\overline{\mbox{MS}}$ charm mass parameter. The
resulting expression reads
\begin{align}
\label{DeltaMncondensate1}
\Delta M_n^{\langle G^2\rangle}
& = \dfrac{1}{(4M_c^2)^{n+2}}\Big\langle\frac{\alpha_s}{\pi}
 G^2\Big\rangle_{\rm RGI}  \left[
   a^{0,0}_{n}+\dfrac{\alpha_{s}(\mu_\alpha)}{\pi}\,a^{1,0}_{n}\right]\,,\\
M_c & = \overline{m}_c(\mu_m)\left\{1 + \dfrac{\alpha_{s}(\mu_\alpha)}{\pi} 
\left[\dfrac{4}{3} - 
\ln\left(\frac{\overline{m}^2_c(\mu_m)}{\mu^2_\mu}\right)\right]\right\}\,,\nonumber
\end{align}
using the renormalization
group invariant (RGI) scheme for the gluon condensate \cite{Narison:1983kn}.
Numerical values for the coefficients $a_n^{i,j}$ are given in
Tab.~\ref{tabgluoncondensate} for $n=1,2,3,4$. For the RGI gluon condensate we
adopt~\cite{Ioffe:2005ym}
\begin{eqnarray}
\label{condensatevalue1}
\Big\langle\frac{\alpha_s}{\pi} G^2\Big\rangle_{\rm RGI}
& = & 0.006\pm0.012\;\mathrm{GeV}^4\,.
\end{eqnarray}
The overall contribution of the gluon condensate correction in
Eq.~(\ref{DeltaMncondensate1}) in the charm quark mass analysis is quite
small. Its contribution to the moments amounts to around $0.4\%$, $2\%$, $5\%$, and $9\%$
for the first four moments, respectively. For $n=1$ it leads to a correction in
the $\overline{\rm MS}$ charm quark mass at the level of $2$~MeV and is an order
of magnitude smaller than our perturbative uncertainty. 
We therefore ignore the condensate correction for the discussion of the perturbative uncertainties in
Sec.~\ref{subsectionmcerror}. Its contribution is, however, included
for completeness in the final charm mass results presented in Secs.~\ref{sectionanalysis} and 
\ref{sectionanalysishigher}.

\begin{table}[t!]\begin{center}
\begin{tabular}{|c|cccc|}
\hline 
 & $n=1$ & $n=2$ & $n=3$ & $n=4$\tabularnewline
\hline
$a_{n}^{0,0}$ & $-16.042$ & $-26.7367$ & $-38.8898$ & $-52.3516$ \tabularnewline
$a_{n}^{1,0}$ & $-143.364$ & $-272.186$ & $-439.820$ & $-646.690$ \tabularnewline
\hline
\end{tabular}\end{center}
\caption{Numerical values for the coefficients of
  Eq.~(\ref{DeltaMncondensate1}). (Gluon condensate
    contribution).\label{tabgluoncondensate}
  }
  \end{table}

\subsection{Running Coupling and Mass}
\label{subsectionrunning}

The analysis of the charmonium sum rules naturally involves renormalization
scales around the charm mass, $\mu\sim {\overline m}_c\sim 1.3$~GeV, which are close to the
limits of a perturbative treatment. In fact, parametrically, the typical scale
relevant for the perturbative computation of the \mbox{$n$-th} moment $M_n^{\rm th}$ is
of order $\mu\sim {\overline m}_c/n$ (see e.g.\ Ref.~\cite{Hoang:1998uv})
because the energy range of the smearing associated to the weight function
$1/s^{n+1}$ in Eq.~(\ref{momentdef2}) decreases with $n$. We will therefore use
$n=1$ for our final numerical analysis. Moreover, it is common practice to quote
the $\overline{\rm MS}$ charm mass $\overline m_c(\overline m_c)$, i.e.\ for the
scale choice $\mu_m=\overline m_c$. 
It is therefore useful to have a look at the quality of the perturbative
behavior of the renormalization group evolution of the strong $\overline{\rm MS}$
coupling $\alpha_s$ and the $\overline{\rm MS}$ charm quark mass.

\begin{figure*}[t]
\subfigure[]{
\includegraphics[width=0.485\textwidth]{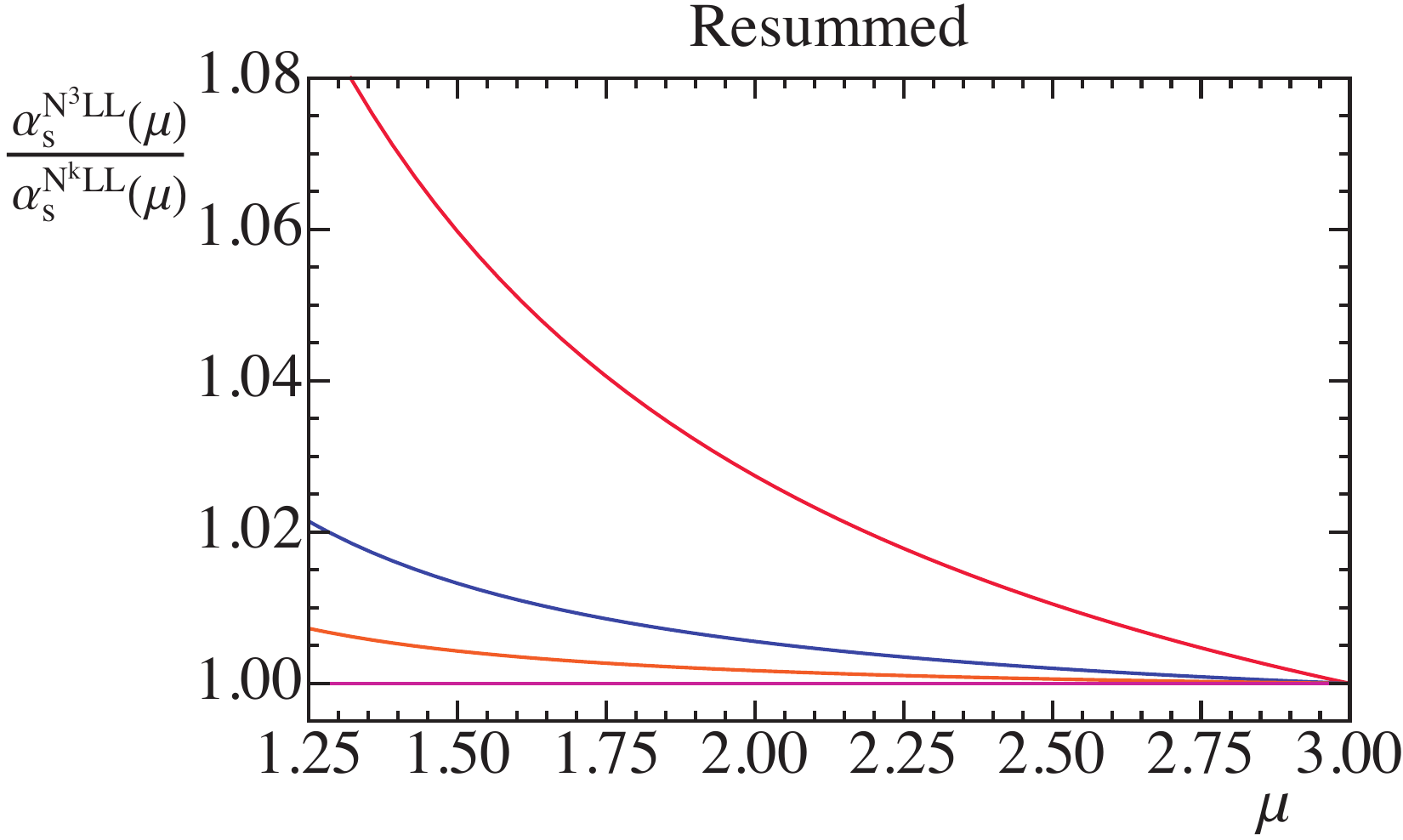}
\label{fig:alphasevolutionexact}
}
\subfigure[]{
\includegraphics[width=0.485\textwidth]{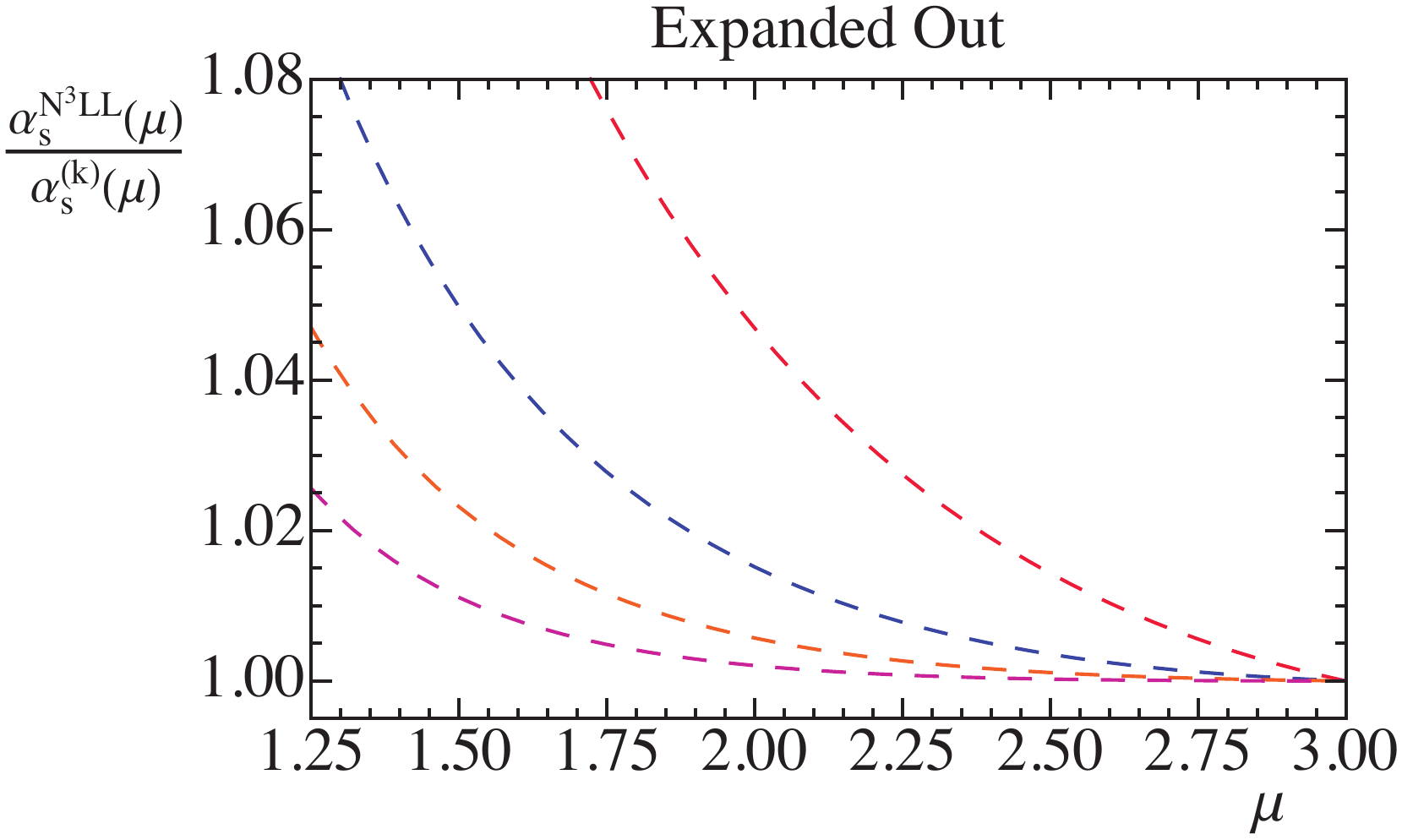}
\label{fig:alphasevolutionexpanded}
}
\vspace{-0.2cm}
\caption{Results for $\alpha^{\rm N^3LL}_s(\mu)/\alpha^{\rm N^kLL}_s(\mu)$ (a)
and $\alpha^{\rm N^3LL}_s(\mu)/\alpha^{(k)}_s(\mu)$ (b), where $\alpha^{\rm N^kLL}_s$
stands for the $(k+1)$-loop running coupling constant and $\alpha^{(k)}_s$ is the corresponding
${\mathcal O}(\alpha_s^{(k+1)})$ fixed-order expression for $\alpha_s$. All orders are run from
the common point $\alpha_s(3~{\rm GeV})=0.2535$.
\label{fig:alphasevolution}}
\end{figure*}

In Fig.~\ref{fig:alphasevolutionexact}\footnote{
In this examination and throughout our other analyses we use the $\overline{\rm
  MS}$ renormalization group equations with $n_f=4$ active running flavors.}
we have displayed $\alpha_s^{\rm N^3LL}(\mu)/\alpha_s^{\rm N^kLL}(\mu)$ using
for $\alpha_s^{\rm N^kLL}$ the \mbox{$(k+1)$-loop} QCD \mbox{$\beta$-function} and the
respective exact numerical solution for  $\alpha_s(3~\mbox{GeV})= 0.2535$ as
the common
reference point. We see that the convergence of the
lower order results towards the 4-loop evolution is very good even down to scales
of around $1$~GeV. The curves indicate that the remaining relative perturbative
uncertainty in the 4-loop evolution might be substantially smaller than $1\%$
for scales down to $\overline m_c\sim 1.3$~GeV. It is also instructive to
examine the evolution using 
a fixed-order expansion. In  Fig.~\ref{fig:alphasevolutionexpanded} we display
$\alpha_s^{\rm N^3LL}(\mu)/\alpha_s^{(m)}(\mu)$ where $\alpha_s^{(m)}(\mu)$ is
the ${\cal O}(\alpha_s^{m+1})$ fixed-order expression for $\alpha_s(\mu)$ using
the reference value $\alpha_s(3~\mbox{GeV})=0.2535$ as the expansion parameter. 
The convergence of the fixed-order expansion for $\alpha_s(\mu)$ towards the
exact N${}^3$LL numerical solution $\alpha_s^{\rm N^3LL}(\mu)$ is somewhat worse
compared to the renormalization group resummed results since the deviation of
the ratio from one is in general larger. However, convergence is clearly
visible. In particular there are not any signs of instabilities.
It therefore seems to be safe to use renormalization scales down to the
charm mass and associated renormalization scale variations as an instrument to
estimate the perturbative uncertainties. 

\begin{figure*}[t]
\subfigure[]{
\includegraphics[width=0.485\textwidth]{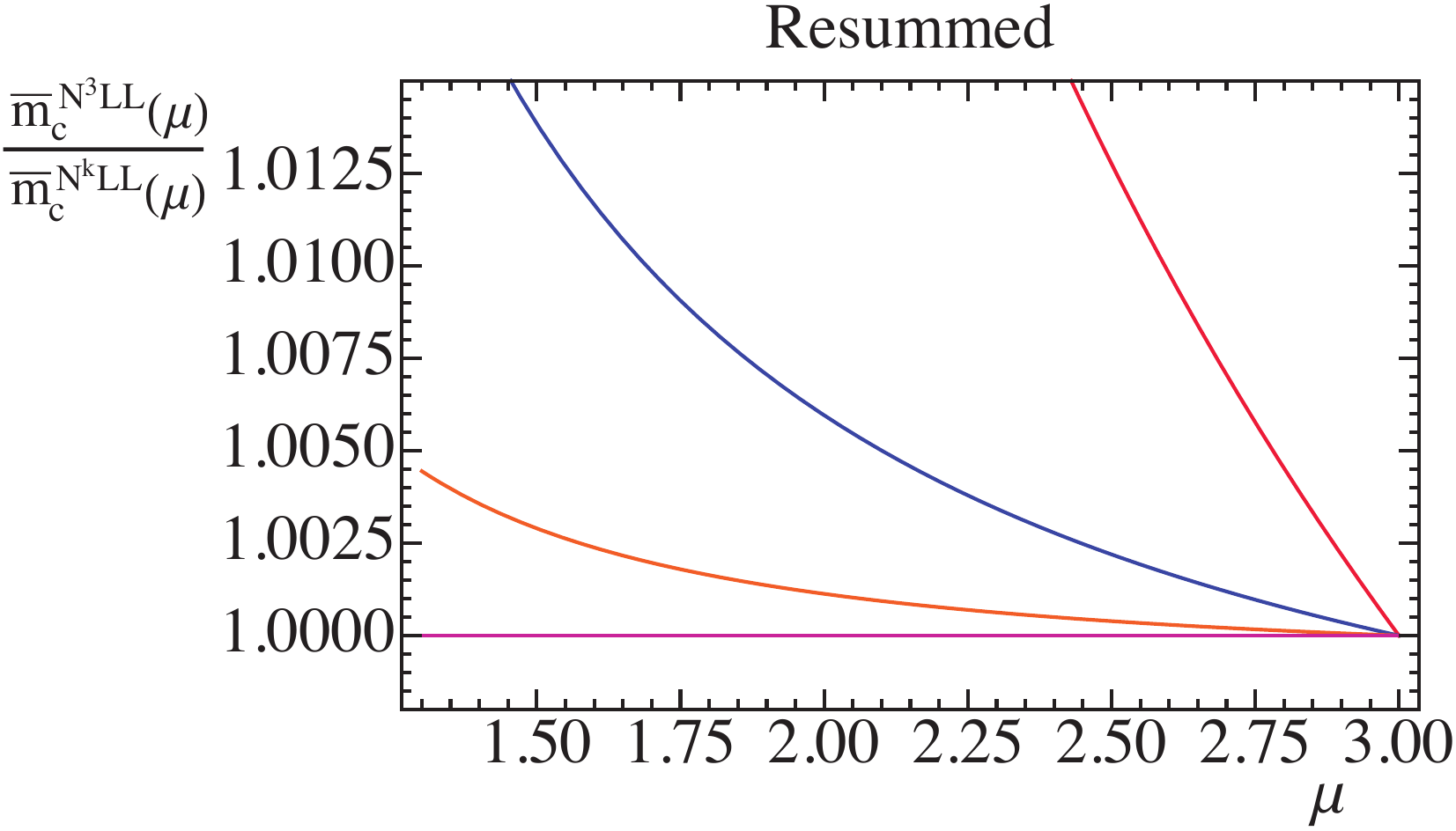}
\label{fig:mcevolutionexact}
}
\subfigure[]{
\includegraphics[width=0.485\textwidth]{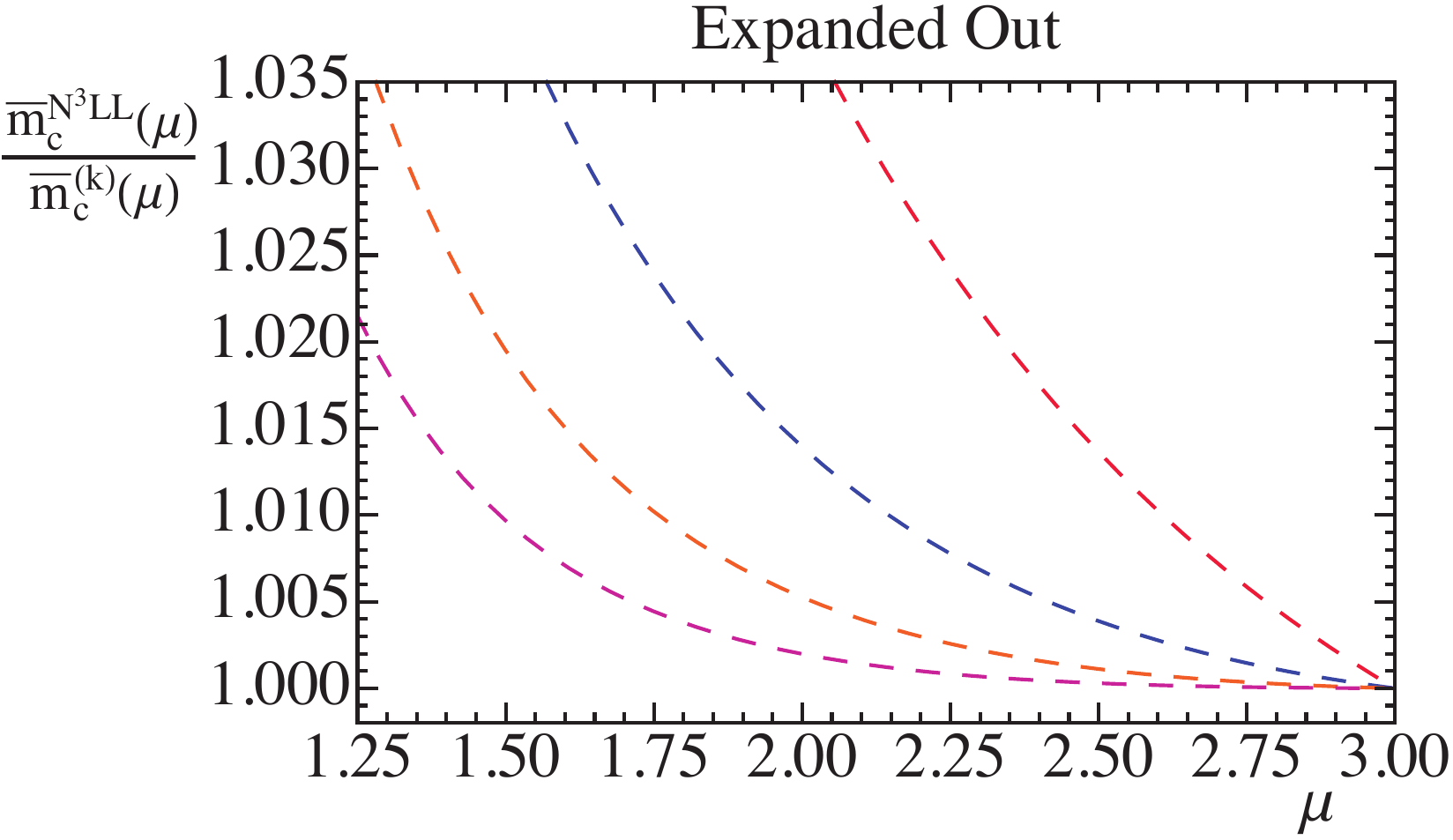}
\label{fig:mcevolutionexpanded}
}
\vspace{-0.2cm}
\caption{Results for $\overline{m}^{\rm N^3LL}_c(\mu)/\overline{m}^{\rm N^kLL}_c(\mu)$ (a) and 
$\overline{m}^{\rm N^3LL}_c(\mu)/\overline{m}^{(k)}_c(\mu)$ (b),  where $\overline{m}^{\rm N^kLL}_c$
stands for the $(k+1)$-loop running $\overline{\rm MS}$ charm mass and $\overline{m}^{(k)}_c$ is the
${\mathcal O}(\alpha_s^{(k+1)})$ fixed-order expression. \label{fig:mcevolution}}
\end{figure*}

In Figs.~\ref{fig:mcevolutionexact} and~\ref{fig:mcevolutionexpanded} an
analogous analysis has been carried out 
for the $\overline{\rm MS}$ charm quark mass. In Fig.~\ref{fig:mcevolutionexact}
$\overline m_c^{\rm N^3LL}(\mu)/\overline m_c^{\rm N^kLL}(\mu)$ is plotted for
$k=0,1,2,3$ using the exact numerical solutions of the \mbox{$(k+1)$-loop}
renormalization group equations and $\overline m_c(\mu=3~\mbox{GeV})$
as the respective reference value.\footnote{The numerical value of $\overline m_c(\mu=3~\mbox{GeV})$
is actually irrelevant, since the running involves $\overline m_c$ only in a
linear way and exactly cancels in the ratio.}
Compared to the  Fig.~\ref{fig:alphasevolution} we
observe a very similar convergence.
In  Fig.~\ref{fig:mcevolutionexpanded},
finally, we show $\overline m_c^{\rm N^3LL}(\mu)/\overline m_c^{(m)}(\mu)$,
where $\overline m_c^{(m)}(\mu)$ is the ${\cal O}(\alpha_s^m)$ fixed-order
expression for $\overline m_c(\mu)$ using 
$\alpha_s(3~\mbox{GeV})=0.2535$ as the expansion parameter. Again,
the convergence towards the exact N${}^3$LL evolved result is very similar to the
corresponding results for the strong coupling, and we find again no evidence for 
perturbative instabilities. Of course the corrections are somewhat larger when
the fixed-order expansion is employed. We therefore conclude that perturbative
evolution and renormalization scale variations for the $\overline{\mbox{MS}}$
charm quark mass can be safely used down to scales
above $\overline m_c\sim 1.3$~GeV. One should of course mention that the lines in
Fig.~\ref{fig:mcevolutionexpanded} also give an indication about the expected
size of scale variations depending on the range of the variations. Scales above
$2$~GeV can lead to sub-MeV variations, while scales down to the charm mass will
result in percent precision (i.e. ${\cal O}(10~\mbox{MeV})$). 

\subsection{Perturbative Uncertainties in the $\overline{\rm MS}$ Charm Mass}
\label{subsectionmcerror}

In this section we discuss in detail the perturbative series for the
determination of the $\overline{\rm MS}$ charm mass $\overline m_c$ and how to
set up an adequate scale variation to estimate the perturbative uncertainty. 
In the previous subsections we have discussed four different ways to carry out
the perturbative expansion and we presented the corresponding order-by-order
analytic expressions. As described there, we can determine at each order of the
perturbative expansion for the moments $M_n^{\rm th}$ a value for $\overline
m_c(\mu_m)$ which also has a residual dependence on $\mu_\alpha$, the
renormalization scale used for $\alpha_s$. To compare the different mass
determinations we then evolve $\overline m_c(\mu_m)$ to obtain $\overline
m_c(\overline m_c)$ using the 4-loop renormalization group equations for the mass
and the strong coupling~\cite{Tarasov:1980au, Larin:1993tp,
  vanRitbergen:1997va,Chetyrkin:1997dh, Vermaseren:1997fq}.\footnote{
For the discussions in this section
we use $\alpha_s^{(n_f=5)}(m_Z)=0.118$
($\alpha_s^{(n_f=4)}(4.2~\mbox{GeV})=0.2245$) as an input using five-to-four
flavor matching at $4.2$~GeV. For the first moment employed for the charm mass
fits we use $M_1=0.2138\,{\rm GeV}^{-1}$.
}
The obtained value of  $\overline m_c(\overline m_c)$ thus has a residual dependence on
the scales $\mu_m$ and $\mu_\alpha$, on the order of perturbation theory and on
the expansion method.\footnote{Of course the extracted mass depends on the moment considered as well. 
Since in our analysis we focus on the first moment only, for simplicity we drop that label.}
For the results we can therefore use the notation
\begin{eqnarray}
\label{mcnotation1} &&
\overline m_c(\overline m_c)[\mu_m,\mu_\alpha]^{i,n}
\,,
\end{eqnarray}
where $n=0,1,2,3$ indicates perturbation theory at ${\cal O}(\alpha_s^n)$ and 
\begin{eqnarray}
\label{methodnotation1}
i & = & \left\{
\begin{array}[c]{ll}
a \qquad & \mbox{(fixed-order expansion),} \\
b & \mbox{(linearized expansion),} \\
c & \mbox{(iterative expansion),} \\
d & \mbox{(contour improved expansion).}
\end{array}\right.
\end{eqnarray}

\begin{figure}
\includegraphics[width=14cm]{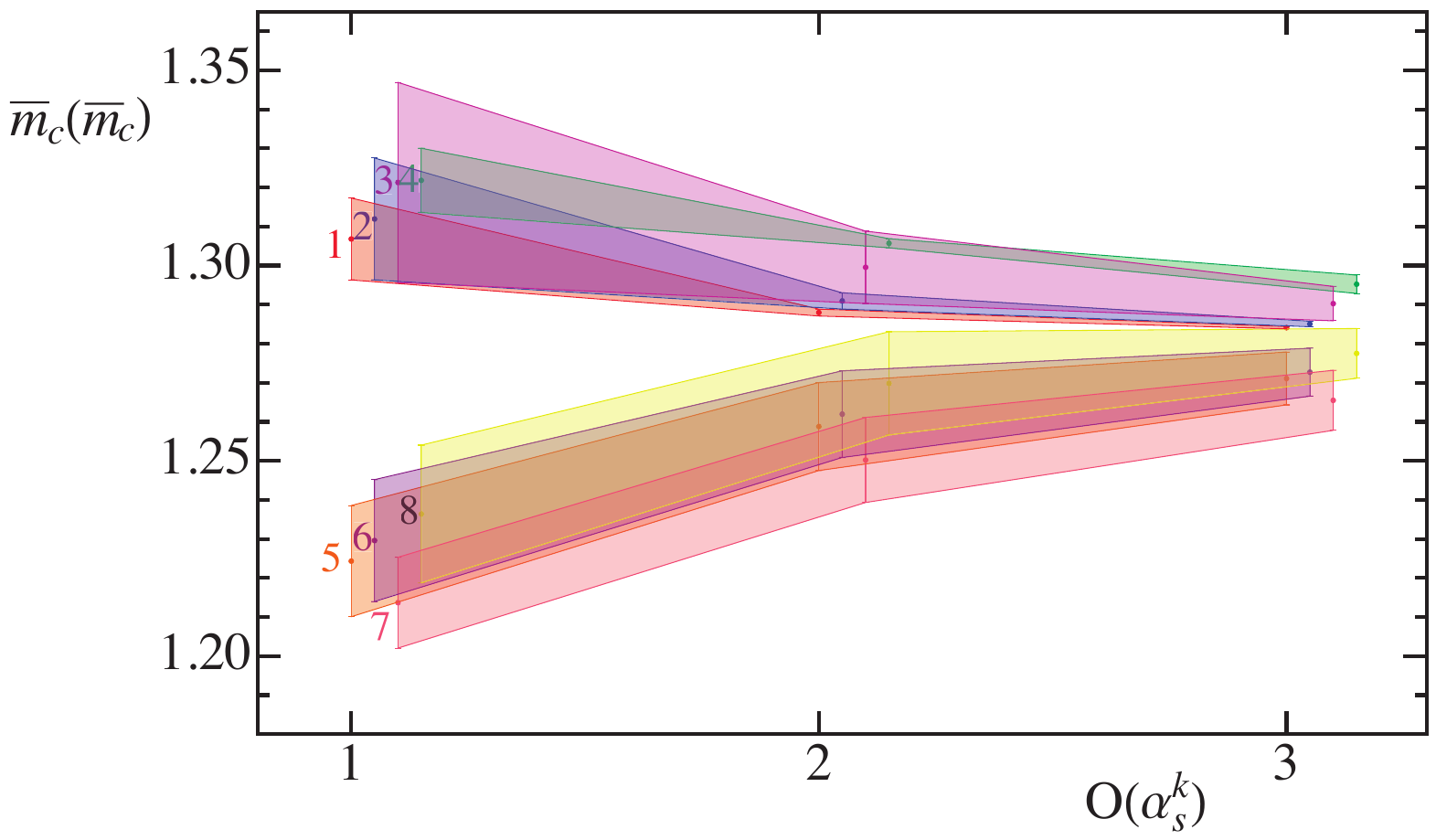}
\caption{Results for $\overline{m}_c(\overline{m}_c)$ at various orders, for methods a (graphs 1 and 5), b (2,6), c (3,7), and d (4,8), 
setting $\mu_\alpha=\mu_m$ (graphs 1-4) and setting $\mu_m=\overline{m}_c(\overline{m}_c)$ (5-8).
The shaded regions arise from the variation $2\,{\rm GeV}\le\mu_\alpha\le4\,{\rm
  GeV}$. \label{fig:trumpet1}} 
\end{figure}

\begin{figure}
\includegraphics[width=13cm]{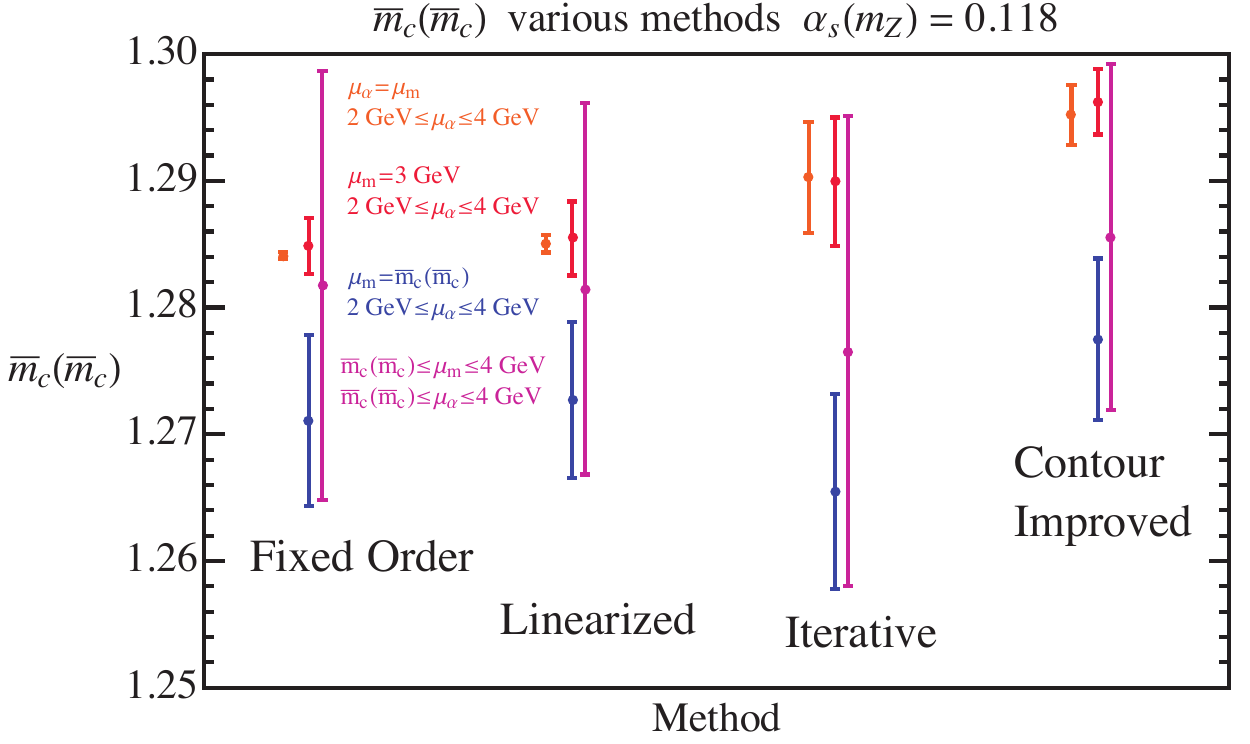}
\caption{Estimates of the perturbative error at ${\mathcal O}(\alpha_s^3)$.
We show the correlated $\mu_m=\mu_\alpha$ variation,
(orange),  setting $\mu_m=\overline{m}_c(\overline{m}_c)$ and setting $\mu_m=3~$GeV,
(blue and red, respectively), and the double scale variation (magenta).\label{fig:trumpet2}} 
\end{figure}

To initiate the discussion, we show in Fig.~\ref{fig:trumpet1} results for
$\overline m_c(\overline m_c)$ at ${\cal O}(\alpha_s^n)$ for expansions a--d
using $\mu_m=\mu_\alpha$ (upper four graphs) and using $\mu_m=\overline
m_c(\overline m_c)$ (lower four graphs). For each method and order we have
displayed the range of $\overline m_c(\overline m_c)$ values for a variation of 
$2~\mbox{GeV}\le \mu_\alpha \le 4~\mbox{GeV}$, which corresponds to the scale
variation employed in Refs.~\cite{Chetyrkin:2006xg,Boughezal:2006px,Chetyrkin:2009fv,Kuhn:2007vp}. 
Their analysis used the fixed-order expansion with the setting
$\mu_m=\mu_\alpha$ and is represented by graph~1. We make several observations: 
\begin{itemize}
\item[(i)] Choosing $\mu_m$ and $\mu_\alpha$ both larger than $2$~GeV makes the
  $\overline m_c(\overline m_c)$ value decrease with the order of perturbation
  theory.
\item[(ii)] Choosing $\mu_m$ smaller than $1.5$~GeV and $\mu_\alpha$ larger than $2$~GeV
  makes the $\overline m_c(\overline m_c)$ value increase with the order of
  perturbation theory. 
\item[(iii)] For most choices of the scale setting and the expansion method the
  spread of the $\overline m_c(\overline m_c)$ values from the variation of
  $\mu_\alpha$ does not decrease in any substantial way with the order. However,
  viewing all methods and scale setting choices collectively a very good
  convergence is observed.
\end{itemize}
We have checked that these statements apply also in general beyond the specific cases displayed
in Fig.~\ref{fig:trumpet1}. 

Quite conspicuous results are obtained for the scale choice $\mu_m=\mu_\alpha$
for the fixed-order (graph 1) and linearized 
expansions (graph 2). Here, extremely small
variations in $\overline m_c(\overline m_c)$ are obtained. They amount to 
$1.8$~MeV ($4$~MeV) and $0.6$~MeV ($1.4$~MeV) at order $\alpha_s^2$ and
$\alpha_s^3$, respectively, for the fixed-order expansions (linearized
expansions). We note that our scale variation for the fixed-order expansion at
${\cal O}(\alpha_s^3)$ is consistent with the corresponding numbers quoted in
Ref.~\cite{Chetyrkin:2006xg,Boughezal:2006px}, where the ${\cal O}(\alpha_s^3)$
corrections to the first moment were computed,\footnote{
We are grateful to Thomas Schutzmeier for confirming agreement of the results of
our ${\cal O}(\alpha_s^3)$ fixed-order code with theirs of
Ref.~\cite{Boughezal:2006px}. }
but differs from the scale
variations given in Refs.~\cite{Chetyrkin:2009fv,Kuhn:2007vp} which also
quoted numerical results from the fixed-order expansion. 
Interestingly the ${\cal O}(\alpha_s^2)$ and ${\cal O}(\alpha_s^3)$
variations we find do not overlap and the ${\cal O}(\alpha_s^3)$ ranges appear
to be highly inconsistent with the ${\cal O}(\alpha_s^3)$ results from the
iterative method (graph 3). A visual display of scale variations obtained from the four
expansion methods at ${\cal O}(\alpha_s^3)$  with different types of variation
methods is given in Fig.~\ref{fig:trumpet2}.

\begin{figure*}[t!]
\subfigure[]
{
\includegraphics[width=0.48\textwidth]{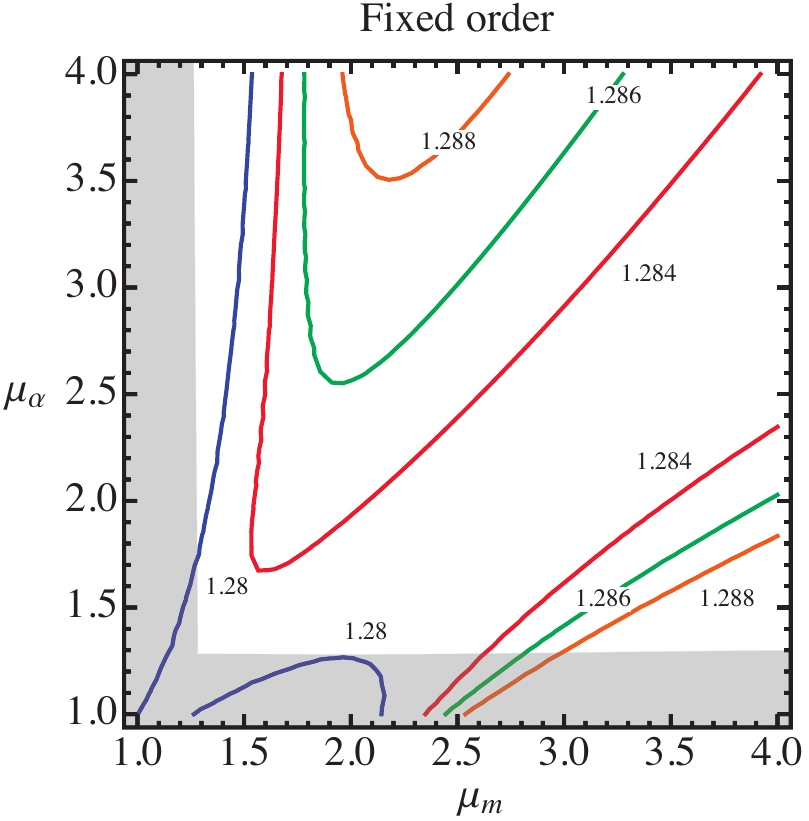}
\label{fig:mccontour1fixed}
}
\subfigure[]{
\includegraphics[width=0.48\textwidth]{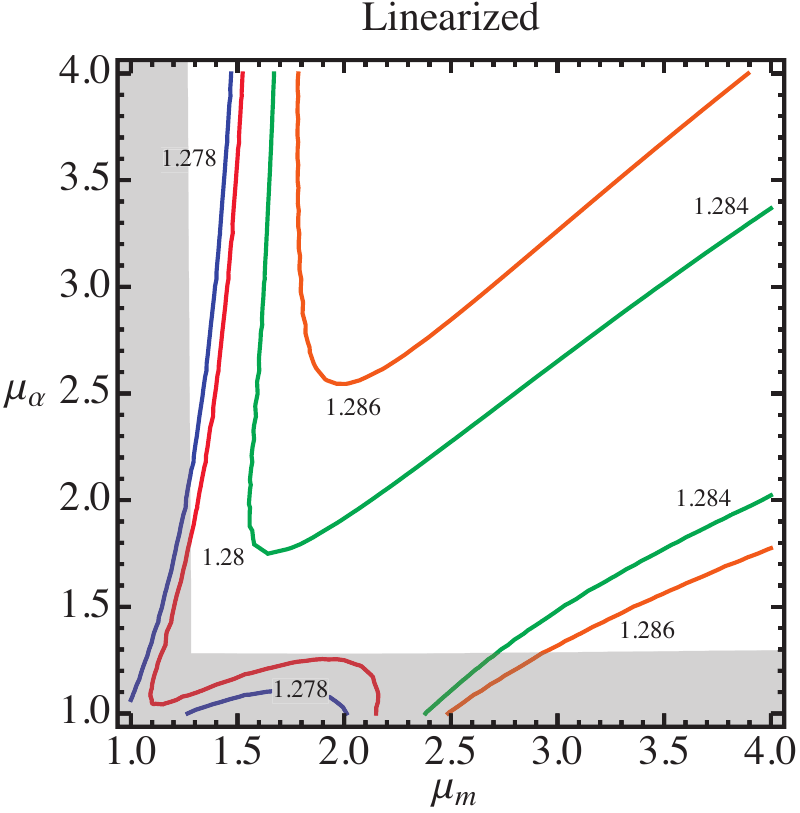}
\label{fig:mccontour1expanded}
}
\subfigure[]{
\includegraphics[width=0.48\textwidth]{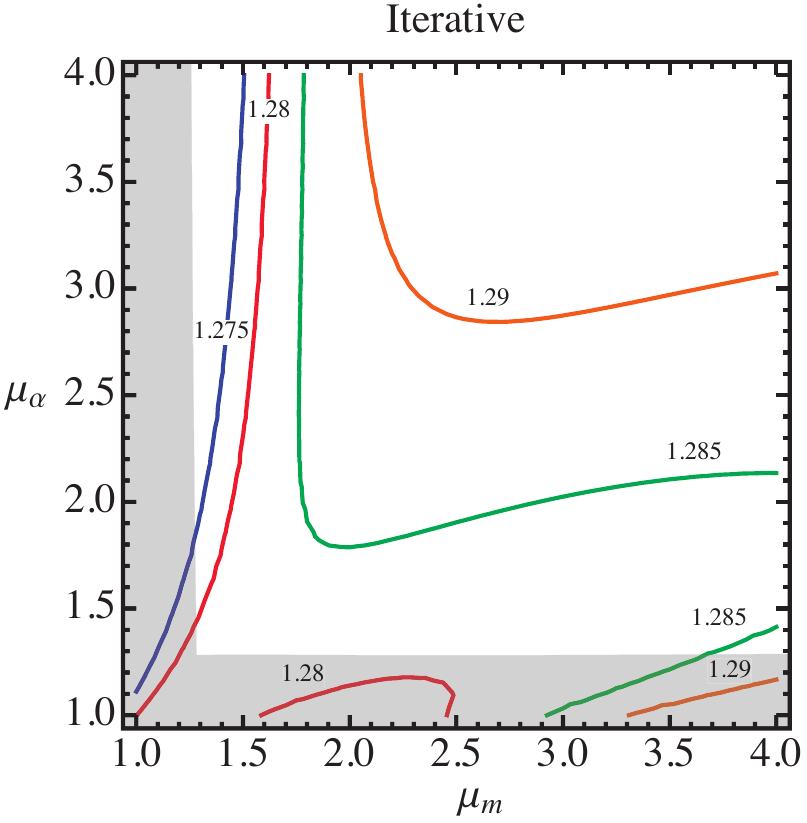}
\label{fig:mccontour1iterative}
}
\subfigure[]{
\includegraphics[width=0.48\textwidth]{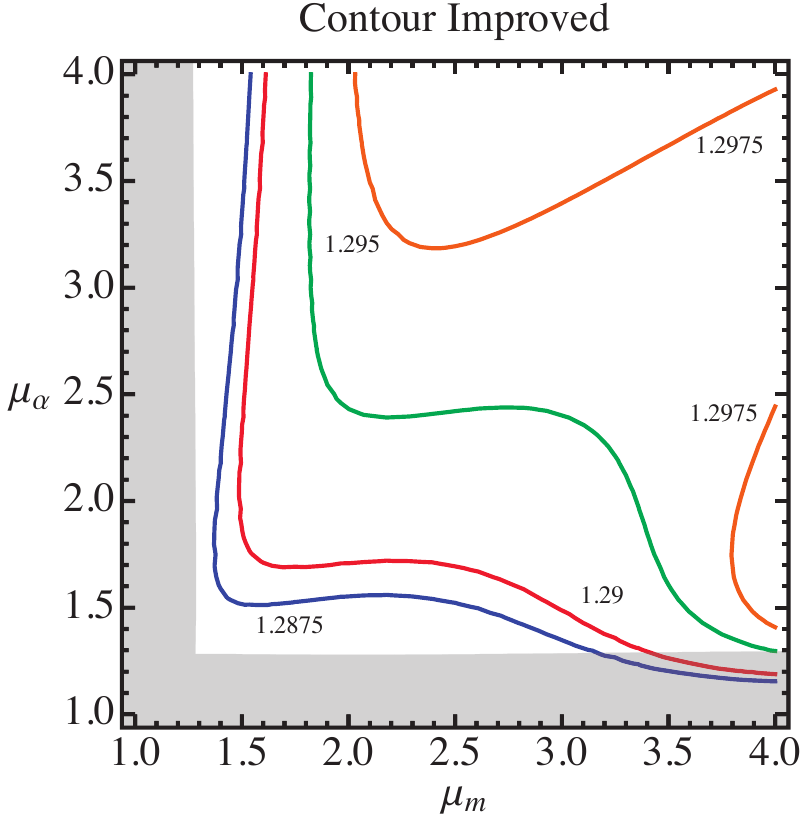}
\label{fig:mccontour1improved}
}
\caption{Contour plots for $\overline{m}_c(\overline{m}_c)$ as a function of
  $\mu_\alpha$ and $\mu_m$ at ${\mathcal O}(\alpha_s^3)$, for methods
  (a)--(d). The shaded areas represent regions with
$\mu_m,\mu_\alpha < \overline{m}_c(\overline{m}_c)$, and are excluded of our analysis.
\label{fig:mccontour1}} 
\end{figure*}

An illustrative way to demonstrate how a small scale variation can arise is
given in Fig.~\ref{fig:mccontour1fixed} - \ref{fig:mccontour1improved}. For all
four expansion methods contour curves of constant $\overline m_c(\overline m_c)$
are displayed as
a function of (the residual dependence on) $\mu_m$ and $\mu_\alpha$. For the
fixed-order (a) and the linearized expansions (b) we see that there are contour
lines closely along the diagonal $\mu_m=\mu_\alpha$. For the fixed-order
expansion (a) this feature is almost exact and thus explains the extremely small
scale-dependence seen in graph~1 of Fig.~\ref{fig:trumpet1}. For the linearized
expansion (b) this feature is somewhat less exact and reflected in the slightly
larger scale variation seen in graph~2 of Fig.~\ref{fig:trumpet1}. On the other
hand, the contour lines of the iterative (c) and contour improved (d) expansions
have a large angle with respect to the diagonal $\mu_m=\mu_\alpha$ leading to
the much larger 
scale variations of $9$ and $5$~MeV, respectively, at ${\cal O}(\alpha_s^3)$
visible in graphs~3 and~4 of Fig.~\ref{fig:trumpet1}. 
The contour plots shown in Fig.~\ref{fig:mccontour1}
also show the variation along the line $\mu_m$ or $\mu_\alpha=\overline
m_c(\overline m_c)$ 
(border of gray shaded areas). Here, the contour lines are relatively dense leading to scale
variation of around $15$~MeV at ${\cal O}(\alpha_s^3)$. 

Overall, we draw the following conclusions:
\begin{itemize}
\item[(1)] The small scale variations observed for the fixed-order and
  linearized expansions for $\mu_m=\mu_\alpha$ result from strong cancellations
  of the individual $\mu_m$ and $\mu_\alpha$ dependences that arise for this
  correlation. 
\item[(2)] Other correlations between $\mu_m$ and $\mu_\alpha$ that do not
  generate large logarithms do not lead to such cancellations. One therefore has
  to consider the small scale variations observed for $\mu_m=\mu_\alpha$ in the
  fixed-order and linearized expansions as accidental. 
\item[(3)] For an adequate estimate of perturbative uncertainties specific
  correlations 
  between $\mu_m$ and $\mu_\alpha$ that are along contour lines of constant
  $\overline m_c(\overline m_c)$ have to be avoided. Moreover, adequate
  independent variations of $\mu_m$ and $\mu_\alpha$ should not
  induce large logarithms. 
\end{itemize}
As the outcome of this discussion we adopt for our charm mass analysis an
independent and uncorrelated variation of $\mu_m$ and $\mu_\alpha$ in the range
\begin{eqnarray}
\label{muammvariation}
\overline m_c(\overline m_c) & \le & \mu_m,\mu_\alpha \, \le \, 4~\mbox{GeV}\,,
\end{eqnarray}
in order to estimate perturbative uncertainties. The excluded region
$\mu_m,\mu_\alpha<\overline{m}_c(\overline{m}_c)$ in the 
\mbox{$\mu_m$-$\mu_\alpha$} plane in the contour plots of Fig.~\ref{fig:mccontour1} is
indicated by the gray shaded areas.
This two-dimensional variation
avoids accidental cancellations from correlated \mbox{$\mu_m$-$\mu_\alpha$} variations,
large logarithms involving ratios of the scales $\overline m_c(\mu_m)$,
$\mu_m$, $\mu_\alpha$ and remains well in the validity ranges for the
perturbative renormalization group evolution of the $\overline{\rm MS}$ charm
quark mass and $\alpha_s$ (see Sec.~\ref{subsectionrunning}). The range of
Eq.~(\ref{muammvariation}) is also consistent with a scale variation 
$\mu\sim 2\,\overline m_c\times (1/2,1,2)$ one might consider as the standard
choice with respect to the to particle threshold located at $\sqrt{s}=2\,{\overline m}_c$.
As a comparison the range $\mu=(3\pm 1)$~GeV corresponds to $\mu=2\,{\overline m}_c(0.8,1.2,1.6)$.
We emphasize that the lower boundary $\overline m_c(\overline m_c)\sim
1.25$~GeV is also reasonable as it represents the common flavor matching scale  
where gauge coupling evolution remains smooth up to NLL order. We do not
see any evidence for perturbation theory for the first moment $M_1$ being
unstable at the charm mass scale, in contrast to claims made in
Ref.~\cite{Kuhn:2007vp}.  

\begin{figure*}[t]
\subfigure[]{
\includegraphics[width=0.485\textwidth]{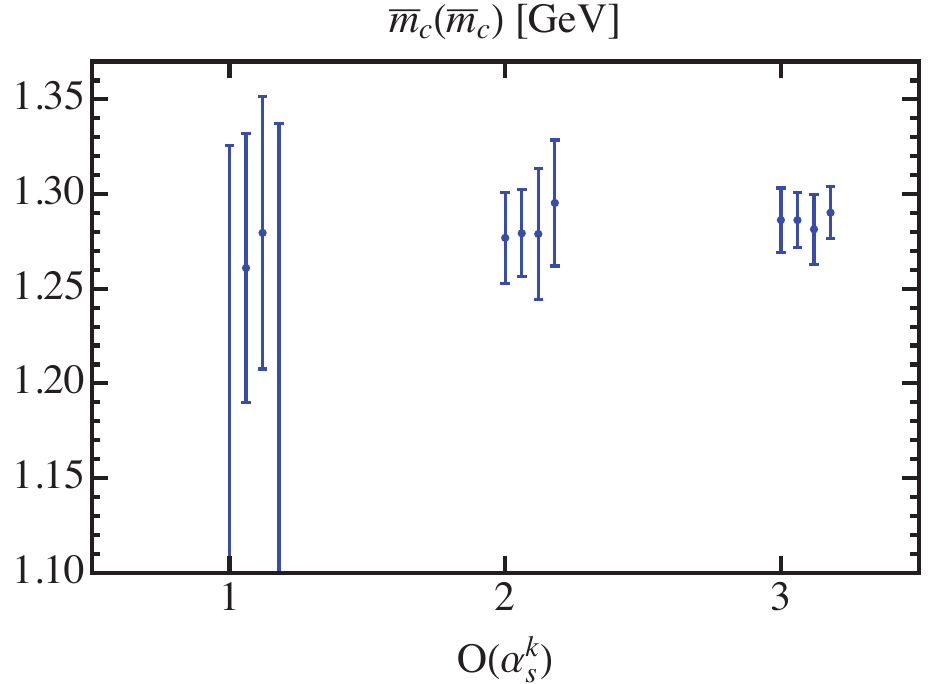}
\label{fig:mmcerror1}
}
\subfigure[]{
\includegraphics[width=0.485\textwidth]{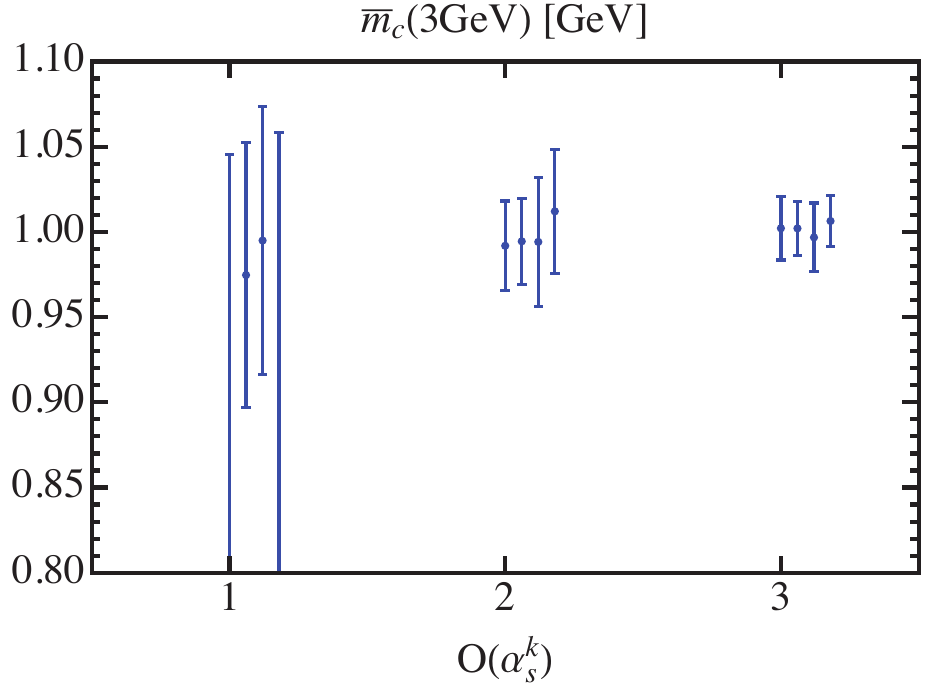}
\label{fig:m3cerror1}
}
\vspace{-0.2cm}
\caption{Results for $\overline{m}_c(\overline{m}_c)$ (a) and $\overline{m}_c(3~{\rm GeV})$ (b), for methods a--d, at orders
${\mathcal O}(\alpha_s^{1,2,3})$. At ${\mathcal O}(\alpha_s^{1})$ the error bars for fixed-order
and contour improved methods extend down to
$0.7~{\rm GeV}$ for $\overline{m}_c(\overline{m}_c)$ and $0.4~{\rm GeV}$ for
$\overline{m}_c(3~{\rm GeV})$.\label{fig:mcerror1}}
\end{figure*}

In Fig.~\ref{fig:mmcerror1} we show the ranges of $\overline m_c(\overline m_c)$
at ${\cal O}(\alpha_s^{1,2,3})$ for the four expansion methods employing the
scale variations of Eq.~(\ref{muammvariation}). We see that all four expansion
methods now lead to equivalent results. The variations are also compatible with
the overall variations shown in Fig.~\ref{fig:trumpet1}. At ${\cal
  O}(\alpha_s^3)$ we obtain a scale variation for $\overline m_c(\overline m_c)$
of around $20$~MeV. This is
an order of magnitude larger than the perturbative uncertainties quoted in
Refs.~\cite{Chetyrkin:2006xg,Boughezal:2006px,Chetyrkin:2009fv,Kuhn:2007vp}. 
In Fig.~\ref{fig:m3cerror1} we have displayed
the corresponding results for $\overline m_c(3~\mbox{GeV})$.  At ${\cal
  O}(\alpha_s^3)$ they also exhibit a scale variation of around $20$~MeV.

\section{Experimental Data}
\label{sectiondata}

\subsection{Data Collections}
\label{subsectioncollections}

\vskip 1mm
\noindent
{\bf Narrow resonances}\\[1mm]
Below the open charm threshold there are the $J/\psi$ and $\psi^\prime$ narrow
charmonium resonances. Their masses, and electronic widths are taken from the
PDG~\cite{Beringer:1900zz}\,\footnote{We actually use the most up to date
values corresponding to: J. Beringer et al. (Particle Data Group), Phys. Rev. D86, 010001 (2012)
and 2013 partial update for the 2014 edition.}
and are collected in Tab.~\ref{tabpsidata}  together with the value
of the pole-subtracted effective electromagnetic coupling at their masses. The
total widths are 
not relevant since we use the narrow width approximation for their contributions to
the moments. The uncertainty for the contribution to the moments coming from the
masses and the effective electromagnetic coupling can be neglected.

\begin{table}[t!]\begin{center}
\begin{tabular}{|c|cc|}
\hline 
 & $J/\psi$ & $\psi^\prime$\tabularnewline
\hline
$M$ (GeV) & $3.096916(11)$ & $3.686109(13)$\tabularnewline
$\Gamma_{ee}$ (keV) & $5.55(14)$ & $2.37(4)$\tabularnewline
$(\alpha/\alpha(M))^{2}$ & $0.9580(3)$ & $0.9557(6)$\tabularnewline
\hline
\end{tabular}\end{center}
\caption{
Masses and electronic widths \cite{Beringer:1900zz} of the narrow
charmonium resonances with total uncertainties and effective electromagnetic coupling, where
$\alpha=1/137.035999084(51)$ is the fine structure constant, 
and $\alpha(M)$ stands for the pole-subtracted effective electromagnetic
coupling at the scale $M$~\cite{Teubnerprivate}. We have also given the
uncertainties in $\alpha(M)$ due to its hadronic contributions. Compared to the
uncertainties of the widths, the uncertainties on $\alpha(M)$ and the masses are negligible.
 \label{tabpsidata}}
\end{table}

\begin{figure*}[t!]
\subfigure[]{
\includegraphics[width=0.48\textwidth]{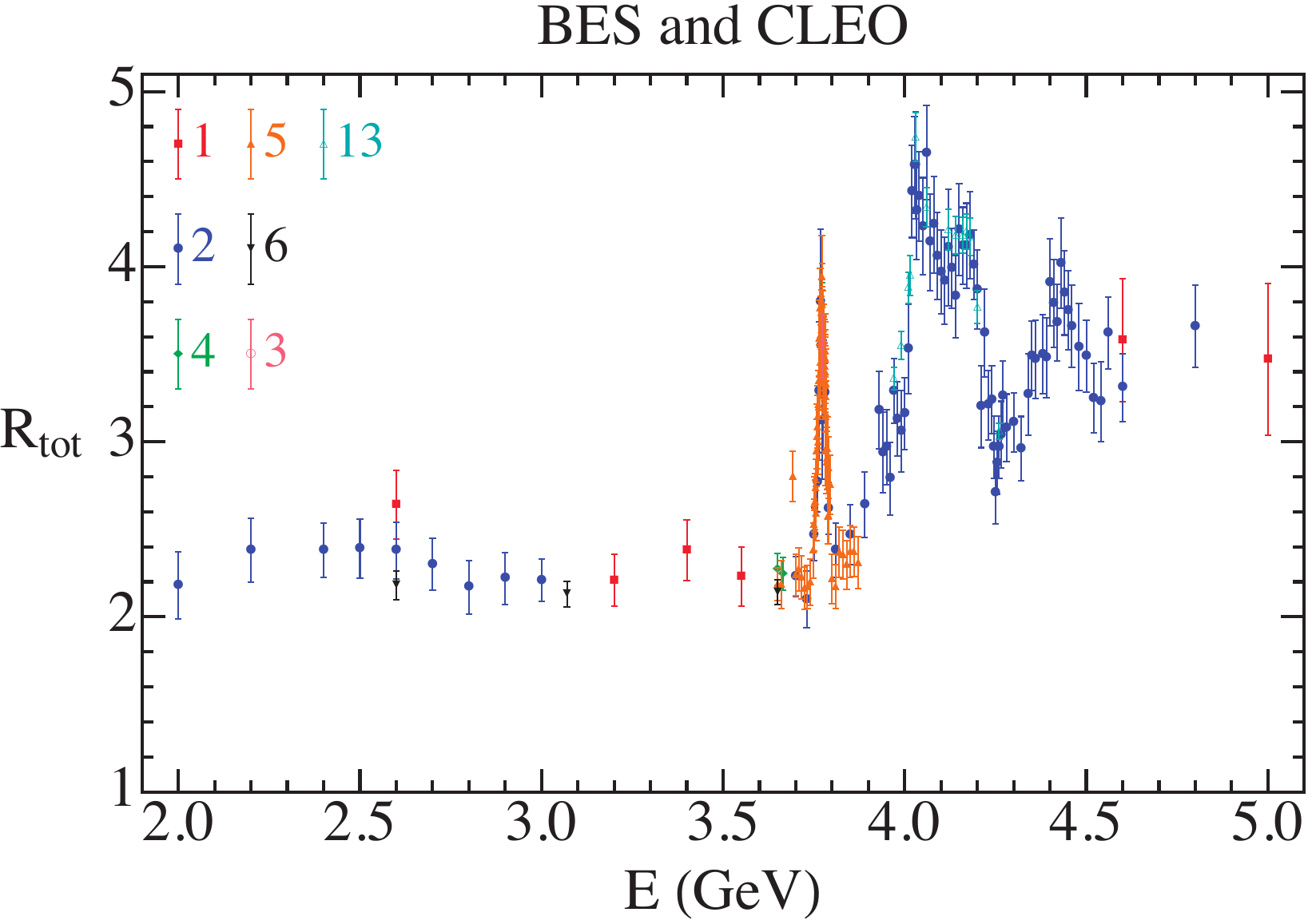}
\label{data-nocharm}
}
\subfigure[]{
\includegraphics[width=0.48\textwidth]{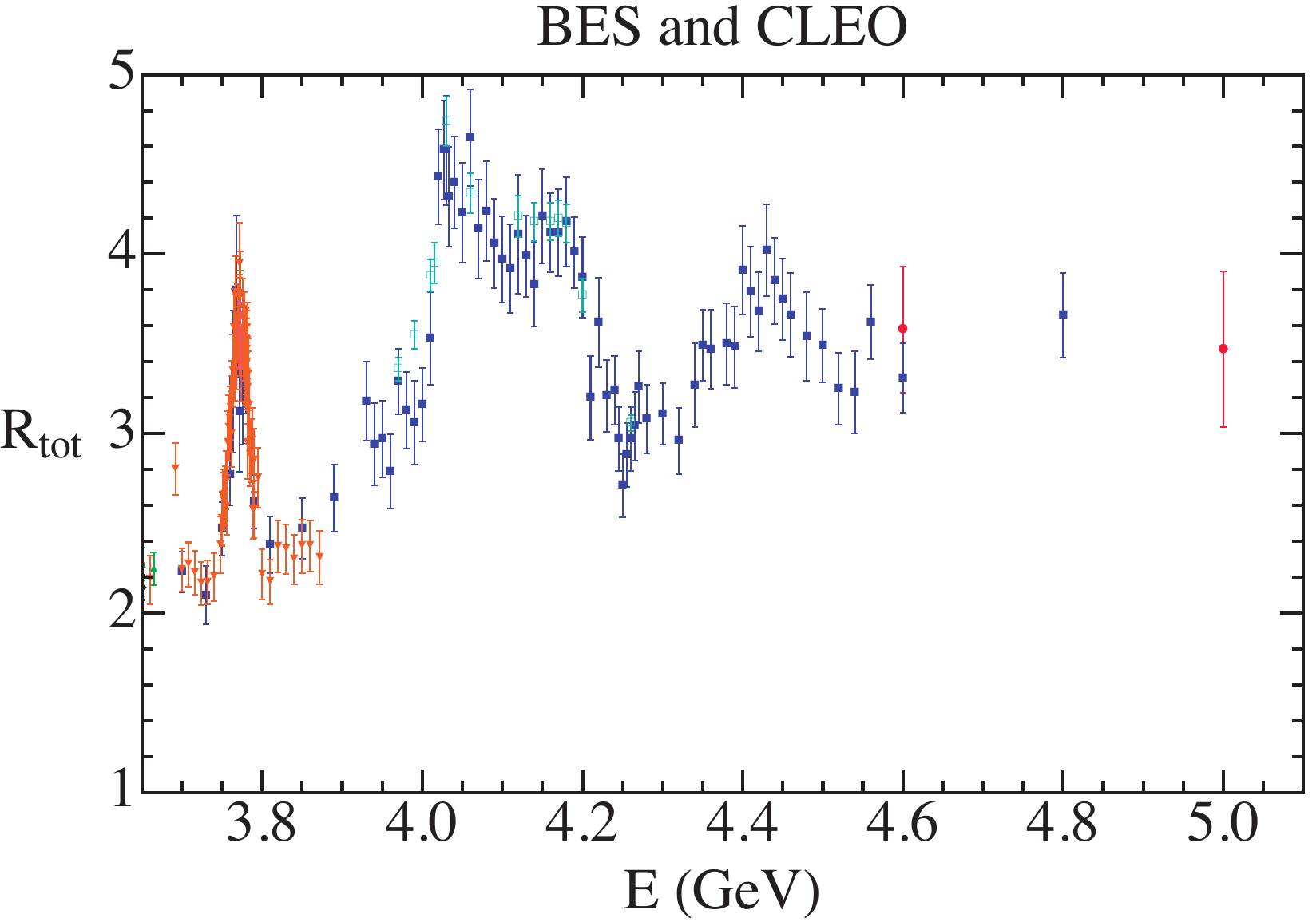}
\label{datathreshold}
}
\subfigure[]{
\includegraphics[width=0.48\textwidth]{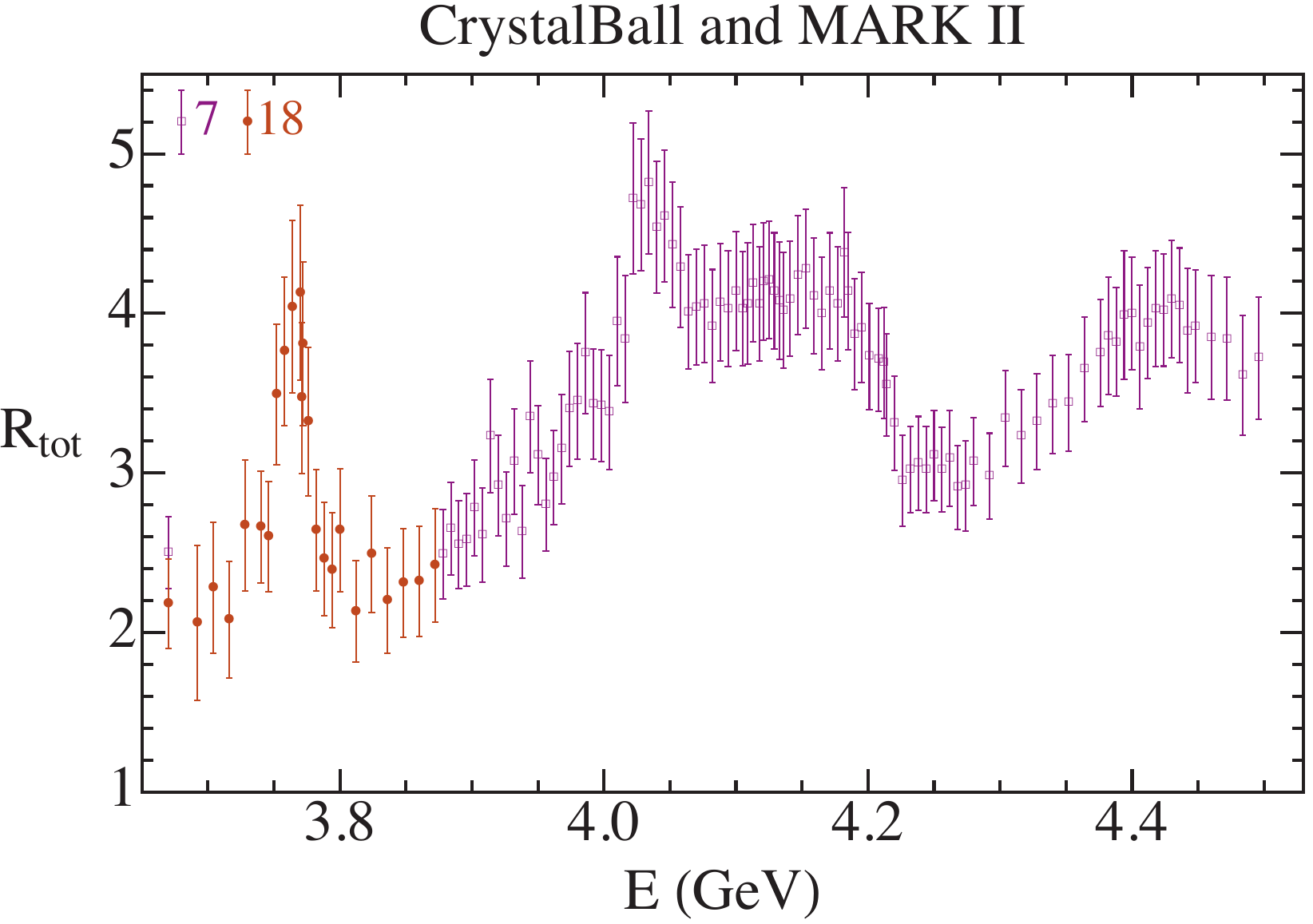}
\label{dataCB}
}
\subfigure[]{
\includegraphics[width=0.48\textwidth]{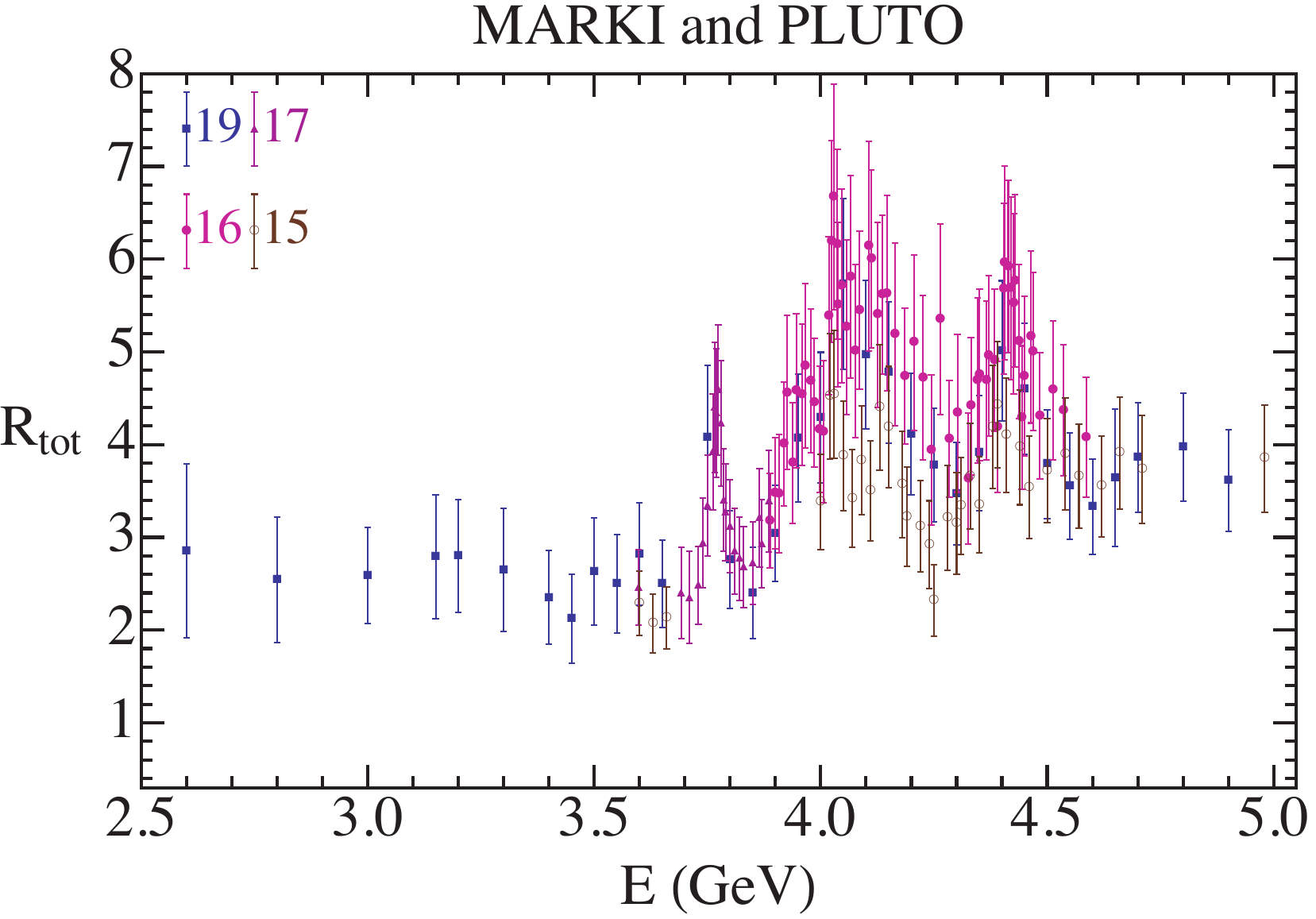}
\label{dataMARK}
}
\subfigure[]{
\includegraphics[width=0.48\textwidth]{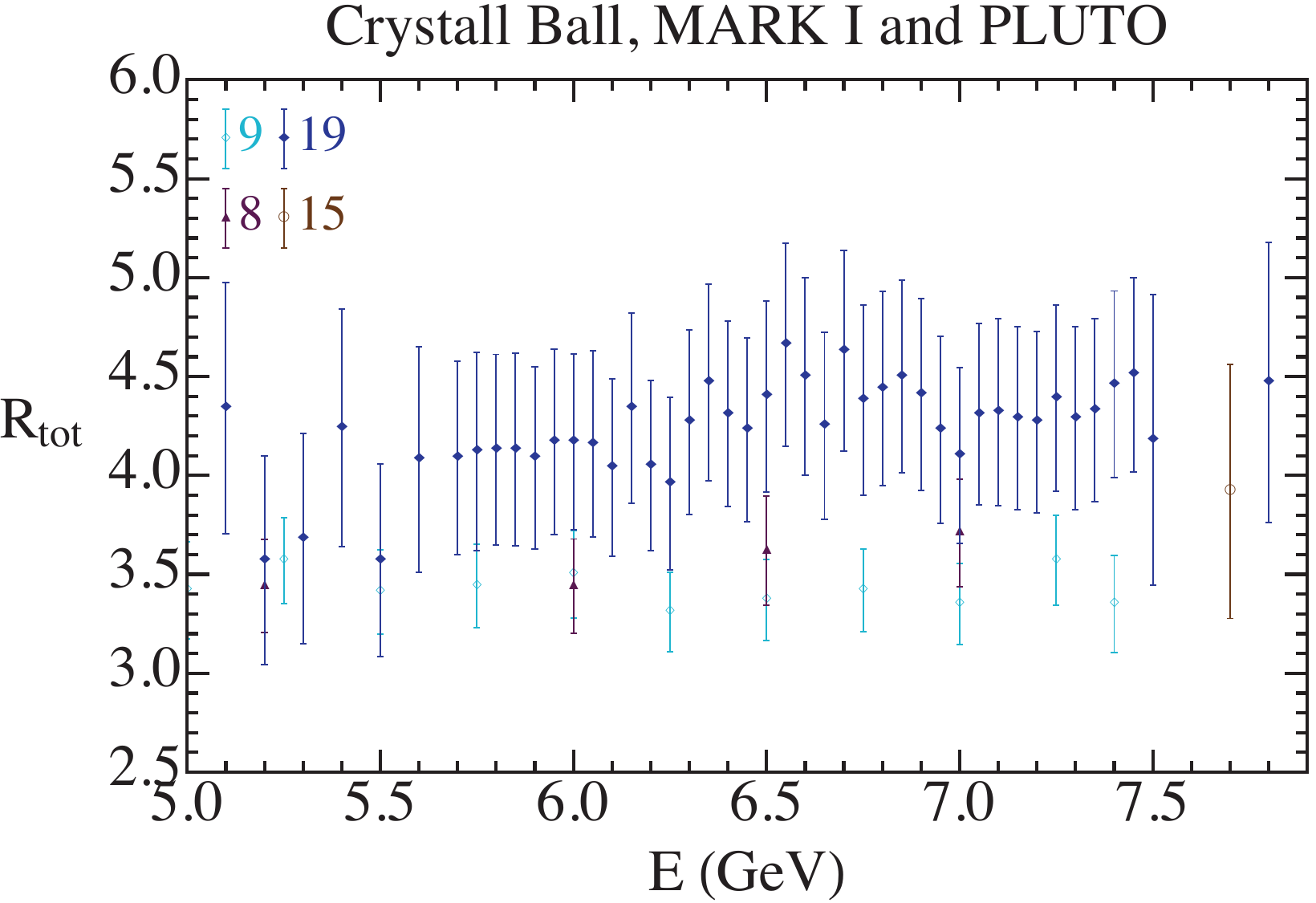}
\label{datagap}
}
\subfigure[]{
\includegraphics[width=0.48\textwidth]{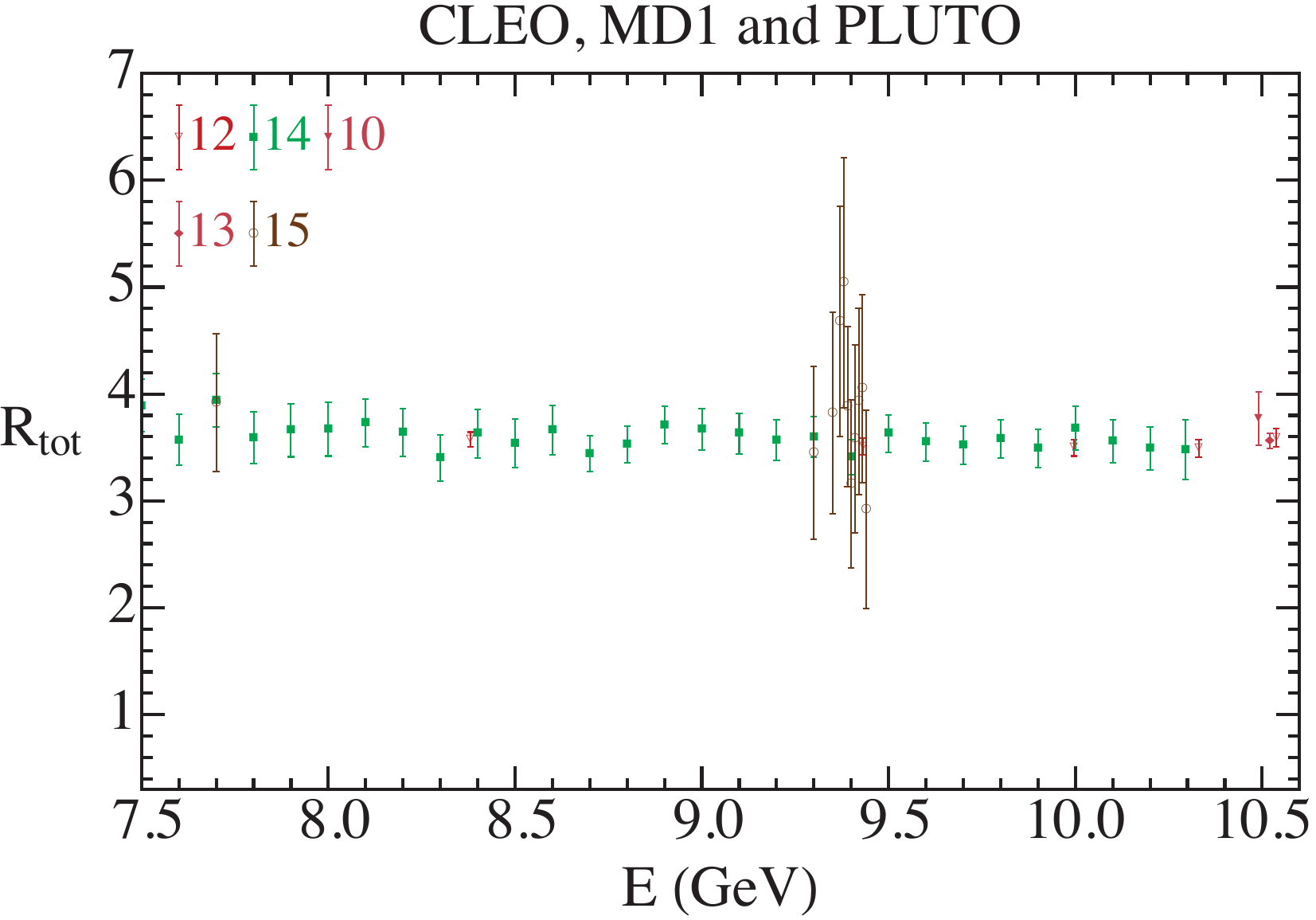}
\label{datacontinuum}
}
\caption{Experimental data. In the two top figures we show BES and CLEO data in
  the non-charm, low charm and threshold regions (a) and low charm region and
  threshold region (b). The two pictures in the middle show less precise
  data in the threshold region: CrystalBall and MARKII (c) and MARKI and
  PLUTO (d). The two pictures on the bottom show data in the continuum
  region: below $8$~GeV data from CrystalBall, MARKI and PLUTO (e) and
  above $8$~GeV data from CLEO, MD1 and PLUTO (f). 
\label{figdatacompilation}}
\end{figure*}

\vskip 2mm
\noindent
{\bf Threshold and data continuum region}\\[1mm]
The open charm threshold is located at $\sqrt{s}=3.73$~GeV. We call the energies
from just below the threshold and up to $5$~GeV the threshold region, and the
region between $5$~GeV and $10.538$~GeV, where the production rate is dominated
by multiparticle final states the data continuum region. In these regions quite
a variety of measurements of the total hadronic cross section exist from 
BES~\cite{Bai:1999pk,Bai:2001ct,Ablikim:2004ck,Ablikim:2006aj,Ablikim:2006mb,:2009jsa},
CrystalBall~\cite{Osterheld:1986hw,Edwards:1990pc},
CLEO~\cite{Ammar:1997sk,Besson:1984bd,:2007qwa,CroninHennessy:2008yi},
MD1~\cite{Blinov:1993fw},
PLUTO~\cite{Criegee:1981qx}, and
MARKI and II~\cite{Siegrist:1976br,Rapidis:1977cv,Abrams:1979cx,Siegrist:1981zp}.
Taken together, the entire energy region up to $10.538$~GeV is densely covered
with total cross section measurements from these 19 data sets.\footnote{There are
  18 references quoted since Ref.~\cite{Edwards:1990pc} provides results from
  two independent runs that we treat as two different data sets.}
The measurements from BES and CLEO have the smallest uncertainties. They do,
however, not cover the region between $5$ and $7$~GeV. Here CrystalBall and
MARKI and~II have contributed measurements albeit with somewhat larger
uncertainties. The statistical and total systematical uncertainties of the
measurements can be extracted from the respective publications.
For some data sets the amount of uncorrelated and correlated systematical
uncertainties is given separately
(BES~\cite{Bai:2001ct,Ablikim:2006aj,Ablikim:2006mb},
CrystalBall~\cite{Osterheld:1986hw,Edwards:1990pc}, 
CLEO~\cite{:2007qwa},
MARKI and II~\cite{Siegrist:1981zp,Abrams:1979cx}, 
MD1~\cite{Blinov:1993fw})
while for all the other data
sets only combined systematical uncertainties are quoted. All these data sets
are shown in Figs.~\ref{figdatacompilation},
where the displayed error bars represent the (quadratically) combined
statistical and systematical uncertainties.

\begin{table}[t!]\begin{center}
\begin{tabular}{|c|ccccc|}
\hline
Numbering   & Reference &         Experiment    &  Year  &  Systematical &  Data points \tabularnewline
\hline
  1 &  \cite{Bai:1999pk} &                      BES                         &  2000   & no splitting given  &     6 \tabularnewline
  2 &  \cite{Bai:2001ct} &                        BES                         &  2002   & splitting given       &    85 \tabularnewline
  3 &  \cite{Ablikim:2004ck} &               BES                         &  2004   & only one point      &      1 \tabularnewline
  4 &  \cite{Ablikim:2006aj} &                BES                         &  2006   & splitting given      &      3\tabularnewline
  5 &  \cite{Ablikim:2006mb} &                BES                         &  2006   & splitting given    &    68\tabularnewline
  6 &  \cite{:2009jsa} &                          BES                         &  2009   & no splitting given &      3\tabularnewline
  7 &  \cite{Osterheld:1986hw} &           CrystalBall             &  1986   & splitting given     &     98\tabularnewline
  8 &  \cite{Edwards:1990pc} &              CrystalBall (Run 1) &  1990   & splitting given     &      4\tabularnewline
  9 &  \cite{Edwards:1990pc} &               CrystalBall            &  1990   & splitting given     &     11\tabularnewline
10 &  \cite{Ammar:1997sk} &                 CLEO                      & 1979    & only one point    &       1\tabularnewline
11 &  \cite{Besson:1984bd} &                 CLEO                      & 1998   & only one point     &       1\tabularnewline
12 &  \cite{:2007qwa} &                          CLEO                     &  2007   & splitting given    &        7\tabularnewline
13 &  \cite{CroninHennessy:2008yi} &     CLEO                     &  2008   & no splitting given&     13 \tabularnewline
14 &  \cite{Blinov:1993fw} &                    MD1                      &  1996   & splitting given     &    31  \tabularnewline
15 &  \cite{Criegee:1981qx} &                 PLUTO                  & 1982   & no splitting given   &     45\tabularnewline
16 &  \cite{Siegrist:1976br} &                  MARKI                   &  1976   & no splitting given &    59\tabularnewline
17 &  \cite{Rapidis:1977cv} &                  MARKI                   &  1977   & no splitting given &    21\tabularnewline
18 &  \cite{Abrams:1979cx} &                 MARKII                  &  1979    & splitting given     &    24\tabularnewline
19 &  \cite{Siegrist:1981zp} &                 MARKI                   &   1981   & splitting given     &    78\tabularnewline
\hline
\end{tabular}\end{center}
\caption{Complete data set of measurements of the total hadronic \mbox{$R$-ratio} used in
  this work. Each
  data set is assigned a number to simplify referencing in discussions and
  figures of this work. Also given is information on the name of the
  experimental collaboration, the year of the publication, on whether the
  splitting of the systematical error in correlated
and uncorrelated contributions is provided, and on the number of data points.
\label{tab:datasets}}
\end{table}

Interestingly, none of the previous charm mass analyses, to the best of our
knowledge, ever used the complete set of available data. As examples, 
Bodenstein~et~al. \cite{Bodenstein:2010qx} used data sets
\cite{Bai:1999pk,Bai:2001ct,Ablikim:2006mb,CroninHennessy:2008yi} from BES and CLEO.
Jamin and Hoang~\cite{Hoang:2004xm} used the data sets of Refs.~\cite{Bai:2001ct},
\cite{Blinov:1993fw} and \cite{Ammar:1997sk} from  
BES, MD1 and CLEO, covering the regions $2~\mbox{GeV} \leq E \leq 4.8~\mbox{GeV}$
and $6.964~\mbox{GeV} \leq E \leq 10.538~\mbox{GeV}$. 
Boughezal~et~al. \cite{Boughezal:2006px}, Kuhn~et~al. \cite{Kuhn:2001dm}, and Narison 
\cite{Narison:2010cg} use only one data set from BES \cite{Bai:2001ct}. 
Kuhn~et~al.~\cite{Chetyrkin:2009fv, Kuhn:2007vp} used the data sets of
Refs.~\cite{Bai:2001ct,Ablikim:2006mb} from BES covering 
the energy region $2.6~\mbox{GeV} \leq E \leq 4.8~\mbox{GeV}$.

We consider three different selections of data sets to study the dependence of
the experimental moments on this choice:
\begin{itemize}
\item[1.] The {\it minimal selection} contains all data sets necessary to cover
  the whole energy region between $2$ and $10.538$~GeV without any gaps and
  keeping only the most accurate ones. These 8 data sets are from BES~\cite{Bai:1999pk,Bai:2001ct,Ablikim:2006aj,:2009jsa},
  CrystalBall~\cite{Edwards:1990pc}, CLEO~\cite{CroninHennessy:2008yi,:2007qwa}
  and MD1~\cite{Blinov:1993fw} corresponding to the data sets 1, 2, 5, 6, 9, 12,
  13, and 14 (see Tab.~\ref{tab:datasets} for references).
\item[2.] The {\it default selection} contains all data sets except for the
  three ones with the largest uncertainties. It contains 16 data sets and fully
  includes the minimal selection. It contains all data sets 
  except for Mark I and II data sets 16, 17 and 19 from
  Refs.~\cite{Siegrist:1981zp,Rapidis:1977cv,Siegrist:1976br}.  
\item[3.] The {\it maximal selection} contains all 19 data sets. 
\end{itemize}
We use the default selection as our standard choice for the charm mass analysis,
but we will also quote results for the other data selections.

\vskip 2mm
\noindent
{\bf Perturbative QCD region}\\[1mm]
Above $10.538$~GeV there are no experimental measurements of the total hadronic
\mbox{$R$-ratio} that might be useful for the experimental moments.
In this energy region we will therefore use perturbative QCD to
provide estimates for the charm production \mbox{$R$-ratio}. As a penalty for not using
experimental data we assign a $10$\% total relative uncertainty to the
contribution of the 
experimental moments coming from this region, which we then treat like an
uncorrelated experimental uncertainty for the combination with the moment
contribution from lower energies. 
We stress that this error is not related to the
  theoretical uncertainty of perturbative QCD in this region (which amounts to
  less than $0.5\%$), but represents a
  conservative error assignment that allows to trace the impact of this region
  in the analyses. We have fixed the value of $10$\% as twice the
  overall offset between the combined data and the perturbative QCD prediction
  for the charm cross section in the energy region between $4.55$ and
  $10.538$~GeV, see the discussion at the end of Sec.~\ref{subsectioncombination}.
As we see in Sec.~\ref{subsectionmoments} the
energy region above $10.538$~GeV contributes only $(6, 0.4, 0.03, 0.002)\%$ to $M^{\rm
  exp}_{1,2,3,4}$. In the first moment $M_1^{\rm exp}$ the total contribution is about
three times larger than the combined statistical and systematical (true)
experimental uncertainties from the other energy regions. So the 10\% penalty we
assign to this approach represents a subleading component of the final quoted
uncertainty.\footnote{
The situation is quite different if experimental data from the
  region above $4.8$~GeV is discarded and  perturbative QCD is used already from
$4.8$~GeV. Here the contribution to the moments 
$M_{1,2,3,4}^{\rm exp}$ from energies above $4.8$~GeV amounts to $(30, 10, 3,
1)\%$. For the first moment the $10$\% penalty would then represent the
largest source of uncertainty and correspond to an uncertainty in the charm mass
of around $18$~MeV.} 
In the second
and higher moments $M_{n\ge 2}^{\rm exp}$ the contributions from above
$10.538$~GeV and the corresponding uncertainty are negligible compared to the
uncertainties from the lower energy regions. This means that our experimental
moments are completely free from any theory-driven input or potential bias.

As the theoretical formula to determine the moment contribution from the
perturbative QCD region we use the ${\cal O}(\alpha_s)$ \cite{Kallen:1955fb},
${\cal O}(\alpha_s^2)$ \cite{Chetyrkin:1979bj,Dine:1979qh,Celmaster:1979xr},
and ${\cal O}(\alpha_s^3)$ \cite{Gorishnii:1988bc, Gorishnii:1991se}
nonsinglet massless quark cross section including charm mass corrections up to 
${\cal O}(\overline m_c^4/s^2)$ \cite{Bernreuther:1981sp,Gorishnii:1986pz,
Chetyrkin:1993hi,Chetyrkin:1994ex,Chetyrkin:2000zk}:
\begin{eqnarray}
R_{cc}^{\rm th}(s) & = &
N_c \,Q_c^2\,R^{\rm ns}(s,\overline m_c^2(\sqrt{s}), n_f=4,\alpha_s^{n_f=4}(\sqrt{s}),\sqrt{s})\,,
\label{Rcchighdef1}
\end{eqnarray}
where
\begin{eqnarray}
\label{Rnsdef1}
\lefteqn{R^{\rm ns}(s,\overline m_c^2(\sqrt{s}), n_f=4,\alpha_s^{n_f=4}(\sqrt{s}),\sqrt{s})}\\
& = &1 + \dfrac{\alpha_{s}}{\pi} + 1.52453\left(\dfrac{\alpha_{s}}{\pi}\right)^2
-11.52034\left(\dfrac{\alpha_{s}}{\pi}\right)^{3} \nonumber\\
 & + & \dfrac{\overline m_c^{2}(\sqrt{s})}{s}
\Big[\, 12\,\dfrac{\alpha_{s}}{\pi} + 109.167 \left(\dfrac{\alpha_{s}}{\pi}\right)^{2}
 + 634.957 \left(\dfrac{\alpha_{s}}{\pi}\right)^{3}\Big] \nonumber\\
 & + & \dfrac{\overline m_c^4(\sqrt{s})}{s^2}\Big[ - 6 - 22\,\dfrac{\alpha_{s}}{\pi}
  + 140.855 \left(\dfrac{\alpha_{s}}{\pi}\right)^{2} 
  +  \left(\dfrac{\alpha_{s}}{\pi}\right)^{3}\left(
3776.94 + 10.3333\, L_m^2\right)\Big]
\,,\nonumber
\end{eqnarray}
with
\begin{equation}
L_m \, \equiv \, \ln\Big(\dfrac{\overline m_c^{2}(\sqrt{s})}{s}\Big)\,.
\end{equation}
For the computation of the contribution to the experimental moments we determine
$\overline m_c(\sqrt{s})$ and $\alpha_s(\sqrt{s})$ appearing in
Eq.~(\ref{Rcchighdef1}) using $\overline m_c(\overline m_c)=1.3$~GeV and
$\alpha_s(m_Z)=0.118$ as initial conditions. 

It is instructive to examine for the moment contributions from
$\sqrt{s}>10.538$~GeV terms related to charm production that we do not account
for in Eq.~(\ref{Rnsdef1}). In Tab.~\ref{tabneglected} the relative size with
respect to the full first four moments (in
percent) of the most important neglected contributions are given.
In the second column the size of the mass corrections up to
order $\overline m_c^4$, which we have included in $R^{\rm th}_{cc}$, are shown
as a reference. The third column shows the contributions coming from 
secondary $c\bar c$ radiation through gluon splitting. The fourth column depicts
the contributions from the ${\cal O}(\alpha_s^3)$ singlet corrections (including
the mass corrections up to order $\overline m_c^4$), which one
can take as an rough estimate for the actual contributions from the charm cut.
Finally in the last column we show the size of the Z-boson exchange
terms integrated from threshold to $10.538$~GeV. This contribution represents the
Z-exchange contribution that is contained in the data, but - by definition - not
accounted for in the theory moments. 
We see that at least for the first two moments, the contributions neglected are
much smaller than the charm mass corrections we have accounted for in the
nonsinglet production rate, which are already constituting a very small effect. 
Overall the numerical effect on the charm mass of all these contributions is
tiny considering the scaling $\overline m_c\sim M_n^{1/2n}$.
Since we assign a $10$\% error on the moments' contribution from the energy
region $\sqrt{s}>10.538$~GeV where we use theory input, our approach to neglect
subleading effects is justified.

\begin{table}[t!]\begin{center}
\begin{tabular}{|c|cccc|}
\hline 
$n$ & Mass corrections & Secondary Radiation & Singlet & Z-boson\tabularnewline
\hline
$1$ & $0.02$ & $0.038$ & $3\times10^{-4}$ & $0.006$\tabularnewline
$2$ & $0.001$ & $9\times10^{-4}$ & $2\times10^{-5}$ & $0.004$\tabularnewline
$3$ & $1\times10^{-4}$ & $4\times10^{-5}$ & $2\times10^{-6}$ & $0.003$\tabularnewline
$4$ & $8\times10^{-6}$ & $3\times10^{-6}$ & $1\times10^{-7}$ & $0.003$\tabularnewline
\hline
\end{tabular}\end{center}
\caption{Relative size (in percent) of some subleading contributions for the
first four moments originating from energies $\sqrt{s}>10.538$~GeV. The second
column shows the charm mass corrections  
contained in $R_{cc}^{\rm th}$ as a reference. The third column shows the
contributions from ${\cal O}(\alpha_s^2)$ secondary $c\bar c$ radiation through
gluon splitting.  
The effects from secondary charm radiation accounting all energies from threshold
to infinity are $(0.042,0.002,3\times 10^{-4},7\times 10^{-5})$\% for the
first four moments.  The fourth column depicts 
the contribution from the ${\cal O}(\alpha_s^3)$ singlet gluon corrections 
accounting for the mass corrections up to order $\overline m_c^4$.
Integrating the known singlet corrections from threshold to infinity the
contributions amount to $(0.005,0.003,0.002,0.001)$\% for the first four
moments. The last column shows the relative corrections from the Z-boson
exchange integrated from threshold to $10.538$~GeV.\label{tabneglected}}
\end{table}

\vskip 2mm
\noindent
{\bf Non-charm background}\\[1mm]
Experimentally only the total hadronic cross section is available. 
Although charm-tagged rate measurements are in principle
possible~\cite{CroninHennessy:2008yi} they have not been provided in
publications. On the other hand, they would also exhibit sizable additional
uncertainties 
related to the dependence on simulations of the decay of charmed mesons
into light quark final states.
So to obtain the charm production cross section from the data we have to
subtract the non-charm background using a model based on perturbative QCD
related to the production of $u$, $d$ and $s$ quarks. A subtle point is related
to the secondary radiation of $c\bar c$ pairs off the $u$, $d$ and $s$ quarks
from gluon splitting and to which extent one has to account theoretically for
the interplay between real and virtual secondary $c\bar c$ radiation which
involves infrared sensitive terms~\cite{Hoang:1997ca}. Since in this work we
define the moments from primary $c\bar c$ production (see Eq.~(\ref{momentdef1})),
secondary $c\bar c$ production is formally counted as non-charm background. Thus
for the model for the non-charm background for $\sqrt{s}>2\, \overline m_c$ we
employ the expression
\begin{eqnarray}
\label{Rudsdef}
R_{uds}(s) & = &
N_c(Q_u^2 + 2 Q_d^2)\,\Bigg[ R^{\rm ns}(s,0,n_f=3,\alpha_s^{n_f=4}(\sqrt{s}),\sqrt{s})\\
 &&+\,
\left(\dfrac{\alpha_s^{n_f=4}(\sqrt{s})}{\pi}\right)^{2}\dfrac{2}{3}\left(\rho^{V}+\rho^{R}+\dfrac{1}{4}\,\log\dfrac{\overline m_{c}^{2}(\overline m_{c})}{s}\right) \Bigg]
\,.\nonumber
\end{eqnarray}
The second term on the RHS describes the contributions from real and virtual
secondary $c\bar c$ radiation.
The analytic expressions for $\rho^R$ and $\rho^V$ can be found in Eqs.~(2) and (6) of Ref.~\cite{Hoang:1994it}. We have checked that the numerical impact of real ($\rho^R$) and virtual
($\rho^V$) secondary radiation individually as well as the complete second term
on the RHS of Eq.~(\ref{Rudsdef}) on the moments is negligible, see
Tab.~\ref{tabneglected}.
We use Eq.~(\ref{Rudsdef}) and fit the non-charm background including also data
in the region $2~\mbox{GeV}\le E\le 3.73~\mbox{GeV}$ via the ansatz
$R_{{\rm non}-c\bar c}(s) = n_{\rm ns}\,R_{uds}(s)$, where the constant $n_{\rm
  ns}$ represents an additional fit parameter. We determine $n_{\rm ns}$ from a
global combined fit including many data sets, as explained below.
This is similar in spirit to Refs.~\cite{Kuhn:2007vp}, where
an analogous constant $n_-$ was determined. Their approach, however, differs
from ours as they fitted $n_-$ separately for the two considered BES
data sets~\cite{Bai:2001ct} and \cite{Ablikim:2006mb} accounting only energies
below $3.73$~GeV.

\subsection{Data Combination}
\label{subsectioncombination}
Combining different experimental measurements of the hadronic cross section one
has to face several issues: (a) the measurements are given at individual
separated energy points, (b) the set of measurements from different publications
are not equally spaced, cover different, partly overlapping energy regions and
have different statistical and systematical uncertainties, (c) the correlations
of systematical errors are only known (or provided) for the data sets within
each publication, 
(d) there are a number of very precise measurements at widely separated energies.

In this section we discuss the combination of the experimental data from the
threshold and the data continuum regions between $2$ and $10.538$~GeV using a
method based on a fitting procedure used before for determining the hadronic
vacuum polarization effects for $g-2$~\cite{Hagiwara:2003da}. In this work we
extend this approach and also account for the subtraction of the non-charm
background.  

\vskip 2mm
\noindent
{\bf Combination method}\\[1mm]
The method uses the combination of data in energy bins (clusters) assuming that
the \mbox{$R$-value} within each cluster changes only very little and can thus be well
approximated by a constant. Thus clusters for energies where $R$ varies rapidly
need to be small (in this case the experimental measurements are also denser).
The \mbox{$R$-value} in each cluster is then obtained by a
$\chi^2$~fitting procedure. Since each experimental data set from any
publication covers an energy range overlapping with at least one other data set,
the clusters are chosen such that clusters in overlapping regions contain
measurements from different data sets. Through the fitting procedure correlations
are then being communicated among different data sets and very accurate
individual measurements can inherit their precision into neighboring
clusters. Both issues are desirable since the hadronic \mbox{$R$-ratio} is a smooth
function with respect to the sequence of clusters. 

To describe the method we have to set up some notation:
\begin{itemize}
\item All measurements $R(E)$ are distinguished according to the energy $E$
  at which they have been carried out.
\item Each such energy point having a measurement is written as $E_i^{k,m}$,
  where 
  $k=1,\ldots,N_{\rm exp}$ refers to the $N_{\rm exp}$ data sets,
  $m=1,\ldots,N_{\rm cluster}$ runs over the $N_{\rm cluster}$ clusters and
  $i=1,\ldots,N^{k,m}$ assigns the \mbox{$i$-th} of the $N^{k,m}$ measurements.
\item Each individual measurement of the \mbox{$R$-ratio} is then written as 
\begin{eqnarray}
\label{Rexp}
R(E^{k,m}_i) & = & R_i^{k,m} \, \pm \, \sigma_i^{k,m} \, 
\pm \, \Delta_i^{k,m}
\,,
\end{eqnarray}
where $R_i^{k,m}$ is the central value, $\sigma_i^{k,m}$ the combined 
statistical and uncorrelated systematical uncertainty and $\Delta_i^{k,m}$
the correlated systematical experimental uncertainty.
\item For convenience we define $\Delta
  f_i^{k,m}=\Delta_i^{k,m}/R_i^{k,m}$ to be the relative
  systematical correlated uncertainty. 
\end{itemize}
As our standard choice concerning the clusters we use 5~different regions each
having equidistant cluster sizes $\Delta E$. The regions are as follows:
\begin{itemize}
\item[(0)] {\bf non-charm region}: has 1 cluster for $2~\mbox{GeV}\leq E \leq
  3.73~\mbox{GeV}$ ($\Delta E=1.73$~GeV).
\item[(1)] {\bf low charm region}: has 2 clusters for $3.73~\mbox{GeV}< E \leq
  3.75~\mbox{GeV}$ ($\Delta E=10$~MeV).
\item[(2)] {\bf $\psi(3S)$ region/threshold region 1}: has 20 cluster for $3.75~\mbox{GeV}< E \leq
  3.79~\mbox{GeV}$ ($\Delta E=2$~MeV).
\item[(3)] {\bf resonance region 2}: has 20 cluster for $3.79~\mbox{GeV}< E \leq
  4.55~\mbox{GeV}$ ($\Delta E=38$~MeV).
\item[(4)] {\bf continuum region}: has 10 cluster for $4.55~\mbox{GeV}< E \leq
  10.538~\mbox{GeV}$ ($\Delta E=598.8$~MeV).
\end{itemize}
We assign to this choice of $52+1$~clusters the notation (2,20,20,10) and later
also examine alternative cluster choices demonstrating that the outcome for the
moments does within errors not depend on them. The cluster in the non-charm
region is used to fit for the normalization constant $n_{\rm ns}$ of the
non-charm background contribution, see Eq.~(\ref{Rudsdef}). 

Our standard procedure to determine the central energy $E_m$ associated to each
cluster is just the weighted average of the energies of all measurements falling
into cluster $m$,
\begin{eqnarray}
\label{Eclusterdef}
E_m & = & 
\dfrac{\sum_{k,i}\dfrac{E_{i}^{k,m}}
{(\sigma_{i}^{k,m})^2+(\Delta_{i}^{k,m})^2}}
{\sum_{k,i}\dfrac{1}{(\sigma_{i}^{k,m})^2+(\Delta_{i}^{k,m})^2}}
\,.
\end{eqnarray}
The corresponding \mbox{$R$-value} for the charm cross section that
we determine through the fit procedure described below is called 
\begin{eqnarray}
\label{Rclusterdef}
R_m & \equiv & R_{c\bar c}(E_m)
\,. 
\end{eqnarray}
We note that using instead the unweighted average or simply the center of
the cluster has a negligible effect on the outcome for the moments since
the clusters we are employing are sufficiently narrow.

\vskip 2mm
\noindent
{\bf Fit procedure and \mbox{$\chi^2$-function}}\\[1mm]
We determine the charm cross section $R_{c\bar c}$ from a \mbox{$\chi^2$-function} that
accounts for the positive correlation among the systematical uncertainties 
$\Delta_i^{k,m}$ within each experiment $k$ and, at the same time, also for
the non-charm background. To implement the correlation we introduce the
auxiliary parameters $d_k$ ($k=1,\ldots, N_{\rm exp}$) that parametrize the
correlated deviation from the experimental central values $R_i^{k,m}$ in units
of the correlated systematical uncertainty $\Delta_i^{k,m}$, see
Eq.~(\ref{Rexp}). In this way 
we carry out fits to $R_i^{k,m}+d_k \,\Delta_i^{k,m}$ and treat the $d_k$ as
additional auxiliary fit parameters that are constraint to be of order one (one
standard deviation) by adding the term 
\begin{eqnarray}
\label{chi2totald}
\chi^2_{\rm corr}(\{d_k\}) & = &
\sum_{k=1}^{N_{\rm exp}} \, d_k^2\,,
\end{eqnarray} 
to the \mbox{$\chi^2$-function}.
To implement the non-charm background we assume that the relative energy
dependence of the non-charm cross section related to primary production of $u$,
$d$ and $s$ quarks is described properly by $R_{uds}$ given in
Eq.~(\ref{Rudsdef}). We then parametrize the non-charm background cross section
by the relation
\begin{eqnarray}
\label{Rncdef}
R_{{\rm non}-c\bar c}(E) &= & n_{\rm ns}\,R_{uds}(E)
\end{eqnarray}
as already described in Sec.~\ref{subsectioncollections},
where the fit parameter $n_{\rm ns}$ is determined mainly from the
experimental data in the first clusters below $3.73$~GeV by adding the term
\begin{eqnarray}
\label{chi2noncharm}
\chi^2_{\rm nc}(n_{\rm ns},\{d_k\}) & = &
\sum_{k=1}^{N_{\mathrm{exp}}}
\sum_{i=1}^{N^{k,1}}\left(\dfrac{
R_i^{k,1}-(1+\Delta f_i^{k,1}\, d_k)\, n_{ns}\, 
R_{uds}(E_i^{k,1})}{\sigma_i^{k,1}}\right)^2\,,
\end{eqnarray}
to the \mbox{$\chi^2$-function}. The complete \mbox{$\chi^2$-function} then has the form
\begin{eqnarray}
\label{chi2total}
\chi^2(\{R_m\}, n_{\rm ns},\{d_k\}) & = &
 \chi^2_{\rm corr}(\{d_k\}) \, + \,
\chi^2_{\rm nc}(n_{\rm ns},\{d_k\})\,+\,\chi^2_{\rm c}(\{R_m\}, n_{\rm ns},\{d_k\})
\,,
\end{eqnarray}
where\footnote{We have checked that the effect of using $R_{uds}(E_m)$ instead
  of $R_{uds}(E_i^{k,m})$ is totally negligible.} 
\begin{eqnarray}
\label{chi2charm}
&&\chi^2_{\rm c}(\{R_m\}, n_{\rm ns},\{d_k\})  = \\[2mm]
&&\sum_{k=1}^{N_{\mathrm{exp}}}
\sum_{m=2}^{N_{\mathrm{clusters}}}\sum_{i=1}^{N^{k,m}}
\left(\dfrac{R_{i}^{k,m}-(1+\Delta f_i^{k,m}\, d_k)\,
(R_m+n_{\rm ns}\, R_{uds}(E_i^{k,m}))}{\sigma_i^{k,m}}\right)^{2}
\,.\nonumber
\end{eqnarray}
Note that in our approach the non-charm normalization constant $n_{\rm ns}$ is
obtained from a combined fit together with the cluster values $R_m$. 

This form of the \mbox{$\chi^2$-function} is an extended and adapted version of the
ones used in Refs.~\cite{Agostini:1993uj, Takeuchi:1995xe}. A special characteristic of the
\mbox{$\chi^2$-functions} in Eqs.~(\ref{chi2noncharm}) and (\ref{chi2charm}) is that the
relative correlated experimental uncertainties $\Delta f_i^{k,m}$ 
enter the \mbox{$\chi^2$-function} by multiplying the fit value $R_m$
rather than the experimental values $R_i^{k,m}$. This leads to a non-bilinear dependence of
the \mbox{$\chi^2$-function} on the $d_m$ and the $R_m$ fit parameters and avoids
spurious solutions where the best fit values for the $R_m$ are located
systematically below the measurements. Such spurious solutions can
arise for data points with substantial positive correlation when
\mbox{$\chi^2$-functions} with strictly bilinear dependences are employed~
\cite{Agostini:1993uj,Takeuchi:1995xe}.\footnote{ 
We prove in Appendix~B the equivalence of a bilinear \mbox{$\chi^2$-function} with fit
auxiliary parameters to a bilinear \mbox{$\chi^2$-function} without auxiliary fit
parameters, but containing the standard correlation
matrix.}

We also note that the implementation of the non-charm background subtraction
given in Eq.~(\ref{chi2charm}) leads to a partial cancellation of systematical
uncertainties for the $R_m$ best fit values for the charm cross
section. Moreover, it is
interesting to mention that in the limit where each cluster contains
exactly one measurement (except below threshold, in which we always keep one cluster)
the \mbox{$\chi^2$-function} decouples, after performing the change of variables
 $R'_m=R_{m}+n_{\rm ns}\, R_{uds}(E^{k,m})$, into the sum of two independent
\mbox{$\chi^2$-functions}, one containing data below threshold and depending only on
$n_{\rm ns}$ and $d_k$, and another one containing data above threshold $R^{k,m}$ (we
drop the label $i$ because having only one data per cluster it can take only the
value $1$) and depending only on $R'_m$.
After minimizing the first \mbox{$\chi^2$-function} one can obtain the best fit values
for $n_{\rm ns}$ and $d_k$, denoted by 
$n_s^{(0)}$ and $d_k^{(0)}$, respectively. The second $\chi^2$ has a minimal value of $0$ and the best fit
parameters read 
$R_m^{(0)}=R^{k,m}/(1+d_k^{(0)}\Delta^{k,m})-n_{\rm ns}^{(0)}R_{uds}(E^{k,m})$. 

Close to the minimum the \mbox{$\chi^2$-function} of Eq.~(\ref{chi2total}) can be
written in the Gaussian approximation
\begin{eqnarray}
\label{chi2Gauss}
\chi^2(\{p_i\}) & = & \chi^2_{\rm min}\,+\,  \sum_{i,j} (p_i-p_i^{(0)})V_{i,j}^{-1}(p_j-p_j^{(0)})
\, + \, {\cal O}\Big((p-p^{(0)})^3\Big)
\,,
\end{eqnarray}
where $p_i=(
\{R_m\}, n_{\rm ns}, \{d_k\})$ and the superscript $(0)$ indicates the
respective best fit value. After determination of the correlation matrix
$V_{ij}$ by numerically inverting $V^{-1}_{i,j}$ we can
drop the dependence on the auxiliary variables $n_{\rm nc}$ and $d_k$ and
obtain the correlation matrix of the $R_m$ from the \mbox{$R_m$-submatrix} 
of $V_{ij}$ which we call $V_{m m^\prime}^R$. In order to separate 
uncorrelated statistical and
systematical uncertainties from correlated systematical ones we compute the
complete $V_{m m^\prime}^R$ accounting for all uncertainties and a simpler
version of the correlation matrix,  $V_{m m^\prime}^{R,u}$ accounting only 
for uncorrelated uncertainties. The latter is obtained from dropping all 
correlated errors $\Delta_i^{k,m}$ from the 
\mbox{$\chi^2$-function}~(\ref{chi2total}).\footnote{In
a very good approximation, it can be also obtained by dropping in 
$V^{-1}_{i,j}$ the rows and columns corresponding
to $d_k$ and inverting the resulting matrix. After that one also drops the
row and column corresponding to $n_{\rm nc}$. We adopt this simplified 
procedure for our numerical fits.}

\begin{figure*}[t!]
\subfigure[]{
\includegraphics[width=0.48\textwidth]{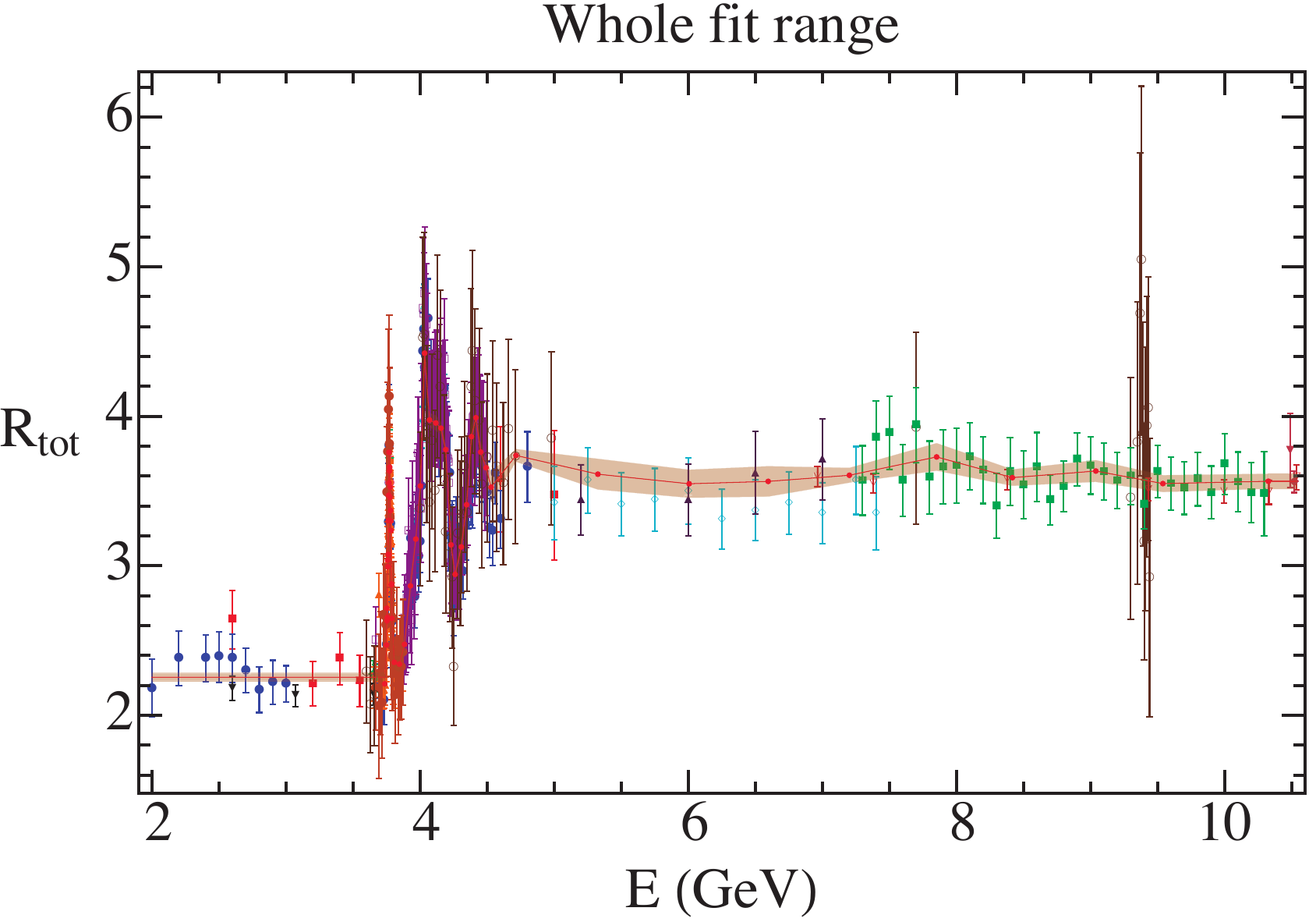}
\label{fig:fitall}
}
\subfigure[]{
\includegraphics[width=0.48\textwidth]{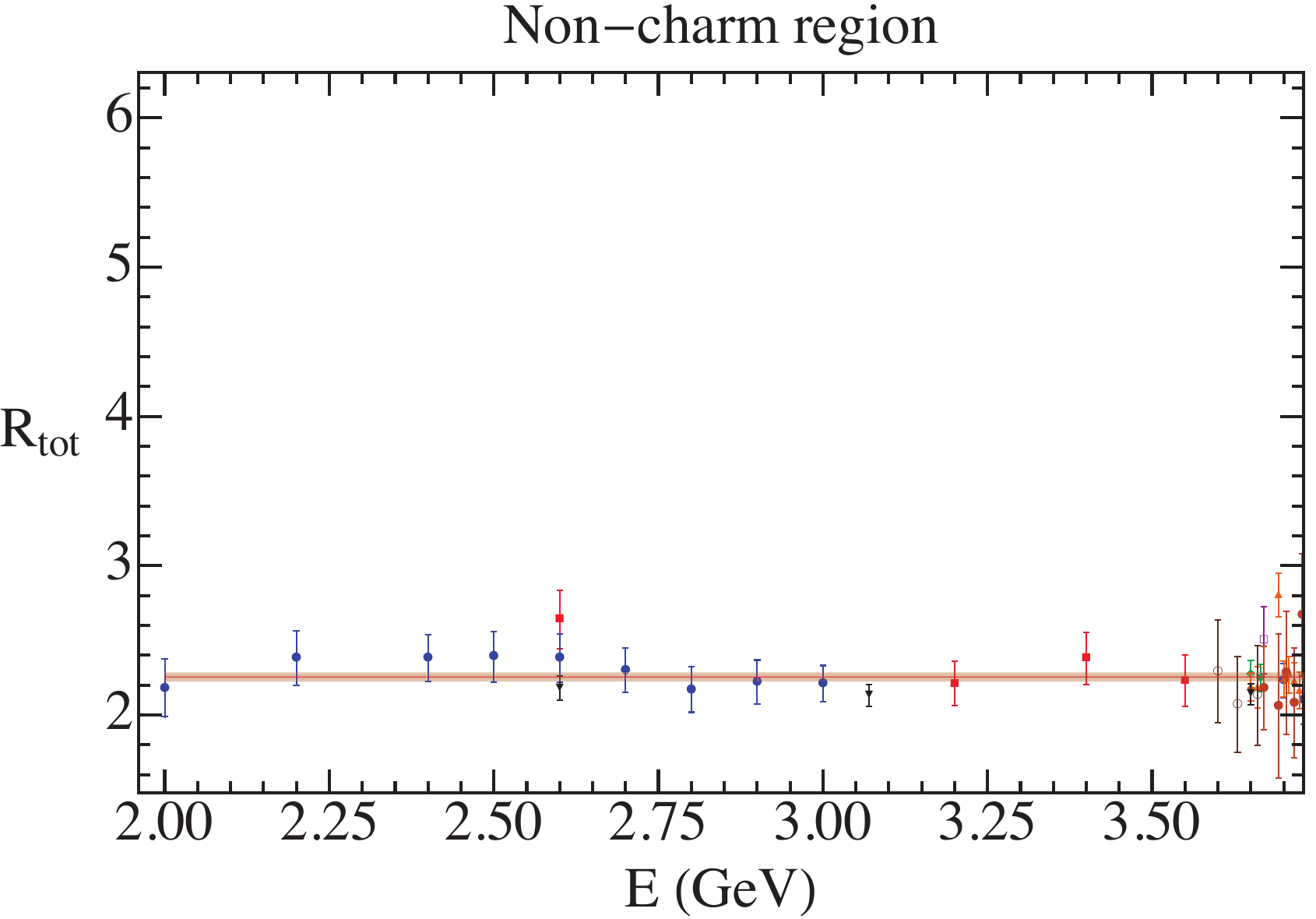}
\label{fig:fitnoncharm}
}
\subfigure[]{
\includegraphics[width=0.48\textwidth]{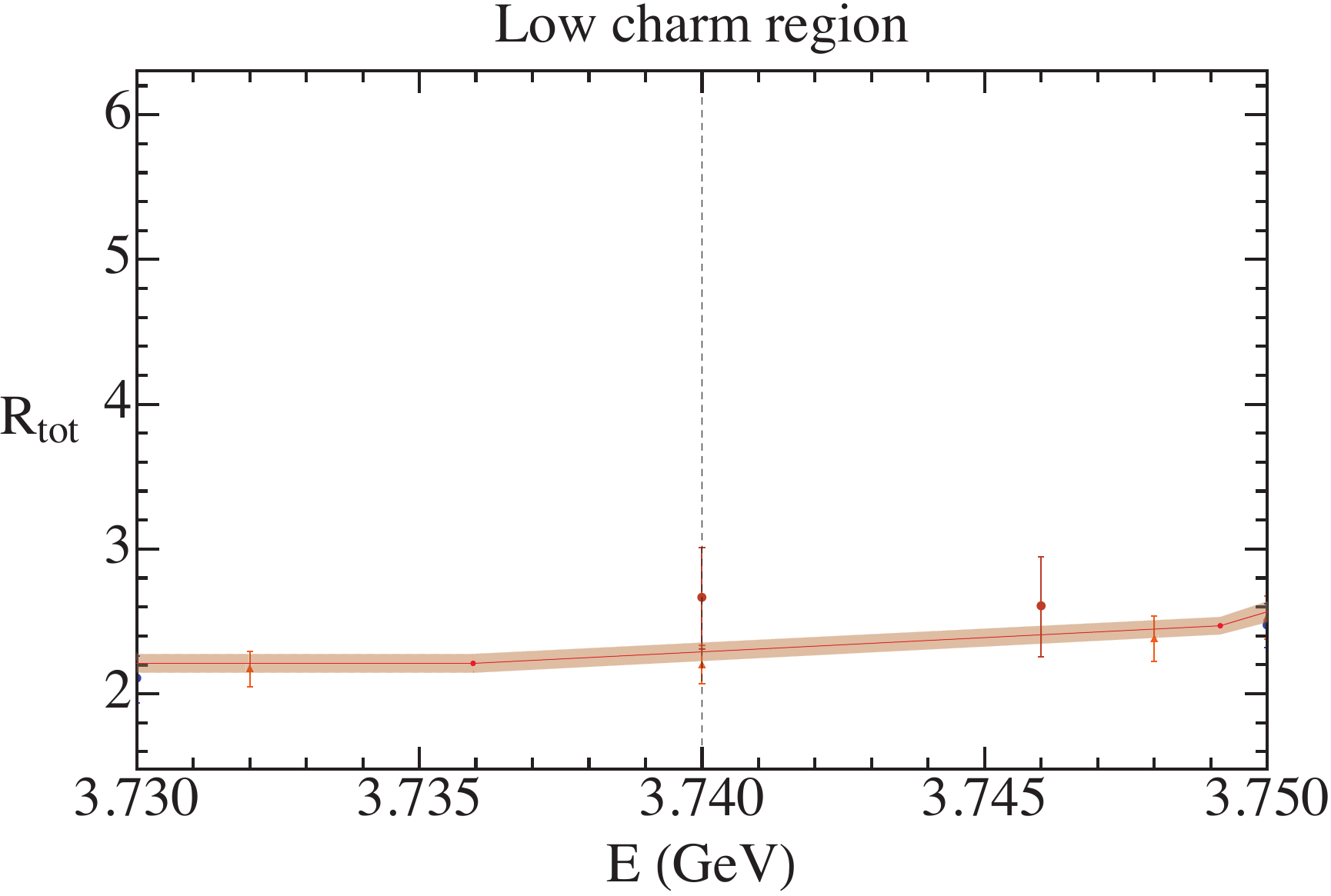}
\label{fig:fitlowcharm}
}
\subfigure[]{
\includegraphics[width=0.48\textwidth]{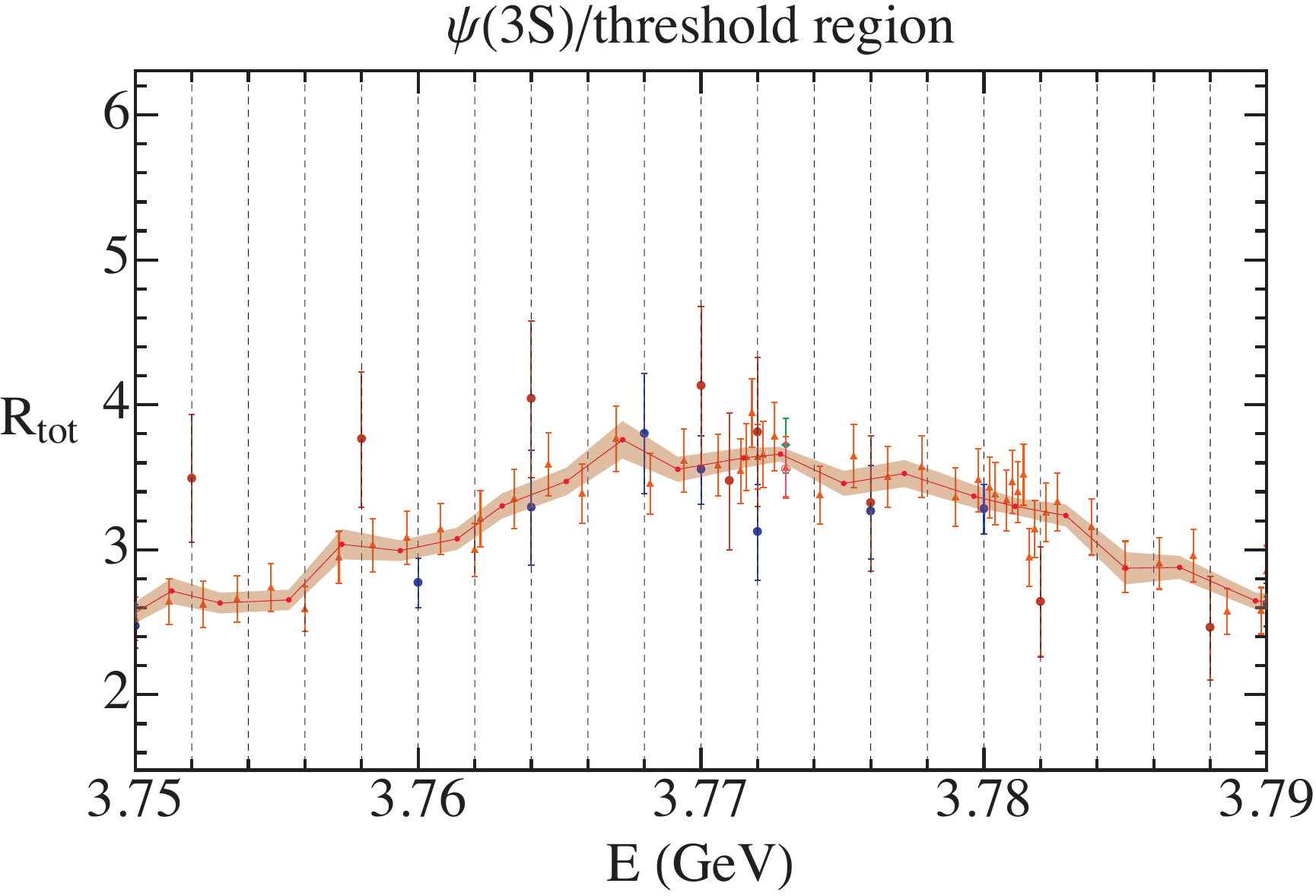}
\label{fig:fitres1}
}
\subfigure[]{
\includegraphics[width=0.48\textwidth]{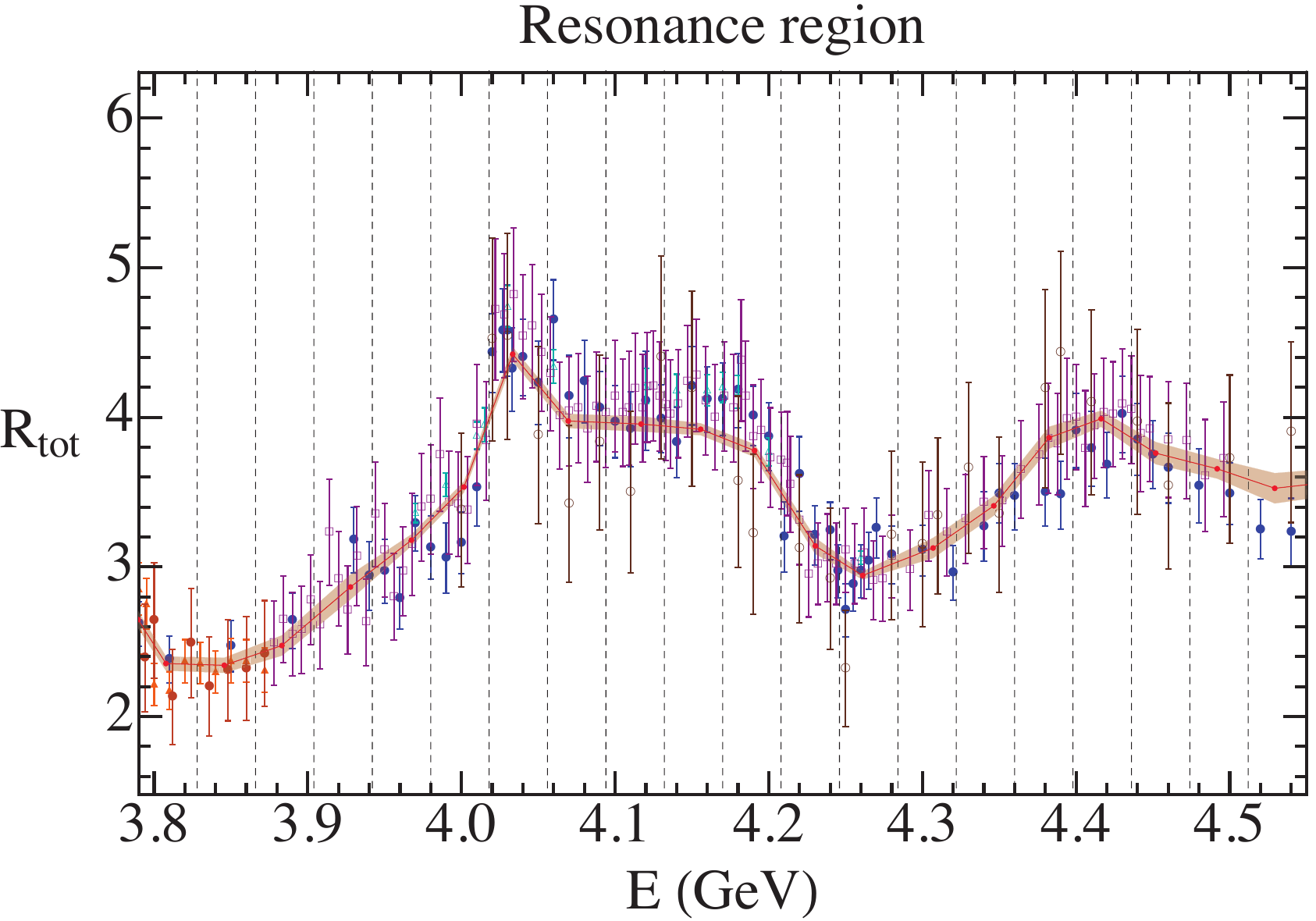}
\label{fig:fitres2}
}
\subfigure[]{
\includegraphics[width=0.48\textwidth]{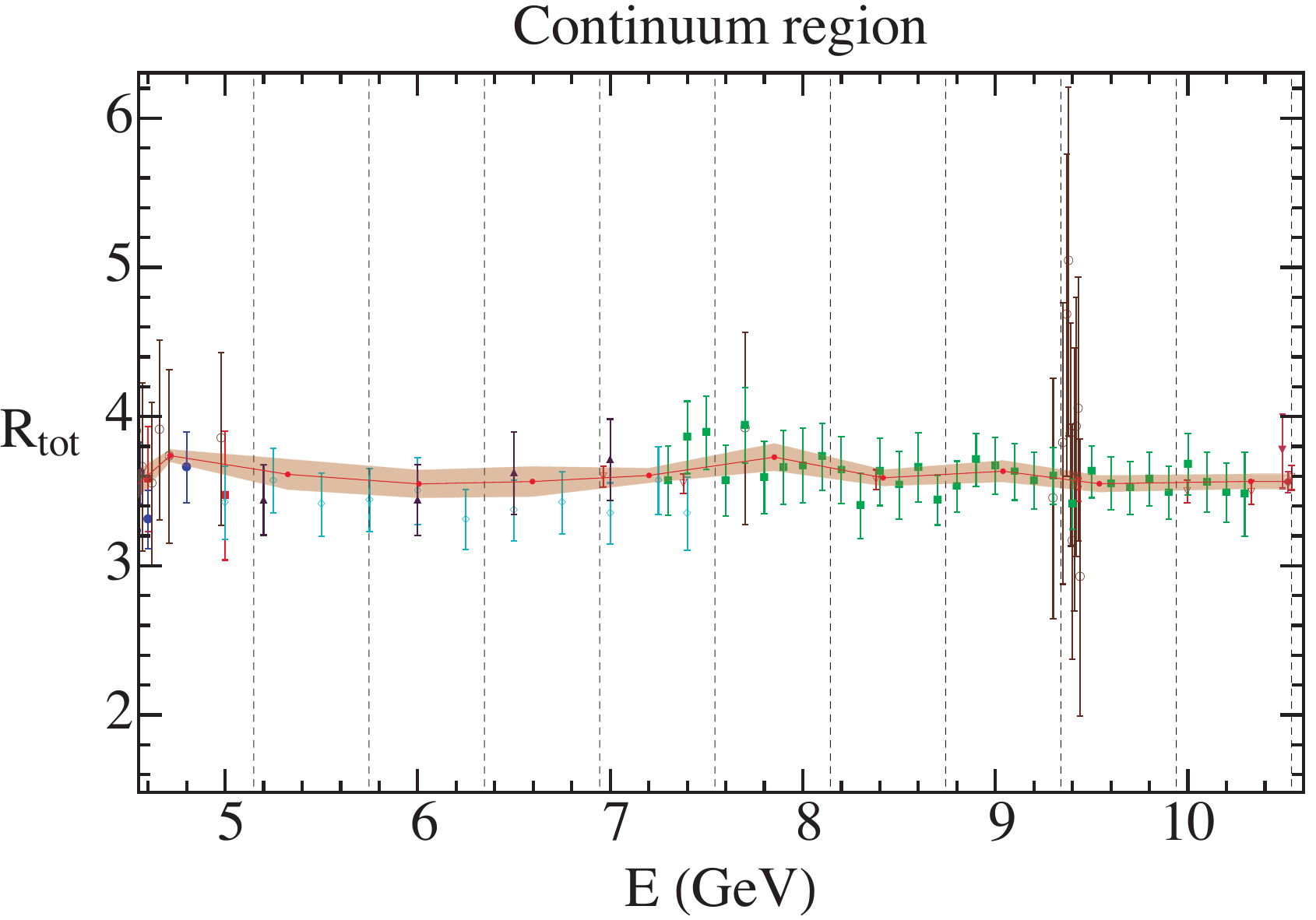}
\label{fig:fitcontinuum}
}
\caption{Result of the fit for the default selection of data sets.
On the top, (a) and (b) show the entire fit region and the non-charm region,
respectively. In the middle row, (c) illustrates the low charm region and
(d) the threshold region 1. In the bottom line (e) and (f) depict threshold 
region 2 and the data continuum region, respectively.  
\label{fig:fitcompilation}}
\end{figure*}

The outcome of the fit  for the sum of the charm and the non-charm  
cross section in the threshold and the data continuum region using the 
standard data set explained above is shown in
Figs.~\ref{fig:fitcompilation}(a)-(f) together with the input data sets.
The red line segments connect the best fit values and the brown band 
represents the combined total uncertainty. The clusters are indicated by dashed 
vertical lines. For completeness we have also given all numerical results for
the $R_m$ values  in Appendix~A. There we also give results for the minimal 
and maximal data set selections.

\begin{figure*}[t!]
\subfigure[]{
\includegraphics[width=0.48\textwidth]{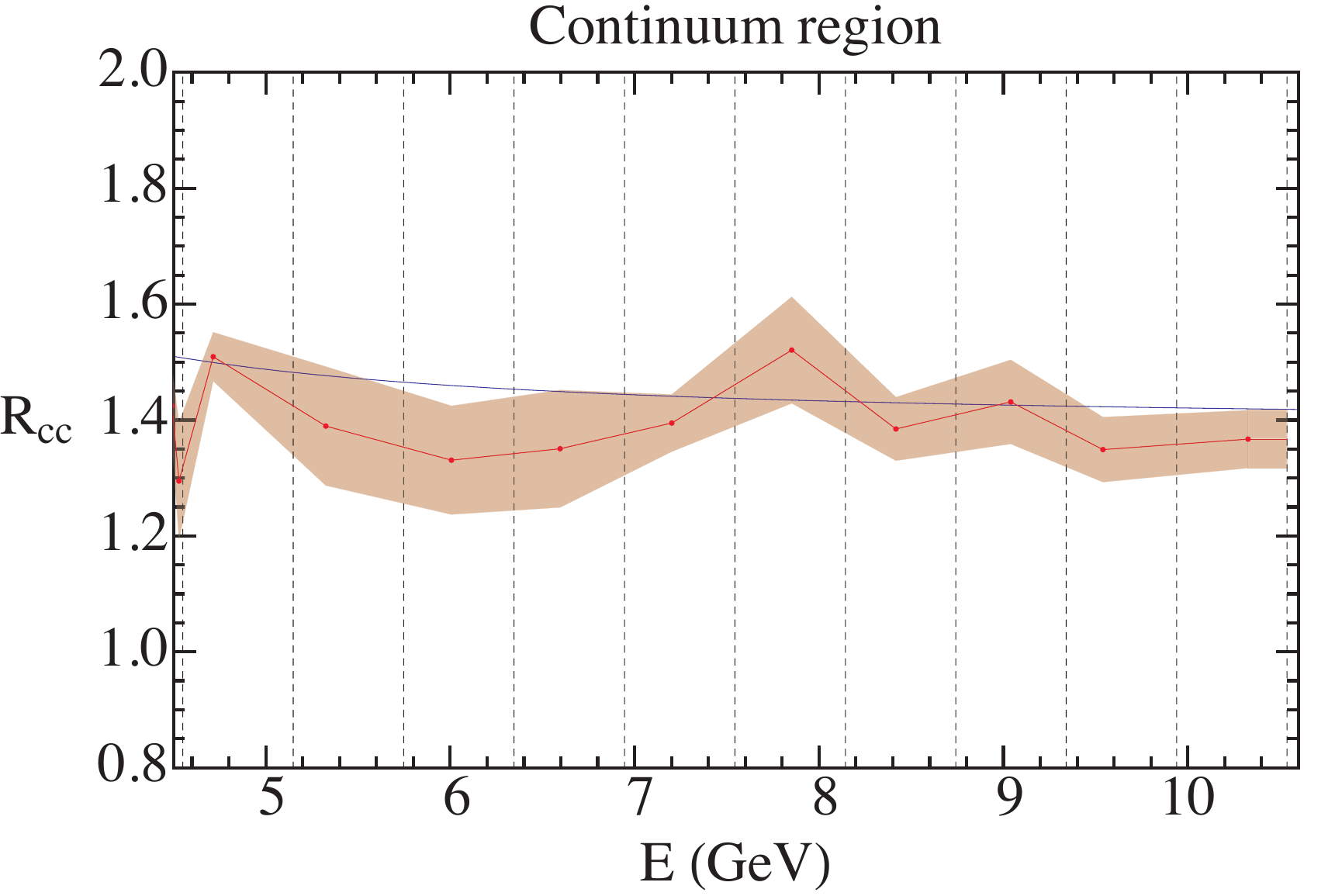}
\label{fig:Rccvsdata}
}
\subfigure[]{
\includegraphics[width=0.48\textwidth]{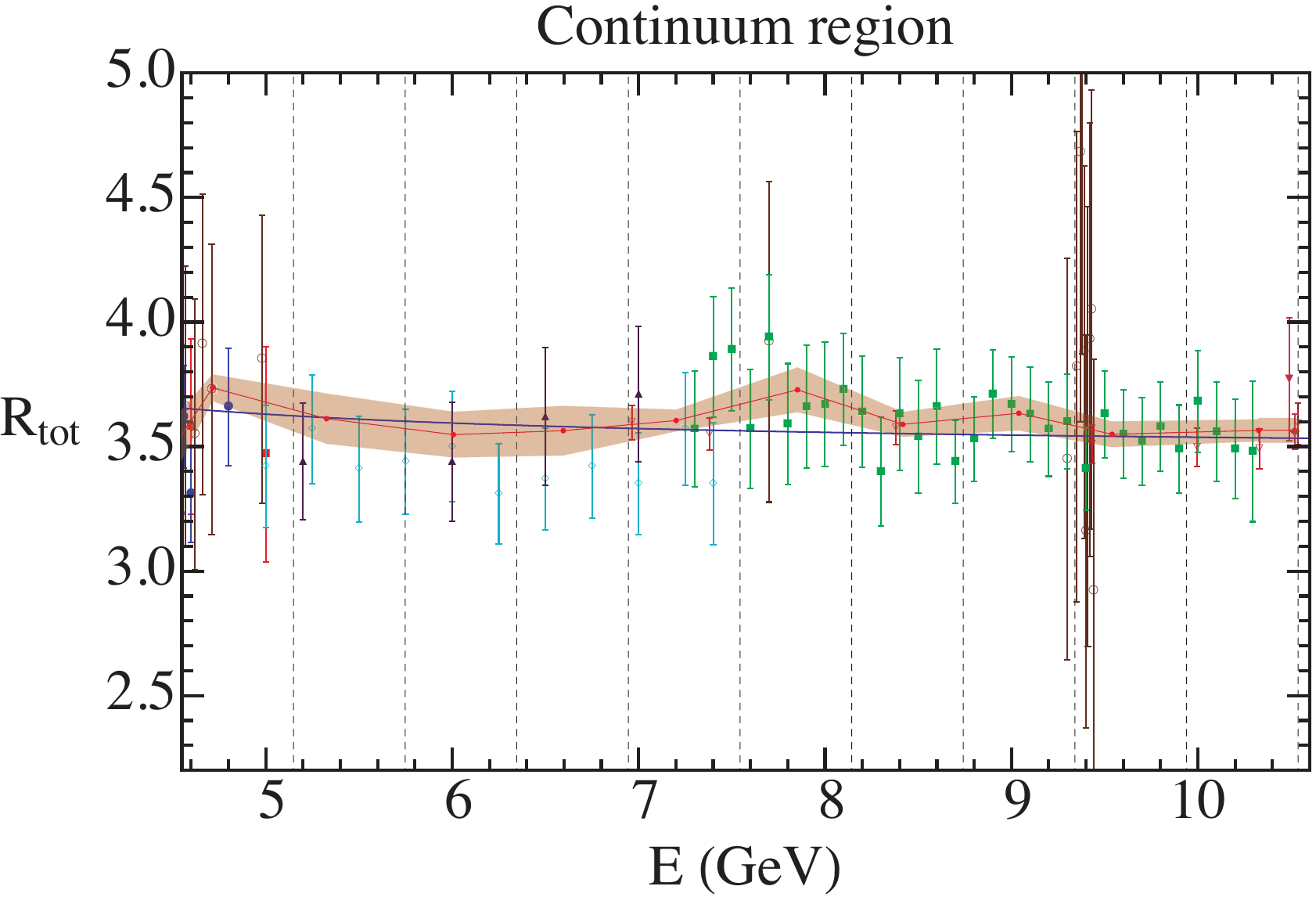}
\label{fig:Rtotvsdata}
}
\caption{Comparison of experimental data (red dots) with uncertainties (brown
  error band) and the predictions from perturbative
  QCD (black solid line) for the charm cross section $R_{c\bar c}$ (left panel) and the total cross
  section $R_{\rm tot}$ (right panel).
\label{fig:Rvsdata}}
\end{figure*}

The fit results in the continuum region for the energies between $4.55$~GeV and
$10.538$~GeV allow for an interesting comparison of the $R$-values
obtained from perturbative QCD and from experimental data based on our combined
fit procedure. In the left panel of Fig.~\ref{fig:Rvsdata} the fit results for
the charm cross section are shown together with the perturbative QCD prediction
from Eq.~(\ref{Rnsdef1}) (black solid line). Within the total experimental errors,
which are around $5\%$, there is good agreement between data and the
perturbative QCD result (which itself has a perturbative error due to scale variations
of less than $1\%$). Interestingly, there appears to be some oscillatory
behavior of the data around the perturbative QCD result, although the
statistical power of the data is insufficient to draw definite conclusions
concerning the physics of these oscillations. Concerning the contributions to
the moments we find $M_{1,{\rm ex}}^{(4.55-10.538){\rm GeV}}= 4.81\,\pm\, 0.18$ from
the data compared to $M_1 ^{(4.55-10.538){\rm GeV}}=5.010 \,\pm\, 0.011$ from perturbative QCD
based on Eq.~(\ref{Rnsdef1}). The central value from perturbative QCD is $4\%$
above the experimental moment. This shows that adopting perturbative QCD
predictions instead of data gives a contribution to the central value of the
moments compatible within the experimental uncertainties. However, using
theoretical uncertainties as an estimate for the experimental ones leads to an
underestimate. In the right panel of Fig.~\ref{fig:Rvsdata} the experimental data
and the perturbative QCD predictions are compared for the total hadronic cross
section. Here the perturbative QCD result is shifted downward with respect to
the data due to the non-charm normalization constant $n_{\rm ns}$ obtained from
our combined fit being slightly larger than unity, see
Eq.~(\ref{eq:ns-results}). Overall, the agreement between 
the combined data and perturbative QCD appears to be even better than for the
charm cross section in particular for the energy region above $9$~GeV.

\subsection{Experimental Moments}
\label{subsectionmoments}

\vskip 2mm
\noindent
{\bf Narrow resonances}\\[1mm]
For the $J/\psi$ and $\psi^\prime$ charmonium contributions to the experimental
moments we use the narrow width approximation,
\begin{eqnarray}
\label{Mnresonances}
M_n^{\rm res} & = & \dfrac{9\,\pi\,\Gamma_{ee}}{\alpha(M)^{2}M^{2n+1}}\,,
\end{eqnarray}
with the input numbers given in Tab.~\ref{tabpsidata}. We neglect the tiny
uncertainties in the charmonium masses as their effects are negligible.

\vskip 2mm
\noindent
{\bf Threshold and data continuum region}\\[1mm]
For the determination  of the moment contributions from the threshold and the
continuum region between $3.73$ and $10.538$~GeV we use the results for the
clustered $c\bar c$ cross section values $R_m$ determined in
Sec.~\ref{subsectioncombination} and the trapezoidal rule. We employ a linear interpolation
for the cross section but keep the analytic form of the integration kernel
$1/s^{n+1}$ exact by including it into the integration measure.
Using the relation ${\rm d}s/s^{n+1} = {\rm d}(E^{-2n}/n)$ we thus obtain
\begin{eqnarray}
\label{Mndata}
M_n^{\rm thr+cont} & = \dfrac{1}{2n}\Bigg[ &\sum_{i=1}^{N_{\mathrm{clusters}}}R_{i}\Bigg(\dfrac{1}{E_{i-1}^{2n}} -
\dfrac{1}{E_{i+1}^{2n}}\Bigg)
+R_{0}\Bigg(\dfrac{1}{E_{0}^{2n}}-\dfrac{1}{E_{1}^{2n}}\Bigg)\\
&+&R_{N_{\mathrm{clusters}}+1}
\Bigg(\dfrac{1}{E_{N_{\mathrm{clusters}}}^{2n}} - \dfrac{1}{E_{N_{\mathrm{clusters}}+1}^{2n}}\Bigg)
\Bigg]\,,\nonumber
\end{eqnarray}
where $R_0$ and $E_0$ are the $R$ and energy values at the lower boundary of the
smallest energy cluster, and $R_{N_{cl}+1}$ and $E_{N_{cl}+1}$ are the
corresponding values of the upper boundary of the highest energy cluster. The
values for $R_0$ and $R_{N_{cl}+1}$ are obtained from linear extrapolation using
the $R_m$ values of the two closest lying clusters\footnote{
For $R_0$ we have $m^\prime=1$ and for $R_{N_{cl}+1}$ we have
$m^\prime=N_{cl}-1$. 
}
$m^\prime$ and $m^\prime+1$ with the formula
\begin{eqnarray}
\label{REextrapol}
R(E) & = & 
\frac{R_{m^\prime+1}-R_{m^\prime}}{E_{m^\prime+1}-E_{m^\prime}}\, (E-E_{m^\prime})+ R_{m^\prime}
\,.
\end{eqnarray}
For the computation of the moment contributions from subintervals within the
range between $3.73$ and $10.538$~GeV we also use Eq.~(\ref{Mndata}) 
using corresponding adaptations for the boundary values at
$m=0$ and $m=N_{cl}+1$. 

\vskip 2mm
\noindent
{\bf Perturbative QCD region}\\[1mm]
For the region above $10.538$~GeV where we use the perturbative QCD input
described in Eqs.~(\ref{Rcchighdef1}) and (\ref{Rnsdef1}) for the charm \mbox{$R$-ratio}
the contribution to the experimental moment is obtained from the defining
equation~(\ref{momentdef2}) with a lower integration limit of $10.538$~GeV:
\begin{eqnarray}
\label{MnQCD}
M_n^{\rm QCD} & = & \gamma \times\int_{(10.538\,\rm GeV)^2}^\infty {\rm d}s \,\dfrac{R_{cc}^{\rm th}(s)}{s^{n+1}}
\,.
\end{eqnarray}
The variable $\gamma$ is an auxiliary variable used to parametrize the $10\%$
uncertainty we assign to the perturbative QCD contribution,
\begin{eqnarray}
\label{gammadef}
\gamma & = & 1.0 \,\pm \, 0.1
\,.
\end{eqnarray}

\vskip 2mm
\noindent
{\bf Correlations}\\[1mm]
The experimental moments are obtained from the sum of the resonance, threshold
plus continuum and perturbative QCD contributions described just above,
\begin{eqnarray}
\label{Mnsum}
M_n^{\rm exp} & = &
M_n^{\rm res} \, + \,
M_n^{\rm thr+cont} \, + \,
M_n^{\rm QCD}
\,.
\end{eqnarray}
To determine the uncertainties we account for the errors in the $e^+e^-$ widths
of $J/\psi$ and $\psi^\prime$ and in the cluster values $R_m$, and for the $10\%$
assigned uncertainty in $M_n^{\rm QCD}$. For the evaluation we use the usual
error propagation based on a $\bar m\times\bar m$ correlation matrix with $\bar
m=N_{cl}+3$. The 
correlation matrix of the experimental moments thus has the form
\begin{eqnarray}
\label{CMcorr}
C_{nn^\prime}^{\rm exp} & = & \sum\limits_{i,j=1}^{N_{cl}+3} 
\Big(\frac{\partial M_n^{\rm exp}}{\partial \bar R_i}\Big)\,
\Big(\frac{\partial M_{n^\prime}^{\rm exp}}{\partial \bar R_j}\Big) 
\, V_{ij}^{\bar R}
\,,
\end{eqnarray}
where we have $\bar R_i =
(\{R_m\},\Gamma_{e^+e^-}(J/\psi),\Gamma_{e^+e^-}(\psi^\prime),\gamma)$. The entries
of $V^{\bar R}$ in the $R_m$ subspace are just the entries of the correlation
matrix $V^R$ obtained from the cluster fit. The diagonal entries in the
$\Gamma_{e^+e^-}$ subspace are the combined statistical and systematical
uncertainties of the $e^+e^-$ widths and the $\delta\gamma=0.1$ for $M_n^{\rm QCD}$, 
respectively.  We treat the latter uncertainty as uncorrelated with
all other uncertainties. So all non-diagonal entries of $V_{ij}^{\bar R}$ for
$i$ or $j=N_{cl}+3$ are zero. For the uncertainty of the $e^+e^-$ widths we
adopt a model where the (quadratic) half of the error is uncorrelated and the
other (quadratic) half is positively correlated among the two narrow resonances,
while we assume no correlation between the narrow resonances and
the $R_n$ cluster values.\footnote{We thank J.\ J. Hern\'andez Rey for pointing
out that, although a coherent study of these correlations does not exist,
treating resonances as uncorrelated to the continuum represents the most appropriate
correlation model.}
Thus for the corresponding non-diagonal entries of $V_{i,j}^{\bar R}$ with 
$i\in\{1,N_{cl}\}$ and \mbox{$j=\{N_{cl}+1,N_{cl}+2\}$} we have the entries
$V_{ij}^{\bar R}=0$, and for $i=N_{cl}+1$ and $j=N_{cl}+2$
we have $V_{ij}^{\bar R}=\delta\Gamma^1_{e^+e^-}\delta\Gamma^2_{e^+e^-}/2$, where
$\delta\Gamma^{1,2}_{e^+e^-}$ are the respective $e^+e^-$ width total uncertainties for $J/\psi$
and $\psi^\prime$, respectively.

The results for the moments showing separately the contributions from the
resonances, various energy subintervals and their total sum using the defaults
data set collection (see Sec.~\ref{subsectioncollections}) are given in
Tab.~\ref{tab:momres1}. 
Using Eq.~(\ref{CMcorr}) it is straightforward to compute the correlation matrix
of the moments, and we obtain
\begin{eqnarray}
\label{CMtotal}
C^{\rm exp} & = &\left(
\begin{array}{cccc}
0.128 & 0.084 & 0.077 & 0.075 \\
0.084 & 0.076 & 0.074 & 0.075 \\
0.077 & 0.074 & 0.075 & 0.077 \\
0.075 & 0.075 & 0.077 & 0.079 
\end{array}
\right)\,,
\end{eqnarray}
for the total correlation matrix accounting for all correlated and uncorrelated
uncertainties. We remind the reader that all numbers related to the moment
$M_n^{\rm exp}$ are given in units of $10^{-(n+1)}\,\mbox{GeV}^{-2n}$. To quote
correlated and uncorrelated uncertainties separately it 
is also useful to show the correlation matrix that is obtained when only
uncorrelated uncertainties are accounted for. 
\begin{eqnarray}
\label{CMuncorr}
C^{\rm exp}_{\rm uc} & = &
\left(
\begin{array}{cccc}
 0.04 & 0.034 & 0.033 & 0.034 \\
0.034 & 0.033 & 0.034 & 0.034 \\
0.033 & 0.034 & 0.034 & 0.036 \\
0.034 & 0.034 & 0.036 & 0.037 
\end{array}\right)
\,.
\end{eqnarray}
These results can be used to carry out combined simultaneous fits to several
of the moments. This is, however, not attempted in this work.

\begin{table}[t!]\begin{center}{\small
\begin{tabular}{|c|cccccc|}
\hline 
n & Resonances & $3.73-4.8$ & $4.8-7.25$  & $7.25-10.538$ & $10.538-\infty$ & Total\tabularnewline
\hline
1 & $11.91(17|20)$ & $3.23(4|6)$   & $3.39(8|13)$  & $1.42(2|5)$     & $1.27(0|13)$ & $21.21(20|30)$\tabularnewline
2 & $11.68(18|20)$ & $1.78(2|3)$   & $1.06(3|4)$   & $0.200(3|7)$    & $0.057(0|6)$ & $14.78(18|21)$\tabularnewline
3 & $11.63(18|20)$ & $1.00(1|2)$   & $0.350(9|13)$ & $0.0294(6|10)$  & $0.0034(0|3)$ & $13.02(19|20)$\tabularnewline
4 & $11.73(19|20)$ & $0.571(7|11)$ & $0.121(3|4)$  & $0.00448(9|15)$ & $2.3(0|2)\times 10^{-4}$ & $12.43(19|20)$\tabularnewline
\hline
\end{tabular}

\caption{Result for the experimental moments for the standard selection of data
  sets. The second column collects the contribution from the narrow resonances
  $J/\psi$ and $\psi^\prime$ 
treated in the narrow width approximation. Third to fifth columns are shown
only as an illustration, but use the outcome of the fit to the entire
fit region, $3.73-10.538$~GeV. The sixth column is obtained using perturbation
theory, and the last column shows the number for the complete moment. For the
moment $M_n^{\rm exp}$ all numbers are given in units of
$10^{-(n+1)}\,\mbox{GeV}^{-2n}$. \label{tab:momres1}}
}\end{center}\end{table}

\begin{table}[t!]\begin{center}{\small
\begin{tabular}{|c|cccccc|}
\hline 
n & Resonances & $3.73-4.8$ & $4.8-7.25$  & $7.25-10.538$ & $10.538-\infty$ & Total\tabularnewline
\hline
1 & $11.91(17|20)$ & $3.14(6|8)$ & $3.24(9|16)$ & $1.38(2|6)$    & $1.27(0|13)$  & $20.95(21|33)$\tabularnewline
2 & $11.68(18|20)$ & $1.75(3|4)$ & $1.01(3|5)$  & $0.195(4|9)$   & $0.057(0|6)$  & $14.69(18|21)$\tabularnewline
3 & $11.63(18|20)$ & $0.99(2|3)$ & $0.33(1|2)$  & $0.0287(6|12)$ & $0.0034(0|3)$ & $12.99(19|20)$\tabularnewline
4 & $11.73(19|20)$ & $0.57(1|1)$ & $0.114(4|5)$ & $0.0044(1|2)$  & $2.3(0|2)\times 10^{-4}$ & $12.42(19|20)$\tabularnewline
\hline
\end{tabular}

\caption{
Results for the experimental moments for the minimal selection of data
sets. Conventions are as in Tab.~\ref{tab:momres1}. All moments are given in units of
$10^{-(n+1)}\,\mbox{GeV}^{-2n}$.
\label{tab:moments-results-minimal}}
}\end{center}\end{table}

\begin{table}[t!]\begin{center}{\small
\begin{tabular}{|c|cccccc|}
\hline 
n & Resonances & $3.73-4.8$ & $4.8-7.25$  & $7.25-10.538$ & $10.538-\infty$ & Total\tabularnewline
\hline
1 & $11.91(17|20)$ & $3.19(3|5)$ & $3.60(6|6)$  & $1.54(2|4)$   & $1.27(0|13)$  & $21.50(19|27)$\tabularnewline
2 & $11.68(18|20)$ & $1.77(2|3)$ & $1.11(2|2)$  & $0.217(4|5)$  & $0.057(0|6)$  & $14.83(18|20)$\tabularnewline
3 & $11.63(18|20)$ & $1.00(1|2)$ & $0.361(7|7)$ & $0.0319(6|8)$ & $0.0034(0|3)$ & $13.03(19|20)$\tabularnewline
4 & $11.73(19|20)$ & $0.57(1|1)$ & $0.123(3|2)$ & $0.0049(1|1)$ & $2.3(0|2)\times 10^{-4}$ & $12.43(19|20)$\tabularnewline
\hline
\end{tabular}

\caption{
Results for the experimental moments for the maximal selection of data sets.
Conventions are as in Tab.~\ref{tab:momres1}. All moments are given in units of
$10^{-(n+1)}\,\mbox{GeV}^{-2n}$.
\label{tab:moments-results-maximal}}
}\end{center}\end{table}

\begin{table}[t!]\begin{center}
\begin{tabular}{|c|cccc|}
\hline 
$n$ & This work & Kuhn~et~al.'07 \cite{Kuhn:2007vp} 
& Kuhn~et~al.'01 \cite{Kuhn:2001dm} & Hoang \& Jamin'04 \cite{Hoang:2004xm}\tabularnewline
\hline
$1$ & $21.21(20|30)$ & $21.66(31)$ & $20.65(84)$ & $20.77(47|90)$\tabularnewline
$2$ & $14.78(18|21)$ & $14.97(27)$ & $14.12(80)$ & $14.05(40|65)$\tabularnewline
$3$ & $13.02(18|20)$ & $13.12(27)$ & $12.34(79)$ & $12.20(41|57)$\tabularnewline
$4$ & $12.43(19|20)$ & $12.49(27)$ & $11.75(79)$ & $11.58(43|53)$\tabularnewline
\hline
\end{tabular}\end{center}
\caption{Comparison of our results for experimental moments to those from
  previous publications with a dedicated data analysis. 
The second column contains our results using the default setting. The third and
fourth columns were determined from the same collaboration and use data from
Refs.~\cite{Bai:2001ct,Ablikim:2006aj} (our datasets 2, 4)
and \cite{Bai:2001ct} (our dataset 2), respectively. The results in the fourth column were also
used in the charm mass analysis of Ref.~\cite{Boughezal:2006px}.
The numbers in the fifth column used data from
Refs.~\cite{Bai:2001ct,Blinov:1993fw,Ammar:1997sk} (our datasets 2, 10, 14). The moments in the third column
use (slightly less precise) experimental data on the narrow resonances from \cite{Nakamura:2010zzi}.
The moments in the last two columns were obtained using (less precise)
experimental data on the narrow resonances \cite{Yao:2006px}. These data have been updated later
\cite{Beringer:1900zz}. All moments are given in units of
$10^{-(n+1)}\,\mbox{GeV}^{-2n}$.
\label{tab:mom-comparison}}
\end{table}

\begin{figure*}[t]
\subfigure[]
{
\includegraphics[width=0.48\textwidth]{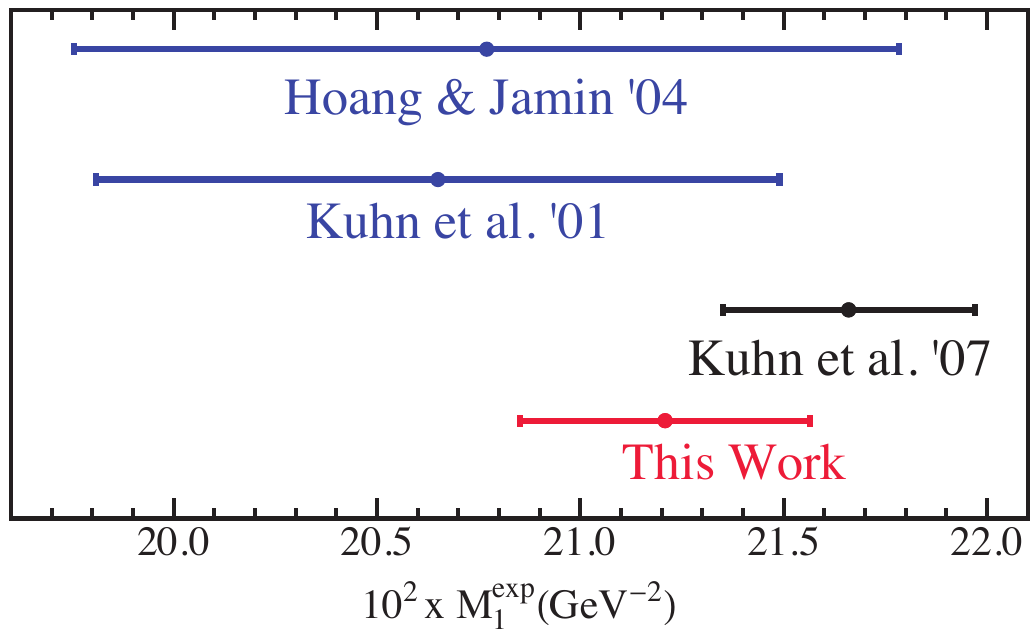}
\label{fig:M1comparison}
}
\subfigure[]{
\includegraphics[width=0.48\textwidth]{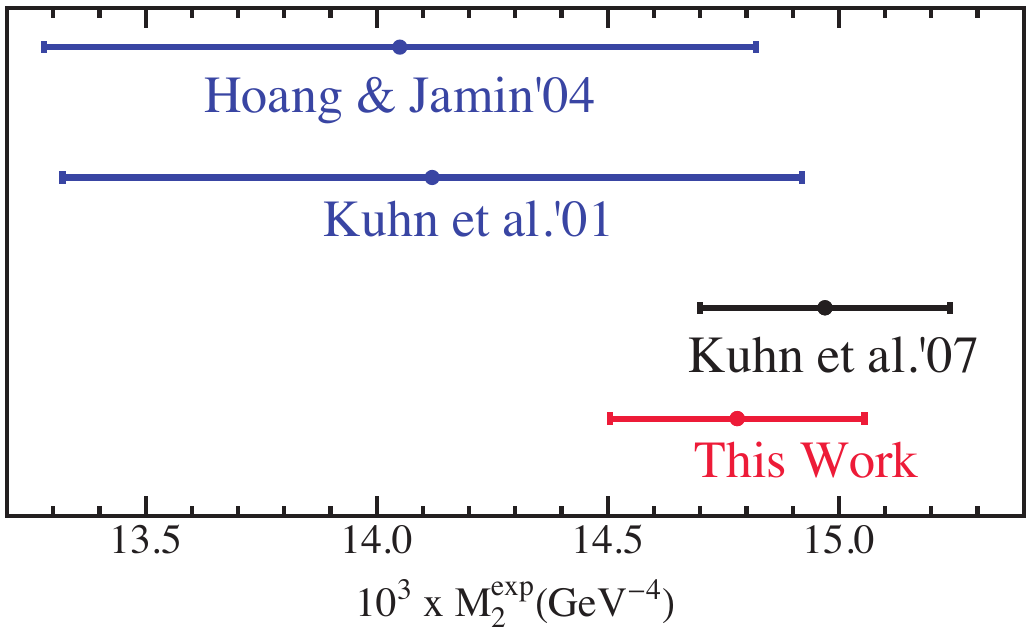}
\label{fig:M2comparison}
}
\subfigure[]{
\includegraphics[width=0.48\textwidth]{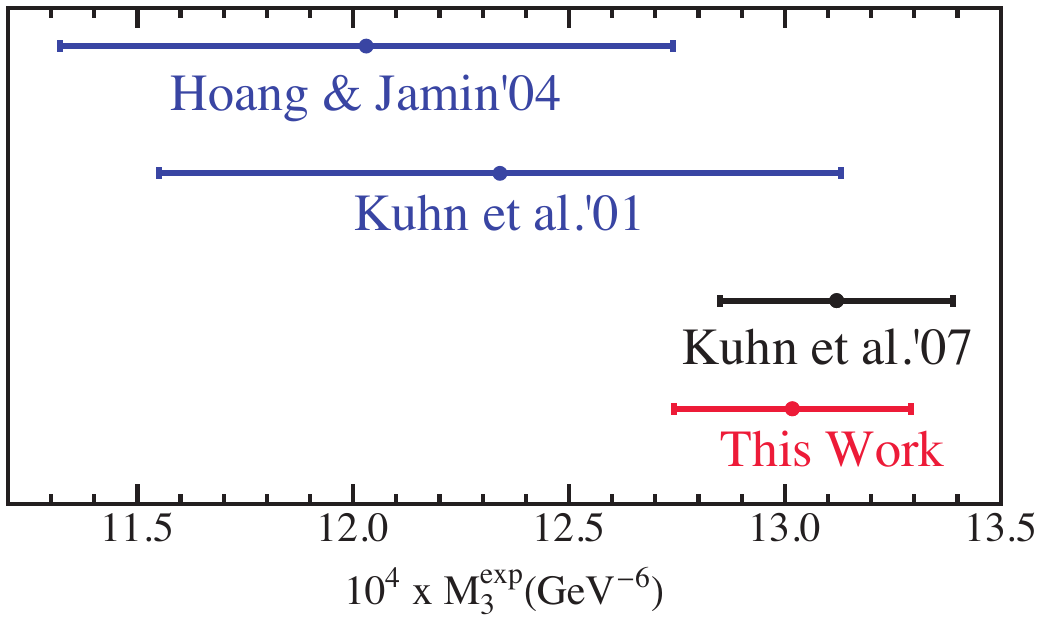}
\label{fig:M3comparison}
}
\subfigure[]{
\includegraphics[width=0.48\textwidth]{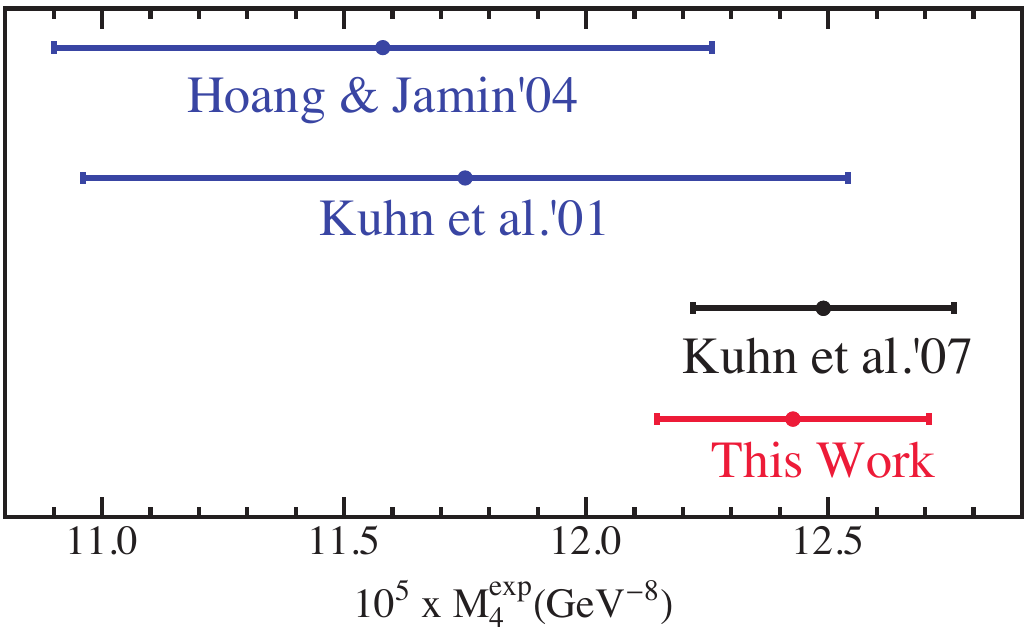}
\label{fig:M4comparison}
}
\caption{Comparison of determinations of the experimental moments. Blue lines
refer to analyses in which 
outdated values for the parameters of the narrow resonances \cite{Yao:2006px} have been used.
\label{fig:Moment-comparison}}
\end{figure*}

\subsection{Examination} 
\label{subsectionexaminations}

We conclude this section with an examination of some of the choices and
assumptions we have implemented for the treatment of the experimental
uncertainties. Our defaults choices include
\begin{itemize}
\item[(i)] treating one (quadratic) half of $J/\psi$ and $\psi^\prime$ $e^+e^-$
  partial width uncertainties as uncorrelated and the other half as positively
  correlated among themselves; assuming no correlations of the $J/\psi$ and
  $\psi^\prime$ partial widths to the $R_m$ cluster values;
\item[(ii)] treating the entire systematical uncertainties of the \mbox{$R$-ratio}
  measurements as correlated for the data sets where only total systematical
  uncertainties were quoted;
\item[(iii)] defining the cluster energies $E_m$ through the weighed average of
  measurement energies falling into the clusters, see Eq.~(\ref{Eclusterdef});
\item[(iv)] using $N_{cl}=52+1$ clusters distributed in groups of (2,20,20,10)
  clusters in the energy ranges bounded by $(3.73, 3.75, 3.79, 4.55, 10.538)$~GeV
  (see Sec.~\ref{subsectioncombination}) and
\item[(v)] using the default data set collection consisting of all data sets
  discussed in Sec.~\ref{subsectioncollections} except for sets 16, 17 and 19 as
  defined in Tab.~\ref{tab:datasets}.
\end{itemize}
In Tab.~\ref{tab:compcorr1} alternative correlation models are being
studied. The second column shows, as a reference, the first four moments
with our default settings and the other columns display the results 
applying changes to the default settings as
explained below. The third column displays the moments treating the 
uncertainties of the $J/\psi$ and $\psi^\prime$ partial widths as
uncorrelated. This decreases the total experimental error in $M^{\rm
  exp}_{1,2,3,4}$ by $(6,7,5,2)\%$. The fourth column displays the moments
treating the uncertainties of the $J/\psi$ and $\psi^\prime$ partial widths
being minimally correlated with the $R_m$ values.\footnote{We use a modified
  version of the minimal correlation model. The non-diagonal entries of the
  correlation matrix are filled in with $\Gamma_i\, R_m\, {\rm Min}^2\left\lbrace
    {\Delta{\Gamma_i}/\Gamma_i, \Delta R_m/R_m}\right\rbrace
  $. Here $\Delta\Gamma_i$ and $\Delta R_m$ represent the systematical
  uncertainties of the width of the narrow resonance and the $R$ value of the
  \mbox{$m$-th} cluster, respectively, and $i=1,2$ refer to $J/\psi$ and
  $\psi^\prime$.} 
Compared to the default setting this increases the total experimental error in
$M^{\rm exp}_{1,2,3,4}$ by $(14,6,3,3)\%$. 
In the fifth column we display the moments treating, for data sets
1, 6, 13, 15, 16 and 17, one (quadratic) half of systematical uncertainties for the
\mbox{$R$-values} uncorrelated and the other half correlated.
In the sixth column all $R_m$ values of all data sets are treated as 
completely uncorrelated. We see
that the central values depend only weakly on the correlation model for those
data were the corresponding information is unknown. In particular, for
the determination of the uncertainties the ignorance about the composition of
the systematical uncertainties in
the \mbox{$R$-values} from data sets 1, 6, 13, 15, 16 and 17 is not essential. 
However, for quoting the final uncertainties it is important to account for all (known) 
correlations since they can affect the outcome significantly.

In Tab.~\ref{tab:compcorr2} we examine the impact of modifying the definition of
the cluster energy $E_m$ and of changes to the default cluster distribution
(2,20,20,10). In the second column we display the resulting moments of the default
setting. In the third and fourth columns we have shown the moments using
for $E_m$ simply the mean of the energies and the center of the cluster,
respectively. The resulting differences to the default definition is an order of
magnitude smaller than the uncertainties and thus negligible.
The fifth, sixth and seventh columns display the moments using some alternative
cluster distributions. The deviations for the default choice illustrated in the
table are much smaller than the uncertainties and typical for all modifications
that satisfy the guidelines for viable cluster definitions we have formulated in
Sec.~\ref{subsectioncombination}. This demonstrates that the choice of the
cluster distribution does not result in a bias for the resulting experimental
moments.

\begin{table}[t!]\begin{center}
{\small
\begin{tabular}{|c|ccccc|}
\hline 
 $n$ & Default & $J/\psi$ and $\psi^\prime$ uncorr.  & Min. overlap to cont. & 50\% correlation & Uncorrelated\tabularnewline
\hline
$1$ & $21.21(20|30)$ & $21.21(26|22)$ & $21.21(20|36)$ & $21.08(22|31)$ & $20.93(28|24)$\tabularnewline
$2$ & $14.78(18|21)$ & $14.78(25|06)$ & $14.78(18|23)$ & $14.76(19|21)$ & $14.72(20|20)$\tabularnewline
$3$ & $13.02(19|20)$ & $13.02(26|02)$ & $13.02(19|21)$ & $13.02(19|20)$ & $13.02(19|20)$\tabularnewline
$4$ & $12.43(19|20)$ & $12.43(27|01)$ & $12.43(19|20)$ & $12.44(19|20)$ & $12.44(19|20)$\tabularnewline
\hline
\end{tabular}
\caption{
Dependence on the correlation model. In the second column we show again
the results for our default set up. In third column we treat the width of the narrow resonances
as uncorrelated. In the fourth column we depict the moments when treating
the widths of the narrow resonances as minimally correlated to the values of $R_m$.
In the fifth column we treat the quadratic half of the systematical errors of data sets 
1, 6, 13, 15, 16 and 17 as correlated. In the sixth column we show the results
when all systematical errors of all data sets are treated as
uncorrelated. (Note that in the latter case there is still correlation
coming from the narrow resonances and the contributions from the perturbative
QCD region). All moments are given in units of
$10^{-(n+1)}\,\mbox{GeV}^{-2n}$.
\label{tab:compcorr1}}
}\end{center}\end{table}

\begin{table}[t!]\begin{center}
{\footnotesize
\begin{tabular}{|c|cccccc|}
\hline 
 $n$& Default & Regular average & Middle point & $(2,20,40,10)$ & $(2,10,20,10)$ & $(2,20,20,20)$ \tabularnewline
\hline
$1$ & $21.21(20|30)$ & $21.21(20|30)$ & $21.24(20|29)$ & $21.20(20|30)$ & $21.22(20|30)$ & $21.22(20|29)$\tabularnewline
$2$ & $14.78(18|21)$ & $14.78(18|21)$ & $14.80(18|21)$ & $14.78(18|21)$ & $14.78(18|21)$ & $14.79(18|21)$\tabularnewline
$3$ & $13.02(19|20)$ & $13.02(19|20)$ & $13.03(19|20)$ & $13.02(19|20)$ & $13.02(19|20)$ & $13.02(19|20)$\tabularnewline
$4$ & $12.43(19|20)$ & $12.43(19|20)$ & $12.43(19|20)$ & $12.43(19|20)$ & $12.43(19|20)$ & $12.43(19|20)$\tabularnewline
\hline
\end{tabular}
\caption{Stability with clustering. In the second column we show the
results for our default set up. In third and fourth column we depict the
resulting moments when for the cluster
energy we pick the regular average of the energies and the center of the
cluster, respectively. In the last three columns we change the default
clustering to $(2,20,40,10)$, $(2,10,20,10)$ and $(2,20,20,20)$.
All moments are given in units of $10^{-(n+1)}\,\mbox{GeV}^{-2n}$.
\label{tab:compcorr2}}
}\end{center}\end{table}

Finally, we also examine the dependence of the moments on the data set
collections as described in Sec.~\ref{subsectioncollections}. 
In Tabs.~\ref{tab:moments-results-minimal} and \ref{tab:moments-results-maximal}
the results for the moments are displayed using the minimal and the maximal
collections (with default choices for all other settings). We see that the
differences in the central values to the default 
collection are the same size as the systematical correlated uncertainties
for the first moment $M_1^{\rm exp}$. For the higher moments the differences are
much smaller than the uncertainties. Using, instead of the default, the minimal
and maximal collections affects the systematic (statistic) uncertainty of $M_1^{\rm exp}$ by only
about $10\%$ ($5\%$). For the higher moments the differences decrease strongly and
basically disappear for the fourth moment. Again, the results show that having
a slightly increased or decreased redundancy in the data set collection only has a minor
impact on the final numbers for the experimental moments.

To summarize, we find that modifications to the choices and assumptions that go
into the combined treatment of the experimental data from different publications
and experiments lead to changes that are well within the experimental
uncertainties we obtain from our combination method. We therefore consider these
uncertainties as appropriate. An instructive comparison of the moments obtained
in our analysis to those obtained in some previous publications is given in
Tab.~\ref{tab:mom-comparison}. A graphical illustration of the results is shown
in Fig.~\ref{fig:Moment-comparison}.

\begin{figure*}[t!]
\subfigure[]
{
\includegraphics[width=0.48\textwidth]{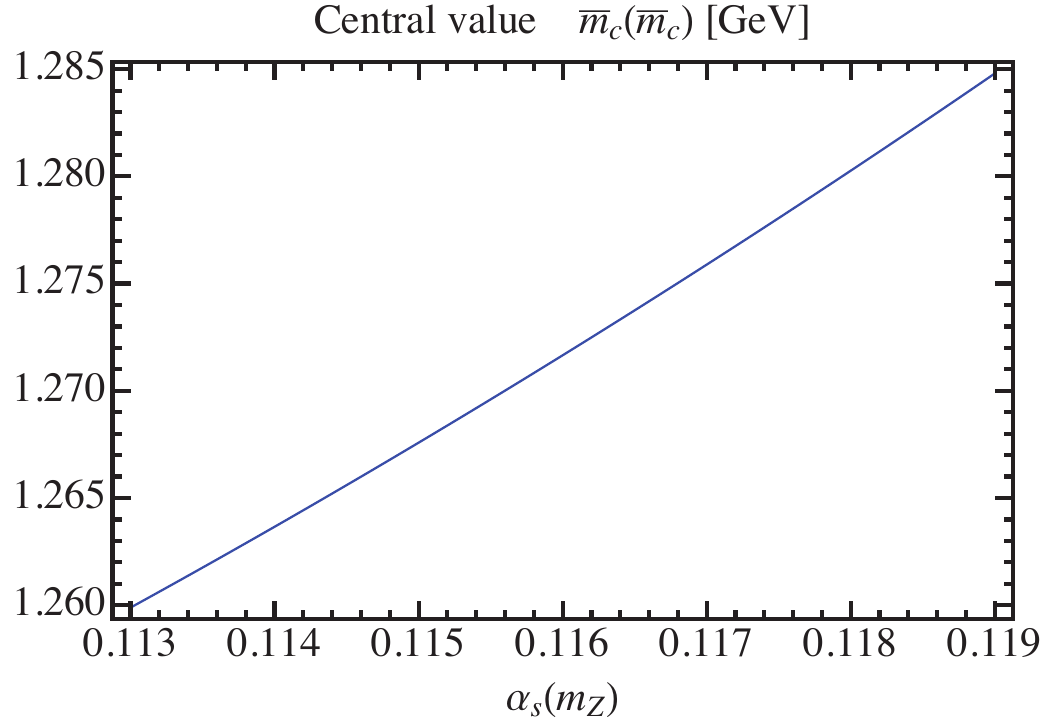}
\label{fig:figmcalphascentral}
}
\subfigure[]{
\includegraphics[width=0.48\textwidth]{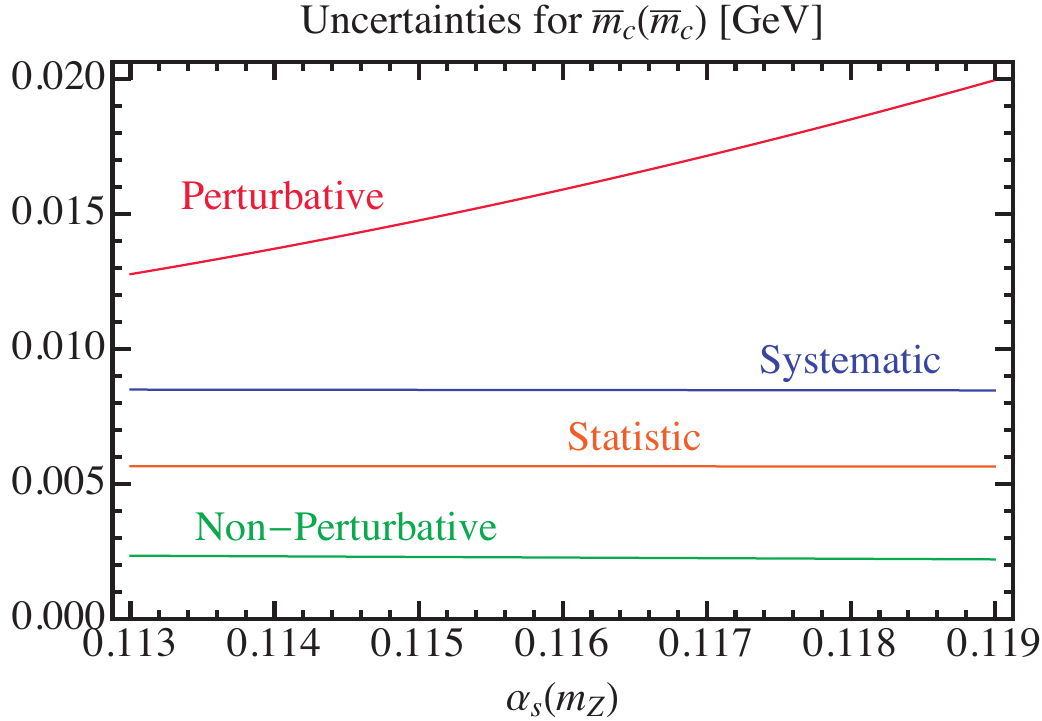}
\label{fig:figmcalphaspert}
}
\caption{
Dependence on $\alpha_s(m_Z)$ of the central values of 
${\overline m_c}({\overline m_c})$ (a) and
its perturbative (b), experimental statistical (c) and experimental systematical (d)
errors. The red dots correspond to the values we actually calculated.
The blue curve shows an interpolation and the red line in (a) and (b) corresponds
to a linear fit.}
\label{figmcalphas}%
\end{figure*}

\begin{figure*}[t!]
\subfigure[]
{
\includegraphics[width=0.48\textwidth]{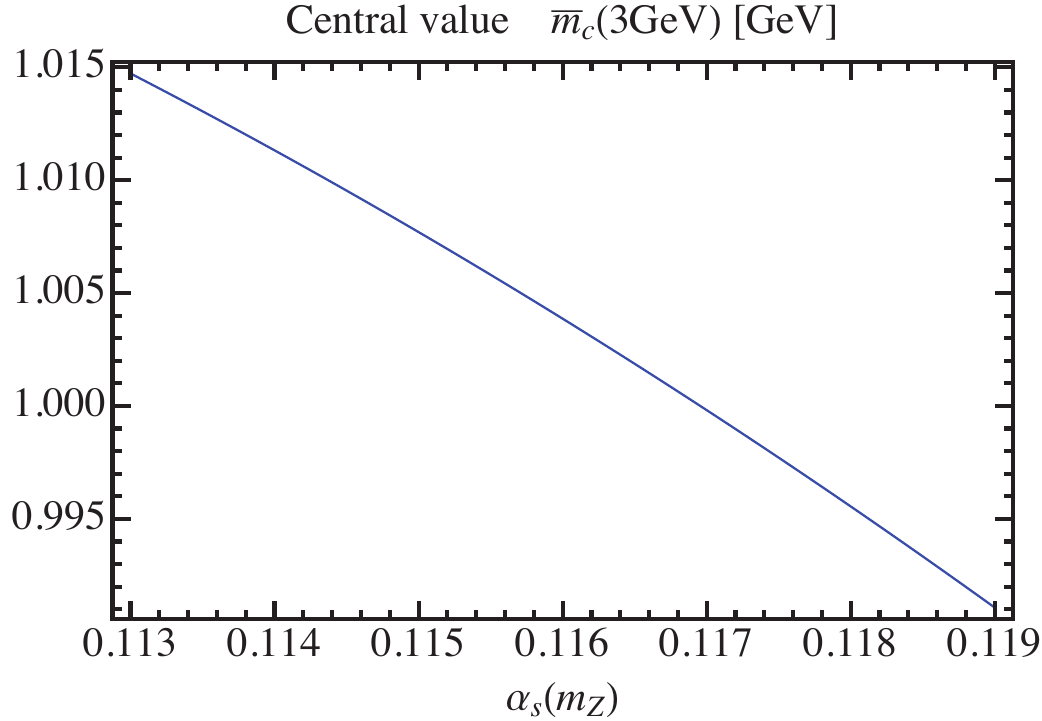}
\label{fig:m3calphascentral}
}
\subfigure[]{
\includegraphics[width=0.48\textwidth]{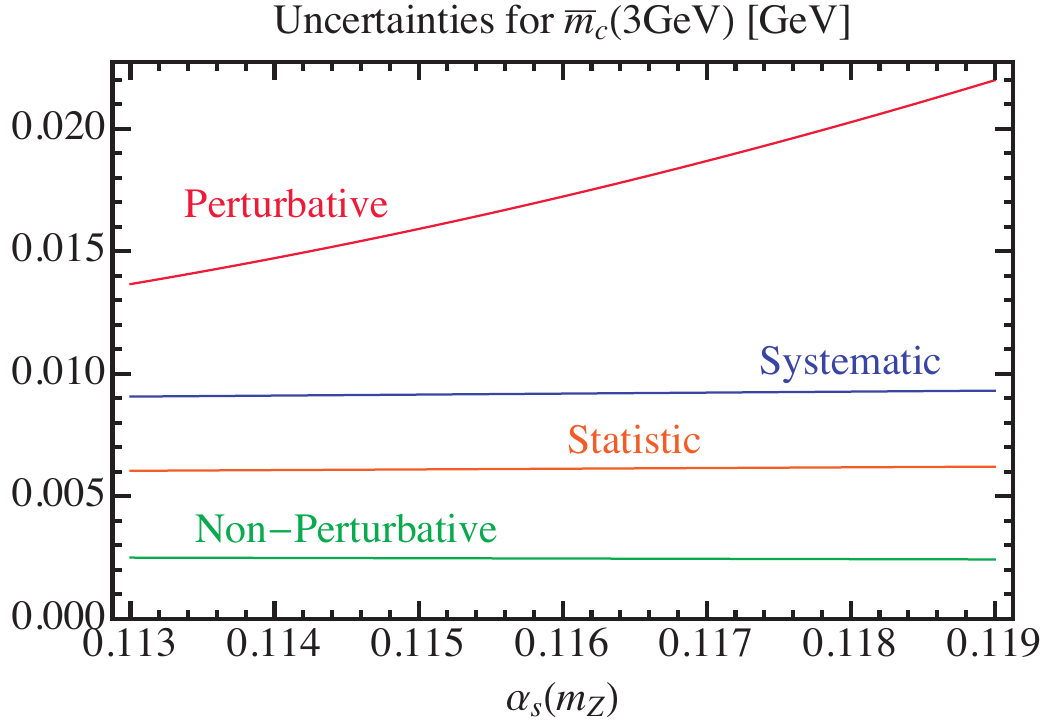}
\label{fig:m3calphaspert}
}
\caption{
Dependence on $\alpha_s(m_Z)$ of the central values of 
${\overline m_c}(3\,{\rm GeV})$ (a) and
its perturbative (b), experimental statistical (c) and experimental systematical (d)
errors. The red dots correspond to the values we actually calculated.
The blue curve shows an interpolation and the red line in (a) and (b) corresponds
to a linear fit.}
\label{fig:m3calphas}%
\end{figure*}

\section{Charm Quark Mass Analysis for the First Moment}
\label{sectionanalysis}

\begin{table}[t!]\begin{center}
\begin{tabular}{|c|cccccccc|}
\hline 
$\alpha_s(m_Z)$ & ${\overline m_c}({\overline m_c})$  & $\Delta_{\rm pert}$ & $\Delta_{\rm stat}$ & 
$\Delta_{\rm syst}$ & ${\overline m_c}(3\,{\rm GeV})$ & $\Delta_{\rm pert}$& $\Delta_{\rm stat}$ & 
$\Delta_{\rm syst}$\tabularnewline
\hline
$0.113$ & $1.260$ & $0.013$ & $0.006$ & $0.009$ & $1.015$ & $0.014$ & \
$0.006$ & $0.009$ \tabularnewline
$0.114$ & $1.264$ & $0.014$ & $0.006$ & $0.008$ & $1.011$ & $0.015$ & \
$0.006$ & $0.009$\tabularnewline
$0.115$ & $1.268$ & $0.015$ & $0.006$ & $0.009$ & $1.008$ & $0.016$ & \
$0.006$ & $0.009$\tabularnewline
$0.116$ & $1.272$ & $0.016$ & $0.006$ & $0.008$ & $1.004$ & $0.017$ & \
$0.006$ & $0.009$ \tabularnewline
$0.117$ & $1.276$ & $0.017$ & $0.006$ & $0.009$ & $1.000$ & $0.019$ & \
$0.006$ & $0.009$\tabularnewline
$0.118$ & $1.280$ & $0.018$ & $0.006$ & $0.009$ & $0.996$ & $0.020$ & \
$0.006$ & $0.009$\tabularnewline
$0.119$ & $1.285$ & $0.020$ & $0.006$ & $0.008$ & $0.991$ & $0.022$ & \
$0.006$ & $0.009$ \tabularnewline
\hline
\end{tabular}\end{center}
\caption{Results for the central values of ${\overline m_c}({\overline m_c})$
  and ${\overline m_c}(3\,{\rm GeV})$  
(second  and sixth columns) and their
perturbative (third and seventh columns), experimental statistical (fourth and eight columns)
and experimental systematical (fifth and ninth columns) errors.}
\label{tabmcalphas}
\end{table}

\begin{figure}[t]
\center
 \includegraphics[width=0.9\textwidth]{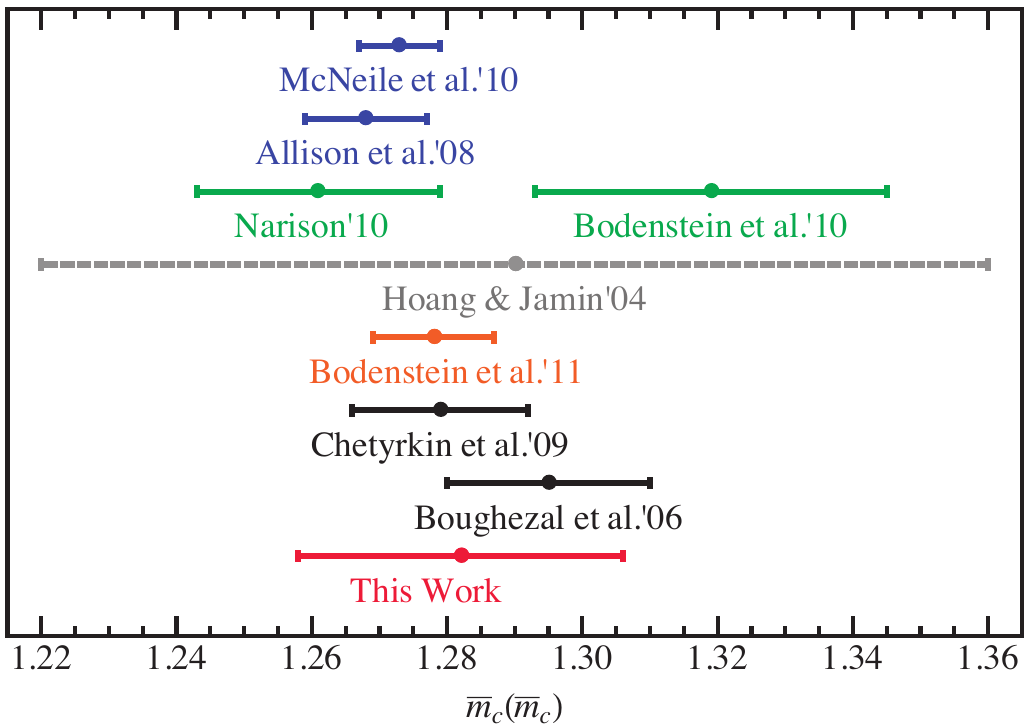}
 \caption{
Comparison of recent determinations of $\overline{m}_{c}(\overline{m}_{c})$. Red
corresponds to our result, black and gray correspond to ${\mathcal O}(\alpha_s^3)$ and
${\mathcal O}(\alpha_s^2)$ charmonimum sum rules analyses, respectively, 
green labels other kind of sum rule analyses (weighted finite energy sum rules
\cite{Bodenstein:2010qx} 
and ratios of \mbox{$Q^2$-dependent} moments \cite{Narison:2010cg}), and blue stands
for lattice simulations.
 \label{fig:comparison-alpha} }
\end{figure}

Since it is theoretically most reliable, we use the first moment $M_1$
as our default for our final numerical charm quark mass analysis. As 
ingredients for the analysis we use
\begin{itemize}
\item[(1)] the iterative expansion method for the perturbative contribution of
  the theoretical moment at ${\cal O}(\alpha_s^3)$, see
  Eq.~(\ref{eq:iterative-general}),
\item[(2)] the gluon condensate correction with its Wilson coefficient
  determined at ${\cal O}(\alpha_s)$ as described in
  Sec.~\ref{subsectioncondensate},
\item[(3)] the first experimental moment
\begin{eqnarray}
\label{M1exp}
M_1^{\rm exp} & = & 0.2121\pm 0.0020_{\rm stat}\pm 0.0030_{\rm syst}~\mbox{GeV}^{-2}\,,
\end{eqnarray}
using our default settings as discussed in Sec.~\ref{sectiondata}.
\end{itemize}
One important source of uncertainty we have not yet discussed is the value of the
strong $\overline{\mbox{MS}}$ coupling $\alpha_s$. Since in the recent
literature~\cite{Gehrmann:2012sc,Abbate:2012jh,Abbate:2010xh,Frederix:2010ne,Gehrmann:2009eh,Blumlein:2006be,Blumlein:2010rn} 
$\alpha_s$ determinations with a spread larger than the
current world average~\cite{Bethke:2009jm} have been obtained, we carry out our numerical
analysis for values of $\alpha_s(m_Z)$ between $0.113$ and $0.119$.\footnote{
As our default we use $\alpha_s^{n_f=5}(m_Z)$ as the input, use the four-loop
QCD beta-function for the renormalization group evolution and three-loop
matching conditions to the $n_f=4$ theory at $\mu=4.2$~GeV.
}
The outcome of our analysis is shown in Tab.~\ref{tabmcalphas}. In
Figs.~\ref{figmcalphas} and \ref{fig:m3calphas} the central values,
perturbative, statistical and systematical
uncertainties are displayed graphically. For the central value and the
perturbative uncertainty, which show a significant dependence on $\alpha_s$, we
can present a linear fit. For the statistical and systematical uncertainties the
variation with $\alpha_s$ is smaller than $1$~MeV and we only quote constant
values. We thus obtain
\begin{eqnarray}
\label{mcfinalalphas}
\overline m_c(\overline m_c)& = & (1.282 + 4.15\,[\alpha_s(m_Z) - 0.1184] ) 
\, \pm \, (0.006)_{\rm stat} 
\, \pm \, (0.009)_{\rm syst}\\ &&
\, \pm \, (0.019 + 1.20\,[\alpha_s(m_Z) - 0.1184])_{\rm pert} 
\, \pm \, (0.002)_{\langle GG\rangle} \, {\rm GeV}
\,,\nonumber
\\[2mm]
\overline m_c(3~\mbox{GeV})& = &(0.994 - 3.94\,[\alpha_s(m_Z) - 0.1184]) 
\, \pm \, (0.006)_{\rm stat} 
\, \pm \, (0.009)_{\rm syst}\\ &&
\, \pm \, (0.021 + 1.39\,[\alpha_s(m_Z) - 0.1184])_{\rm pert} 
\, \pm \, (0.002)_{\langle GG\rangle}\,{\rm GeV}\,.\nonumber
\end{eqnarray}
Taking as an input
\begin{eqnarray}
\label{alphaswa}
\alpha_s(m_Z) & = & 0.1184  \pm 0.0021\,,
\end{eqnarray}
which is the current world Bethke average~\cite{Bethke:2009jm}
with a tripled uncertainty we obtain
\begin{eqnarray}
\label{mcfinalalphaswa}
\overline m_c(\overline m_c) & = & 1.282
\, \pm \, (0.006)_{\rm stat} 
\, \pm \, (0.009)_{\rm syst} 
\, \pm \, (0.019)_{\rm pert}
\, \pm \, (0.010)_{\alpha_s} \\&&
\, \pm \, (0.002)_{\langle GG\rangle}\, {\rm GeV}\nonumber
\,,
\\[2mm]
\overline m_c(3~\mbox{GeV}) & = & 0.994
\, \pm \, (0.006)_{\rm stat} 
\, \pm \, (0.009)_{\rm syst} 
\, \pm \, (0.021)_{\rm pert}
\, \pm \, (0.010)_{\alpha_s}\\&&
\, \pm \, (0.002)_{\langle GG\rangle}\, {\rm GeV}\nonumber
\,,
\end{eqnarray}
which represents, together with Eq.~(\ref{mcfinalalphas}), our final numerical
result for the $\overline{\mbox{MS}}$ charm mass. Our result is in good
agreement with other recent determinations of $\overline{m}_{c}(\overline{m}_{c})$.
A summary of the numerical results is shown in Tab.~\ref{tab:comparison-alpha}
and in Fig.~\ref{fig:comparison-alpha}. Compared to the analysis carried out in
Refs.~\cite{Chetyrkin:2009fv,Kuhn:2007vp} our
experimental uncertainty is larger by $2$~MeV and our perturbative uncertainty
is larger by $17$~MeV, which is
a factor of $10$ larger.
Compared to Refs.~\cite{Chetyrkin:2006xg,Boughezal:2006px}
the discrepancy in the perturbative error estimate is even larger.
Among the most recent high-precision determination of the charm mass based on
${\cal O}(\alpha_s^3)$ input from perturbative QCD, our
analysis has the biggest error mainly due to our more appropriate treatment of
perturbative uncertainties.

\begin{table}[t!]\begin{center}
\begin{tabular}{|l|ccc|}
\hline 
 & $\overline{m}_{c}(\overline{m}_{c})$ & $\alpha_{s}(m_{Z})$ used & $\overline{m}_{c}(\overline{m}_{c})^{\alpha_{s}(m_{Z})=0.1180}$ \tabularnewline
\hline
This work & $1.282\pm0.024$ & $0.1184\pm0.0021$ & $1.280\pm0.023$\tabularnewline
Chetyrkin~et~al. \cite{Chetyrkin:2009fv} & $1.279\pm0.013$ & $0.1189\pm0.0020$ & $1.277\pm0.012$\tabularnewline
Boughezal~et~al. \cite{Boughezal:2006px} & $1.295\pm0.015$ & $0.1182\pm0.0027$ & $ $--\tabularnewline
Hoang \& Jamin \cite{Hoang:2004xm} & $1.29\phantom{0}\pm0.07\phantom{0}$ & $0.1180\pm0.0030$ & $1.29\pm0.07$\tabularnewline
Bodenstein~et~al. \cite{Bodenstein:2010qx} & $1.319\pm0.026$ & $0.1213\pm0.0014$ & $1.295\pm0.026$\tabularnewline
Bodenstein~et~al. \cite{Bodenstein:2011ma} & $1.278\pm0.009$ & $0.1184\pm0.0007$ & --\tabularnewline
Narison \cite{Narison:2010cg} & $1.261\pm0.018$ & $0.1191\pm0.0027$ & $ $--\tabularnewline
Allison~et~al. \cite{Allison:2008xk} & $1.268\pm0.009$ & $0.1174\pm0.0012$ & $ $--\tabularnewline
McNeile~et~al. \cite{McNeile:2010ji} & $1.273\pm0.006$ & $0.1183\pm0.0007$ & $ $--\tabularnewline
\hline
\end{tabular}\end{center}
\caption{Comparison of recent determinations of $\overline{m}_{c}(\overline{m}_{c})$. 
In the second column we show the final result as quoted in the publications,
and in the third the value of $\alpha_{s}(m_{Z})$ used in the analysis. In the fourth column, when possible,
we extrapolate the results to $\alpha_{s}(m_{Z})=0.1180$. For the result of
Ref.~\cite{Bodenstein:2010qx} we perform a linear extrapolation  
using the two values for $\alpha_{s}(m_{Z})=0.1189$ and $0.1213$ quoted in the paper. The last two rows 
correspond to lattice studies. In the analysis of McNeile et al.\ the value of $\alpha_{s}(m_{Z})$ was
simultaneously determined from the analyses. There have been recent determinations of the
charm mass using DIS data, see Refs.~\cite{Alekhin:2012vu,Alekhin:2012un}, which are not included on
this table. The values found are in agreement with our determination, but with significantly larger errors.
updated table\label{tab:comparison-alpha}}
\end{table}

\section{Higher Moment Analysis}
\label{sectionanalysishigher}
\begin{figure}[t]
\center
 \includegraphics[width=0.6\textwidth]{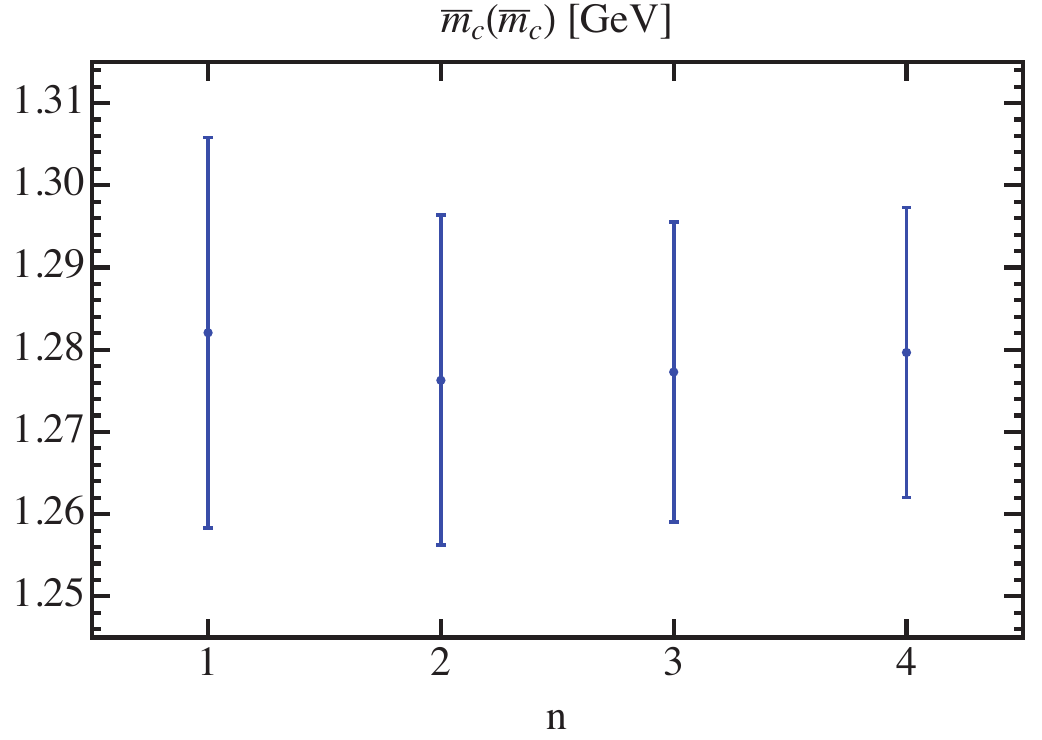}
 \caption{Determination of $\overline{m}_{c}(\overline{m}_{c})$ from the first four moments.
 We use the expanded out iterative method, including the gluon condensate contribution.
 All uncertainties have been added in quadrature to compute the shown error bar.
 \label{fig:higher-moments}}
\end{figure}

To cross check the results for the charm mass obtained from the first moment
$M_1$ we now carry out an analysis of the moments $M_n$ for $n=2,3,4$ using
again the iterative expansion method of the theoretical ${\cal O}(\alpha_s^3)$
moments (see Eq.~(\ref{eq:iterative-general})) and the gluon condensate
correction as discussed in Sec.~\ref{subsectioncondensate}. To obtain the
perturbative error we again vary the renormalization scales $\mu_m$ and
$\mu_\alpha$ as discussed in Sec.~\ref{subsectionmcerror}. The experimental
moments are
\begin{alignat}{8}\label{Mnexp}
& M_2^{\rm exp} &\;=\;&& 0.01478   &&\;\pm\;&& 0.00018_{\rm stat}  &&\;\pm\;&& 0.00021_{\rm syst} & ~\mbox{GeV}^{-4}\,,\\
& M_3^{\rm exp} &\;=\;&& 0.001302  &&\;\pm\;&& 0.000019_{\rm stat}  &&\;\pm\;&& 0.000020_{\rm syst}& ~\mbox{GeV}^{-6}\,, \nonumber\\
& M_4^{\rm exp} &\;=\;&& 0.0001243 &&\;\pm\;&& 0.0000019_{\rm stat}  &&\;\pm\;&& 0.0000020_{\rm syst}& ~\mbox{GeV}^{-8}\,, \nonumber
\end{alignat}
using our default settings as discussed in Sec.~\ref{sectiondata}. Due to the
strong correlation of the experimental moments there is essentially no gain in
statistical power from a combined fit. We therefore carry out individual fits of
the higher moments and consider the results as consistency checks of our first
moment analysis, and we do not intend to average the results. The results for
$\overline m_c(\overline m_c)$ and
$\overline m_c(3~\mbox{GeV})$ keeping $\alpha_s(m_Z)$ as a parameter are given 
in App.~C.
Taking Eq.~(\ref{alphaswa}) as input for $\alpha_s(m_Z)$ we obtain the results
shown in Tab.~\ref{tab:high-n}. The results are in excellent agreement with the
outcome of the first moment analysis. A graphical comparison of the results with
all uncertainties added in quadrature is
given in Fig.~\ref{fig:higher-moments}.
\begin{table}[tb!]
\center
\begin{tabular}{|l|ccccccc|}
\hline 
 $n$ & $\overline{m}_{c}(\overline{m}_{c})$ & $\Delta_{\rm stat}$ & $\Delta_{\rm syst}$
& $\Delta_{\rm pert}$ & $\Delta_{\rm \alpha_s}$ & $\Delta_{\langle GG\rangle}$ & $\Delta_{\rm tot}$\tabularnewline
\hline
1 & $1.282$ & $0.006$ & $0.009$ & $0.019$ & $0.010$ & $0.002$ & $0.024$ \tabularnewline
2 & $1.276$ & $0.004$ & $0.004$ & $0.018$ & $0.007$ & $0.004$ & $0.020$ \tabularnewline
3 & $1.277$ & $0.003$ & $0.003$ & $0.016$ & $0.005$ & $0.004$ & $0.018$ \tabularnewline
4 & $1.280$ & $0.002$ & $0.002$ & $0.016$ & $0.004$ & $0.005$ & $0.018$ \tabularnewline
\hline
\end{tabular}
\caption{Results for $\overline{m}_{c}(\overline{m}_{c})$ for the first four 
moments in GeV units. The first column labels the moment, the central value is shown in the second,
statistical and systematical experimental uncertainties are listed in third and fourth
column, and the last two columns show errors due to uncertainties in $\alpha_s$ and the
gluon condensate. In this table we use the value $\alpha_s(m_Z)=0.1184\pm0.0021$.
\label{tab:high-n}}
\end{table}

\section{Conclusions and Final Thoughts}
\label{sectionconclusions}

In this work we have used state-of-the-art ${\cal O}(\alpha_s^3)$ input from
perturbative QCD to determine the $\overline{\mbox{MS}}$ charm quark mass from
relativistic (low $n$) charmonium sum rules using experimental data on the
total hadronic cross section in $e^+e^-$ annihilation. The main aims were (i) to
carefully reexamine perturbative uncertainties in the charm mass extractions
from the
moments of the charm vector current correlator and (i) to fully exploit the
available experimental data on the hadronic cross section.

We carried out this work having in mind recent ${\mathcal O}(\alpha_s^3)$ sum rule
analyses \cite{Chetyrkin:2006xg,Boughezal:2006px,Chetyrkin:2009fv, Kuhn:2007vp, Bodenstein:2011ma}, 
where perturbative errors of $2$~MeV or smaller were quoted.
Moreover in Refs.~\cite{Chetyrkin:2009fv, Kuhn:2007vp} an experimental uncertainty
of $9$~MeV was quoted. Given these numbers we found it appropriate to reexamine 
this sum rules analysis. In their work the
perturbative uncertainty estimate was achieved using a specific choice to arrange
the perturbative expansion and by setting the renormalization scales in
$\alpha_s$ ($\mu_\alpha$) and of the $\overline{\mbox{MS}}$ charm quark mass
($\mu_m$) equal. We found that this results in an accidental
cancellation of $\mu_\alpha$ and $\mu_m$ scale variations that is not
observed in other alternative ways to treat the perturbative
expansion. Moreover, concerning the experimental input their work relied on
perturbative QCD predictions instead of available data for energies
$\sqrt{s}$ between $4.8$~GeV and $10.538$~GeV
which results in a dependence on the ad-hoc assumption concerning the experimental
uncertainty in particular for the first moment.

Concerning the assessment of perturbative and experimental uncertainties
we implemented in our work the following improvements:
\begin{itemize}
\item[1.]
We demonstrated that for achieving an estimate of perturbative uncertainties
based on scale variations that is independent of the perturbative expansion method
one needs to vary $\mu_\alpha$ and $\mu_m$ independently, albeit with ranges
that avoid large logs. As a result the perturbative uncertainty estimates using
different ways to carry out the expansion in $\alpha_s$ become equivalent, which
is not the case for $\mu_\alpha=\mu_m$. Another important ingredient
of our perturbative error estimate is that we allow renormalization scales at
the charm mass $\overline m_c(\overline m_c)$, where the perturbative expansion
is stable, but which was excluded in some of the previous analyses.
\item[2.]
Using a data clustering method similar to Refs.~\cite{Agostini:1993uj,Takeuchi:1995xe,
Hagiwara:2003da} we combined available
data on the total $e^+e^-$ hadronic cross section from many different
experiments covering energies up to $\sqrt{s}=10.538$~GeV to fully exploit the
existing experimental information for the experimental moments. This avoids a
significant dependence of the experimental moments on ad-hoc assumptions on the 
``experimental'' uncertainty being associated to the
QCD theory input used for energies above $10.538$~GeV. This is because energies
above $10.538$~GeV only have very small contributions to the low-$n$ moments. As
a result we were also able 
to quantify the correlation between different experimental
moments. We also took
the opportunity to include recent PDG updates concerning the data for the $\psi^\prime$ resonance.
\end{itemize}
Using $\alpha_s(m_Z)$ as an unspecified variable and the theoretically more
reliable first moment $M_1$ for the fits we have obtained
\begin{eqnarray}
\overline m_c(\overline m_c)& = &(1.282 + 4.15\,[\alpha_s(m_Z) - 0.1184]) 
\, \pm \, (0.006)_{\rm stat} 
\, \pm \, (0.009)_{\rm syst}\\ &&
\, \pm \, (0.019 + 1.20\,[\alpha_s(m_Z) - 0.1184])_{\rm pert} 
\, \pm \, (0.002)_{\langle GG\rangle}\, {\rm GeV}
\,,\nonumber
\\[2mm]
\overline m_c(3~\mbox{GeV})& = &(0.994 - 3.94\,[\alpha_s(m_Z) - 0.1184]) 
\, \pm \, (0.006)_{\rm stat} 
\, \pm \, (0.009)_{\rm syst}\\ &&
\, \pm \, (0.021 + 1.39\,[\alpha_s(m_Z) - 0.1184])_{\rm pert} 
\, \pm \, (0.002)_{\langle GG\rangle}\, {\rm GeV}\,.\nonumber
\end{eqnarray}
for the $\overline{\mbox{MS}}$ charm mass. At the level of uncertainties
obtained in our work excellent convergence of perturbation theory was observed.
Adopting $\alpha_s(m_Z) = 0.1184 \pm 0.0021$ we then obtain
\begin{eqnarray}
\overline m_c(\overline m_c) & = & 1.282
\, \pm \, (0.006)_{\rm stat} 
\, \pm \, (0.009)_{\rm syst} 
\, \pm \, (0.019)_{\rm pert}
\, \pm \, (0.010)_{\alpha_s}\\&&
\, \pm \, (0.002)_{\langle GG\rangle}\, {\rm GeV}\nonumber
\,,\nonumber
\\[2mm]
\overline m_c(3~\mbox{GeV}) & = & 0.994 
\, \pm \, (0.006)_{\rm stat} 
\, \pm \, (0.009)_{\rm syst} 
\, \pm \, (0.021)_{\rm pert}
\, \pm \, (0.010)_{\alpha_s} \\&&
\, \pm \, (0.002)_{\langle GG\rangle}\, {\rm GeV}
\,.\nonumber
\end{eqnarray}
Our perturbative error of $19$~MeV is a factor of ten larger, and
the experimental uncertainty of $11$~MeV is by $2$~MeV
larger than in the most recent analysis of Ref.~\cite{Kuhn:2007vp}.
For estimating the perturbative
error a range of scale variations between $\overline m_c(\overline m_c)$ and
$4$~GeV was employed. Adding all uncertainties quadratically we obtain
\begin{eqnarray}
\label{conclusionmc}
\overline m_c(\overline m_c) & = & 1.282 
\, \pm \, 0.024~\mbox{GeV}
\,,\nonumber
\\[2mm]
\overline m_c(3~\mbox{GeV}) & = & 0.994\,\pm \, 0.026~\mbox{GeV}
\,,
\end{eqnarray}
giving an uncertainty that is twice the size of the one obtained in
Refs.~\cite{Chetyrkin:2009fv,Kuhn:2007vp}. This difference arises mainly from
the more appropriate estimate of perturbative uncertainties we obtained in our work.

As a final thought one might ask which further improvements might be possible in
the future. As can be seen from Tab.~\ref{tab:momres1}, from the experimental
side the biggest improvement could be made from more accurate measurements of
the $J/\psi$ and $\psi^\prime$ electronic partial widths. The current relative
uncertainties are $2.5$\% and $1.7$\%, respectively. Here some improvement might
be conceivable with dedicated measurements. On theory side, viewing the
uncertainties and the good behavior of the perturbative series, it is not
unreasonable to assume that the computation of ${\cal 
  O}(\alpha_s^4)$ moments of the vector current correlator might further reduce
the perturbative error below the level of $20$~MeV. Using the
$\overline{\mbox{MS}}$ scheme for the charm mass the OPE states that the
remaining perturbative infrared renormalon ambiguity is of order $\Lambda_{\rm
  QCD}^4/\overline m_c^3\sim {\cal O}(5 \mbox{\,-}15~\mbox{MeV})$. This
expectation has also been confirmed by explicit bubble chain (large-$\beta_0$)
computations~\cite{Grozin:2004ez} and indicates that a further reduction of the
perturbative uncertainty is not excluded.

However, to throw in some words of caution, at the level of the present
perturbative uncertainties one should also remind oneself about possible
loopholes still left in the charmonium sum rule method. An issue we would like
to mention concerns the separation of charm and non-charm hadronic production
rates needed to carry out the charmonium sum rule. On the theory side the issue
is conceptually subtle due to the singlet and secondary charm radiation contributions which arise
at ${\cal O}(\alpha_s^3)$ and ${\cal O}(\alpha_s^2)$, respectively. In this work
(as well as in Ref.~\cite{Chetyrkin:2009fv}) both contributions have been considered as
non-charm although they contain terms belonging to the $c\bar c$ final
state. This treatment might be justified since the size of the corresponding
terms are quite small (see Tab.~\ref{tabneglected}) and since it is the common
approach to determine the experimental charm production rate in the continuum
region by subtracting theoretical
results (or models) for the non-charm rate from the measured total hadronic
\mbox{$R$-ratio}. In 
our method to determine the experimental moments this subtraction involves a
normalization constant multiplying the theoretical non-charm \mbox{$R$-ratio} that is
fitted within our clustering method as well accounting for data below and above
the charm threshold. The 
result (see Eq.~(\ref{eq:ns-results})) reveals a disparity of $4$\% between the theoretical
non-charm \mbox{$R$-ratio} and the data. Setting, in contrast, the normalization constant
to unity results in a shift in the charm mass by $-15$~MeV.\footnote{
The same disparity was found in Ref.~\cite{Kuhn:2007vp}. In their analysis the
corresponding effect is $-5$~MeV since experimental data were used in the
experimental moments only for energies $\sqrt{s}<4.8$~MeV. In this approach,
however, the moments are strongly dependent on the uncertainty one assigns to
the theory input used for $\sqrt{s}>4.8$~MeV.
} 
Since this shift is compatible with the overall systematical uncertainty in the
experimental data, we have not treated it as an additional source of
uncertainty. On the other hand, the size of the shift could also be
considered as an inherent conceptual uncertainty related to separating the
charm from the non-charm \mbox{$R$-ratio}, which is based on theory considerations
rather than on experimental methods and which apparently cannot be improved
simply by
higher order perturbative computations. We also refer to
Refs.~\cite{Portoles:2001yu,Portoles:2002rt}  
for related conceptual discussions. 
As an alternative, one might avoid the
separation of the charm and the non-charm contributions altogether and use the
total hadronic cross section for the charm mass fits. Apart from the shift
mentioned above such an approach would, however, also lead to a substantially
larger dependence on the uncertainties in $\alpha_s$. Given these considerations
we believe that a substantial reduction of the uncertainties also relies on a
resolution of the disparity mentioned above. This might certainly involve more 
precise measurements in the charm threshold and below-threshold regions, but
also some deeper conceptual insight. Until
then a substantial reduction of the uncertainties shown in
Eq.~(\ref{conclusionmc}) appears hard to achieve without imposing ad-hoc
assumptions.

\bigskip
\acknowledgments
This work was supported in part by the European Community's Marie-Curie
Research Networks under contract MRTN-CT-2006-035482 (FLAVIAnet),
MRTN-CT-2006-035505 (HEPTOOLS) and PITN-GA-2010-264564 (LHCphenOnet), and by the U.S. Department
of Energy under the grant FG02-94ER40818. 
V.\ Mateu has been partially supported by a
DFG ``Eigenen Stelle'' under contract MA 4882/1-1 and by a Marie Curie
Fellowship under contract PIOF-GA-2009-251174. S.\ Zebarjad thanks the
MPI for hospitality while part of this work was accomplished. S.\ Zebarjad and
V.\ Mateu are grateful to the MPI guest program for partial support.
We thank S.\ Schutzmeier for confirmation of our numerical ${\cal O}(\alpha_s^3)$
fixed-order results.
A.\ Hoang acknowledges discussion with H.\ K\"uhn and C.\ Sturm.
V.\ Mateu acknowledges discussion with J.\ J.\ Hern\'andez Rey.
We thank D~Nomura and T.~Teubner for providing us with the effective
electromagnetic constant.
We thank the {\it Erwin Schr\"odinger International Institute for Mathematical
  Physics} (ESI Vienna), where a part of this work has been accomplished, for partial
support.

\section*{Appendix A: Results of the Fit Procedure}

\addcontentsline{toc}{section}{Appendix A: Results of the Fit Procedure}\newcounter{alpha1} \renewcommand{\thesection}{\Alph{alpha1}} \renewcommand{\theequation}{\Alph{alpha1}.\arabic{equation}} \renewcommand{\thetable}{\Alph{alpha1}.\arabic{table}} \setcounter{alpha1}{1} \setcounter{equation}{0} \setcounter{table}{0}
\label{ap:FitOutcome}

In this appendix we present in some more detail the numerical results of our fit
procedure. In Tabs.~\ref{tab:FitFunction-Standard},
\ref{tab:FitFunction-Minimal} and \ref{tab:FitFunction-Maximal}, 
the results for the cluster energies and the cluster charmed \mbox{$R$-values} are  shown for the
standard, minimal and maximal selection of data sets, respectively, using our
default setting for the correlations. We use the results for the standard data
set selection for our final charm mass analysis. The numbers in the parentheses
correspond 
to the statistical and systematical errors. The correlation matrices for the
\mbox{$R$-values} is available, but cannot be displayed due to lack of space. They can
be obtained by the authors on request.
For the three data selections, the
fit gives the following minimal $\chi^2$ per degree of freedom,
\begin{equation}
\dfrac{\chi_{\rm standard}^2}{\rm dof}=1.89\,, \qquad
\dfrac{\chi_{\rm minimal}^2}{\rm dof}=1.86\,,
\qquad \dfrac{\chi_{\rm maximal}^2}{\rm dof}=1.81\,,
\label{eq:chi2-results}
\end{equation}
and the following normalization constants for the non-charm background
\begin{eqnarray}
\!\!\!\!\!\!n^{\rm standard}_{\rm ns}&=&1.039 \pm 0.003_{\rm stat} \pm
0.012_{\rm syst},\;  
n^{\rm minimal}_{\rm ns}=1.029 \pm 0.003_{\rm stat} \pm 
0.015_{\rm syst},
\label{eq:ns-results} \\
\!\!\!\!\!\!n^{\rm maximal}_{\rm ns}&=&1.023 \pm 0.003_{\rm stat} \pm 0.011_{\rm syst}.\nonumber
\end{eqnarray}
The fit results for the normalization constant $n_{\rm ns}$ is compatible
with the corresponding normalization constant $n_-=1.038$ used in
Ref.~\cite{Kuhn:2007vp} for the subtraction of the non-charm background for the BES
2001 dataset (our data set~2) but is not compatible with the result for the
subtraction constant $n_-=0.991$ concerning the BES 2006 data set (our dataset~5).
Since the minimal $\chi^2/{\rm dof}$ values are not close to unity, one has to
conclude that the fit quality is not really very good. This is not at all visible
from the agreement of the fit and the data for the total cross section (see
Figs.~\ref{fig:fitcompilation}) and thus might be related to the disparity
between the fits of charm versus non-charm production rates described in
Sec.~\ref{sectionconclusions}. 

\begin{figure*}[t]
\subfigure[]
{
\includegraphics[width=0.48\textwidth]{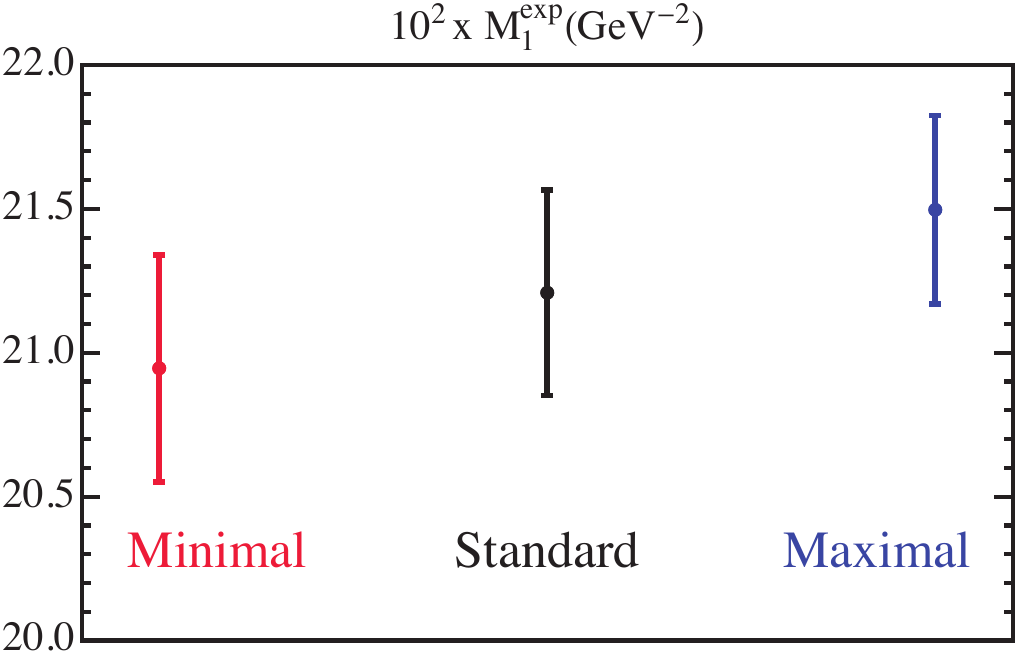}
\label{fig:M1selection}
}
\subfigure[]{
\includegraphics[width=0.48\textwidth]{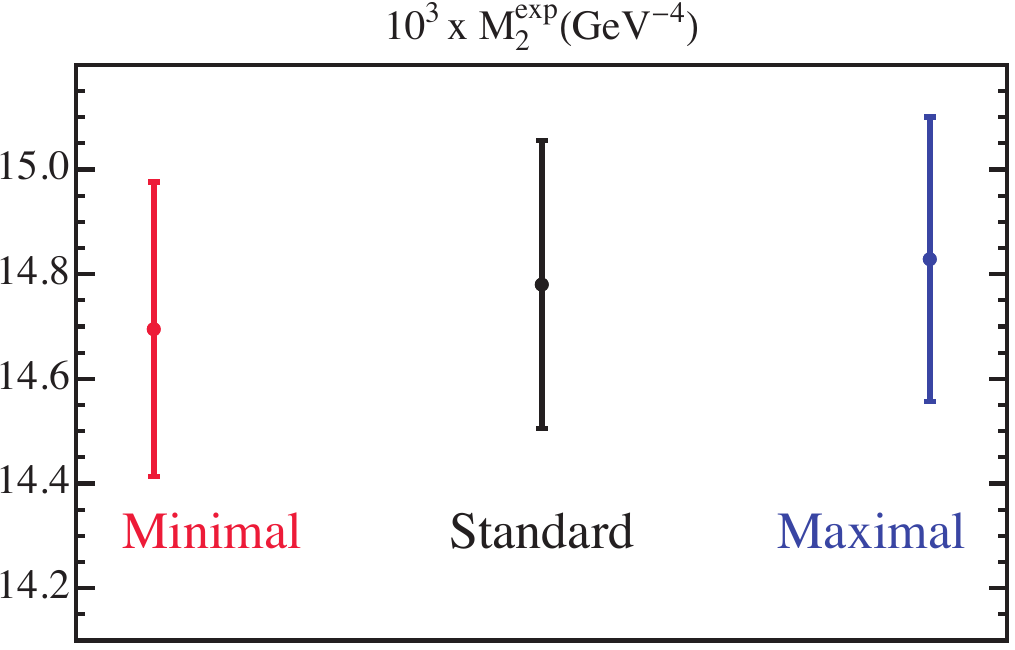}
\label{fig:M2selection}
}
\subfigure[]{
\includegraphics[width=0.48\textwidth]{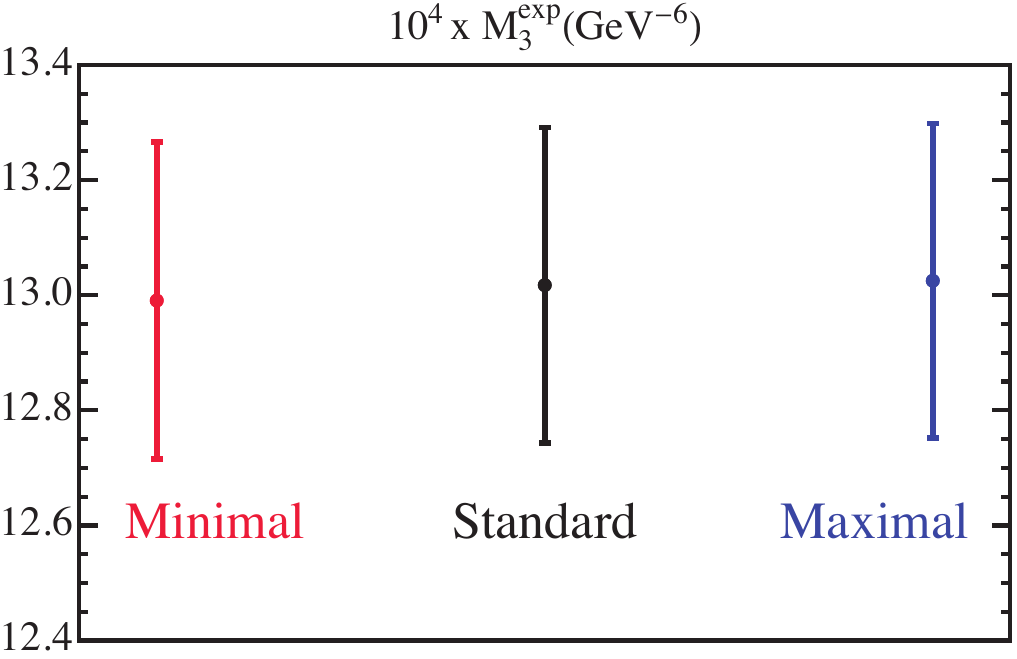}
\label{fig:M3selection}
}
\subfigure[]{
\includegraphics[width=0.48\textwidth]{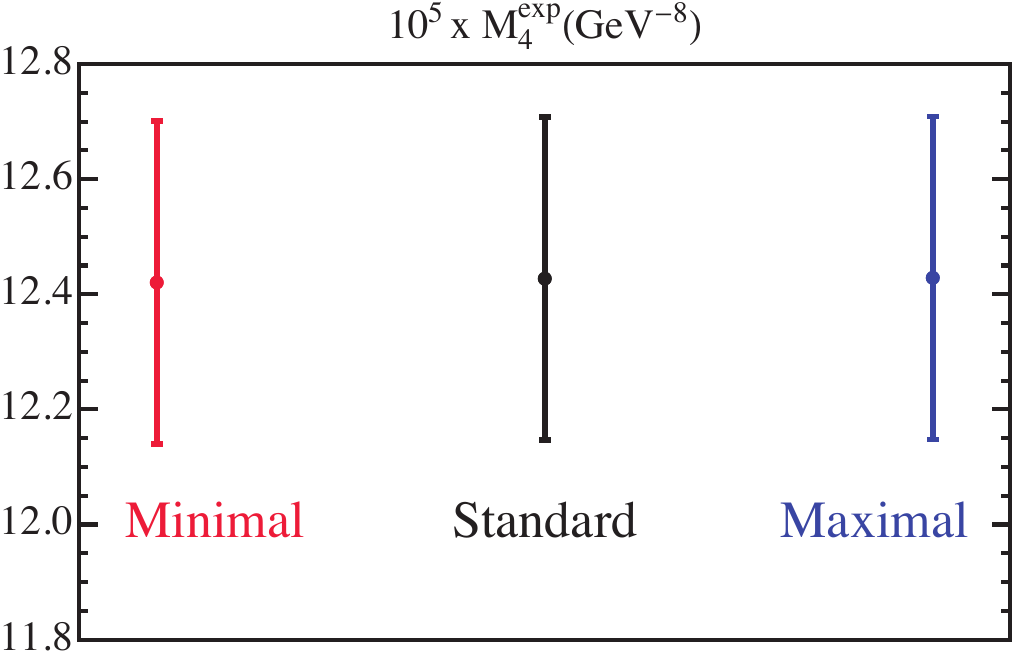}
\label{fig:M4selection}
}
\caption{Comparison of the results for the experimental moments using the three data selections.
\label{fig:Selection-comparison}}
\end{figure*}

In Eq.~(\ref{correlmatrices}) we show for the correlation matrices of the first
four experimental moments for the minimal and the maximal data set
selection. The results for 
our standard selection are given in Eqs.~(\ref{CMtotal}) and (\ref{CMuncorr}).
All numbers are related to moments $M_n^{\rm exp}$ normalized to units of
$10^{-(n+1)}\,\mbox{GeV}^{-2n}$. 
We show the results accounting for the full set of correlated and uncorrelated
uncertainties and the correlation matrices accounting only for
uncorrelated systematical and statistical uncertainties (subscript uc). 

\begin{table}[t!]\begin{center}
\begin{tabular}{|cc|cc|cc|cc|}
\hline
$E$ & $R$ &$E$ & $R$ &$E$ & $R$ &$E$ & $R$ \tabularnewline
\hline
$ 3.736$ & $ -0.03(6|2)$ & $ 3.749$ & $  0.23(6|2)$ & $ 3.751$ & $  0.47(9|3)$ & $ 3.753$ & $  0.39(7|3)$\tabularnewline
$ 3.755$ & $  0.41(6|3)$ & $ 3.757$ & $  0.80(10|3)$ & $ 3.759$ & $  0.75(7|3)$ & $ 3.761$ & $  0.84(7|3)$\tabularnewline
$ 3.763$ & $  1.06(8|3)$ & $ 3.765$ & $  1.23(9|4)$ & $ 3.767$ & $  1.52(12|4)$ & $ 3.769$ & $  1.31(8|3)$\tabularnewline
$ 3.771$ & $  1.39(6|4)$ & $ 3.773$ & $  1.42(2|4)$ & $ 3.775$ & $  1.22(8|3)$ & $ 3.777$ & $  1.29(8|4)$\tabularnewline
$  3.78$ & $  1.13(7|3)$ & $ 3.781$ & $  1.06(5|3)$ & $ 3.783$ & $  1.00(7|3)$ & $ 3.785$ & $  0.63(11|3)$\tabularnewline
$ 3.787$ & $  0.64(7|3)$ & $  3.79$ & $  0.41(5|2)$ & $ 3.808$ & $  0.11(4|2)$ & $ 3.846$ & $  0.10(5|2)$\tabularnewline
$ 3.883$ & $  0.24(7|2)$ & $ 3.928$ & $  0.63(8|2)$ & $ 3.967$ & $  0.94(3|2)$ & $ 4.002$ & $  1.30(3|3)$\tabularnewline
$ 4.033$ & $  2.18(3|4)$ & $ 4.069$ & $  1.74(4|3)$ & $ 4.117$ & $  1.72(4|3)$ & $ 4.156$ & $  1.68(1|4)$\tabularnewline
$ 4.191$ & $  1.54(3|3)$ & $  4.23$ & $  0.91(5|2)$ & $ 4.261$ & $  0.71(2|2)$ & $ 4.307$ & $  0.89(6|2)$\tabularnewline
$ 4.346$ & $  1.18(6|2)$ & $ 4.382$ & $  1.63(7|3)$ & $ 4.416$ & $  1.76(4|3)$ & $ 4.452$ & $  1.53(7|3)$\tabularnewline
$ 4.492$ & $  1.42(6|3)$ & $ 4.529$ & $  1.29(10|3)$ & $ 4.715$ & $  1.51(3|3)$ & $ 5.326$ & $  1.39(7|7)$\tabularnewline
$ 6.006$ & $  1.33(6|7)$ & $ 6.596$ & $  1.35(7|7)$ & $ 7.202$ & $  1.39(2|4)$ & $ 7.852$ & $  1.52(8|5)$\tabularnewline
$ 8.417$ & $  1.38(3|5)$ & $  9.04$ & $  1.43(5|6)$ & $  9.54$ & $  1.35(3|5)$ & $ 10.327$ & $  1.37(1|5)$\tabularnewline
\hline
\end{tabular}
\end{center}
\caption{Best fit values for the standard selection of data sets.
The energy of the cluster is measured in GeV, and for $R$ the first number in brackets is the
statistical error and the second the systematical one.\label{tab:FitFunction-Standard}}
\end{table}

\begin{eqnarray}\label{correlmatrices}
C^{\rm exp}_{\rm min}&=&\left(
\begin{array}{cccc}
0.156 & 0.094 & 0.080 & 0.077\\
0.094 & 0.079 & 0.076 & 0.076\\
0.080 & 0.076 & 0.076 & 0.077\\
0.077 & 0.076 & 0.077 & 0.079\\ 
\end{array}
\right),\,\, C^{\rm exp}_{\rm min, uc}\,=\,\left(
\begin{array}{cccc}
0.046 & 0.037 & 0.034 & 0.034\\
0.037 & 0.034 & 0.034 & 0.035\\
0.034 & 0.034 & 0.035 & 0.036\\
0.034 & 0.035 & 0.036 & 0.037
\end{array}\right),\\\nonumber\\
C^{\rm exp}_{\rm max}&=&\left(
\begin{array}{cccc}
0.107 & 0.079 & 0.075 & 0.075\\
0.079 & 0.074 & 0.074 & 0.075\\
0.075 & 0.074 & 0.075 & 0.077\\
0.075 & 0.075 & 0.077 & 0.079\\
\end{array}
\right),\,\, 
C^{\rm exp}_{\rm max, uc}\,=\,\left(
\begin{array}{cccc}
0.036 & 0.033 & 0.033 & 0.033\\
0.033 & 0.033 & 0.033 & 0.034\\
0.033 & 0.033 & 0.034 & 0.036\\
0.033 & 0.034 & 0.036 & 0.037
\end{array}\right).\nonumber
\end{eqnarray}

\begin{table}[t!]\begin{center}
\begin{tabular}{|cc|cc|cc|cc|}
\hline
$E$ & $R$ &$E$ & $R$ &$E$ & $R$ &$E$ & $R$ \tabularnewline
\hline
$ 3.736$ & $ -0.07(6|2)$ & $ 3.749$ & $  0.22(6|2)$ & $ 3.751$ & $  0.38(9|3)$ & $ 3.753$ & $  0.38(7|3)$\tabularnewline
$ 3.755$ & $  0.40(6|3)$ & $ 3.757$ & $  0.68(11|3)$ & $ 3.759$ & $  0.74(7|3)$ & $ 3.761$ & $  0.82(7|4)$\tabularnewline
$ 3.763$ & $  1.01(8|4)$ & $ 3.765$ & $  1.21(9|4)$ & $ 3.767$ & $  1.49(12|4)$ & $ 3.769$ & $  1.27(8|4)$\tabularnewline
$ 3.772$ & $  1.37(6|4)$ & $ 3.773$ & $  1.38(8|4)$ & $ 3.775$ & $  1.21(8|4)$ & $ 3.777$ & $  1.26(8|4)$\tabularnewline
$  3.78$ & $  1.11(7|3)$ & $ 3.781$ & $  1.06(5|4)$ & $ 3.783$ & $  0.97(7|4)$ & $ 3.785$ & $  0.62(11|3)$\tabularnewline
$ 3.787$ & $  0.66(8|3)$ & $  3.79$ & $  0.40(5|3)$ & $ 3.808$ & $  0.11(5|2)$ & $ 3.845$ & $  0.12(5|2)$\tabularnewline
$ 3.879$ & $  0.18(9|2)$ & $ 3.935$ & $  0.85(14|3)$ & $ 3.969$ & $  1.01(4|3)$ & $ 4.002$ & $  1.39(3|4)$\tabularnewline
$ 4.032$ & $  2.25(5|5)$ & $ 4.066$ & $  1.94(4|5)$ & $ 4.118$ & $  1.78(5|5)$ & $ 4.157$ & $  1.77(1|5)$\tabularnewline
$ 4.191$ & $  1.62(3|5)$ & $ 4.232$ & $  0.99(8|3)$ & $ 4.261$ & $  0.76(2|3)$ & $ 4.309$ & $  0.83(10|3)$\tabularnewline
$  4.35$ & $  1.22(11|3)$ & $ 4.385$ & $  1.28(14|4)$ & $ 4.415$ & $  1.62(10|4)$ & $  4.45$ & $  1.54(11|4)$\tabularnewline
$ 4.492$ & $  1.30(13|4)$ & $ 4.529$ & $  1.03(13|3)$ & $ 4.716$ & $  1.25(8|3)$ & $ 5.378$ & $  1.40(8|8)$\tabularnewline
$ 6.008$ & $  1.31(7|8)$ & $ 6.622$ & $  1.30(8|8)$ & $ 7.206$ & $  1.37(2|6)$ & $ 7.856$ & $  1.47(8|7)$\tabularnewline
$ 8.417$ & $  1.36(3|6)$ & $ 9.037$ & $  1.40(5|7)$ & $ 9.544$ & $  1.31(3|6)$ & $ 10.252$ & $  1.34(3|6)$\tabularnewline
\hline
\end{tabular}
\end{center}
\caption{Best fit values for the minimal selection of data sets. Conventions are as in
Tab.~\ref{tab:FitFunction-Standard}.\label{tab:FitFunction-Minimal}}
\end{table}

\begin{table}[t!]\begin{center}
\begin{tabular}{|cc|cc|cc|cc|}
\hline
$E$ & $R$ &$E$ & $R$ &$E$ & $R$ &$E$ & $R$ \tabularnewline
\hline
$ 3.736$ & $  0.01(6|2)$ & $ 3.749$ & $  0.29(5|2)$ & $ 3.751$ & $  0.48(8|3)$ & $ 3.753$ & $  0.40(7|2)$\tabularnewline
$ 3.755$ & $  0.42(6|3)$ & $ 3.757$ & $  0.80(10|3)$ & $ 3.759$ & $  0.76(7|2)$ & $ 3.761$ & $  0.85(6|3)$\tabularnewline
$ 3.763$ & $  1.06(8|3)$ & $ 3.765$ & $  1.24(8|3)$ & $ 3.767$ & $  1.51(12|4)$ & $ 3.769$ & $  1.31(7|3)$\tabularnewline
$ 3.771$ & $  1.39(6|4)$ & $ 3.773$ & $  1.41(2|4)$ & $ 3.775$ & $  1.21(8|3)$ & $ 3.777$ & $  1.28(8|4)$\tabularnewline
$  3.78$ & $  1.14(7|3)$ & $ 3.781$ & $  1.06(5|3)$ & $ 3.783$ & $  1.00(7|3)$ & $ 3.785$ & $  0.61(9|3)$\tabularnewline
$ 3.787$ & $  0.64(7|3)$ & $  3.79$ & $  0.41(5|2)$ & $ 3.808$ & $  0.13(4|2)$ & $ 3.846$ & $  0.10(4|2)$\tabularnewline
$ 3.884$ & $  0.26(5|2)$ & $ 3.927$ & $  0.66(7|2)$ & $ 3.967$ & $  0.98(3|2)$ & $ 4.002$ & $  1.30(2|2)$\tabularnewline
$ 4.033$ & $  2.18(3|3)$ & $  4.07$ & $  1.74(3|3)$ & $ 4.117$ & $  1.75(4|3)$ & $ 4.156$ & $  1.70(1|4)$\tabularnewline
$ 4.191$ & $  1.54(3|3)$ & $  4.23$ & $  0.91(5|2)$ & $ 4.261$ & $  0.74(2|2)$ & $ 4.307$ & $  0.87(6|2)$\tabularnewline
$ 4.346$ & $  1.15(5|2)$ & $ 4.382$ & $  1.55(6|3)$ & $ 4.416$ & $  1.80(4|3)$ & $ 4.452$ & $  1.52(6|3)$\tabularnewline
$ 4.492$ & $  1.33(5|2)$ & $  4.53$ & $  1.19(8|2)$ & $ 4.722$ & $  1.42(3|2)$ & $  5.36$ & $  1.39(6|4)$\tabularnewline
$ 6.018$ & $  1.47(4|3)$ & $ 6.608$ & $  1.66(4|3)$ & $ 7.202$ & $  1.54(2|3)$ & $ 7.851$ & $  1.63(8|5)$\tabularnewline
$ 8.417$ & $  1.50(3|4)$ & $  9.04$ & $  1.55(5|5)$ & $  9.54$ & $  1.47(3|4)$ & $ 10.327$ & $  1.48(1|4)$\tabularnewline
\hline
\end{tabular}
\end{center}
\caption{Best fit values for the maximal selection of data sets. Conventions are as in
Tab.~\ref{tab:FitFunction-Standard}.\label{tab:FitFunction-Maximal}}
\end{table}

\section*{\boldmath Appendix B: On the Equivalence of $\chi^2$-Functions}

\addcontentsline{toc}{section}{Appendix B: On the Equivalence of $\chi^2$-Functions}
\newcounter{alpha}
\renewcommand{\thesection}{\Alph{alpha}}
\renewcommand{\theequation}{\Alph{alpha}.\arabic{equation}}
\renewcommand{\thetable}{\Alph{alpha}} 
\setcounter{alpha}{2} \setcounter{equation}{0} \setcounter{table}{0}

In this appendix we demonstrate that a \mbox{$\chi^2$-function} in which the auxiliary
fit parameters $d_k$, which describe the correlated deviation off the
experimental central value within experiment $k$, multiplies only the
experimental systematical uncertainties,\footnote{
For the discussion in this appendix we do for simplicity of the presentation not
account for fits of the non-charm contribution. It is, however, straightforward
to generalize the presentation to this case.} 
\begin{equation}
\chi^2 = \sum_k\left[d_k^2+\sum_{i,m}\left(\dfrac{R_i^{k,m}+d_k\,
\Delta_i^{k,m}-R_m}{\sigma_i^{k,m}}\right)^2\right]\,, \label{additive}
\end{equation}
is mathematically equivalent to the well known \mbox{$\chi^2$-function} written solely
in terms of the fit parameters $R_m$ and a correlation matrix,  
\begin{equation}
\bar \chi^2=\sum_{i,j,k,m,n}(R_i^{k,m}-R_m)(V_k^{-1})_{ij}^{mn}(R_j^{k,n}-R_n)\,,
\label{traditional}
\end{equation}
where there is no correlation among the different experiments $k$ and
$k^\prime$, and the correlation matrix within one experiment $k$ has the form
\begin{equation}
(V_k)_{ij}^{mn}=\sigma_i^{k,m\,2}\delta_{ij}\delta^{mn}+\Delta_i^{k,m}\Delta_j^{k,n}
\,.
\label{tradcorr}
\end{equation}
The matrix Eq.~(\ref{tradcorr}) can be inverted analytically giving
\begin{equation}
(V_k^{-1})_{ij}^{mn} = \dfrac{\delta_{ij}\delta^{mn}}{(\sigma_i^{k,m})^2}-
\dfrac{1}{1+\dfrac{\left\langle \Delta^2\right\rangle_k}{\left\langle \sigma\right\rangle_k^2}}
\dfrac{\Delta_i^{k,m}}{(\sigma_i^{k,m})^2}\dfrac{\Delta_j^{k,n}}{(\sigma_j^{k,n})^2}\,,
\label{invtradcorr} 
\end{equation}
where we have defined the mean statistical error and the statistical average of
the systematical error within one experiment as follows:
\begin{equation}
\left\langle \sigma\right\rangle_k^2 \equiv 
\left(\sum_{i,m}\dfrac{1}{(\sigma_i^{k,m})^2}\right)^{-1},\qquad
\left\langle \Delta^2
\right\rangle_k\equiv\left\langle \sigma\right\rangle_k^2\sum_{i,m}
\dfrac{(\Delta_i^{k,m})^2}{(\sigma_i^{k,m})^2}\,.
\label{averages}
\end{equation}

In case of a sizable positive correlation between measurements in the same
experiment (such that $\Delta_i^{k,m}/R_i^{k,m}$ is sizable and approximately
constant) is it known that the form of the \mbox{$\chi^2$-function} in
Eq.~(\ref{additive}) leads to best fit values $R_m$ that are systematically
below the measurements~\cite{Agostini:1993uj}. Our proof demonstrates that the 
standard \mbox{$\chi^2$-function} of Eq.~(\ref{traditional}) has the same property
for correlation matrices with the form of Eq.~(\ref{tradcorr}). This motivates
to use the so-called minimal-overlap correlation model where the second term on
the RHS of Eq.~(\ref{traditional}) is replaced by
$\mbox{min}^2(\Delta_i^{k,m},\Delta_j^{k,n})$. In general, within the
minimal-overlap model, the
correlations are sufficiently reduced such that the unphysical effect described
above does not arise.  

We proceed by showing that one can ``integrate out"
auxiliary parameters $d_k$ in Eq.~(\ref{additive})
obtaining a new function $\tilde \chi^2(R^i)$ which yields the same
results for the best fit for the $R_m$,
as long as one works in the Gaussian approximation. 
The minimum of $\chi^2(R_i,d_k)$ is located at the best fit values (indicated by
superscripts $(0)$) defined by the conditions
\begin{equation}
\left.\dfrac{\partial\chi^{2}}{\partial R_{i}}\right|_{R_i^{(0)},d_k^{(0)}}=\left.\dfrac{\partial\chi^{2}}{\partial d_{j}}\right|_{R_i^{(0)},d_k^{(0)}}=0\,.\label{Mincond}
\end{equation}
To invert the matrix of second derivatives we proceed in blocks 
\begin{equation}
\left(\begin{array}{cc}
\dfrac{\partial^{2}\chi^{2}}{\partial R_{i}\partial R_{j}} & \dfrac{\partial^{2}\chi^{2}}{\partial R_{i}\partial d_{k}}\\
\dfrac{\partial^{2}\chi^{2}}{\partial d_{l}\partial R_{j}} & \dfrac{\partial^{2}\chi^{2}}{\partial d_{l}\partial b_{k}}\end{array}\right)_{\tilde{R},\tilde{b}}\left(\begin{array}{cc}
c_{jm} & b_{jr}\\
b_{mk} & a_{kr}\end{array}\right)=\left(\begin{array}{cc}
\delta_{im} & 0\\
0 & \delta_{lr}\end{array}\right).
\end{equation}
where $c_{ij}$ and $a_{kr}$ are  $N_{\rm cluster}\times N_{\rm cluster}$ and  
$N_{\rm exp}\times N_{\rm exp}$ square matrices, respectively, and
$b_{jr}$ is a  $N_{\rm cluster}\times N_{\rm exp}$ rectangular matrix. 
We find the following four matrix relations
\begin{eqnarray}
&&\!\!\!\!\!\!\!\!\!\!\!\!\!\!\!\!\!\!\!\!\!\!\!\!
\sum_{j}\dfrac{\partial^{2}\chi^{2}}{\partial R_{i}\partial R_{j}}\, c_{jm}+\sum_{k}\dfrac{\partial^{2}\chi^{2}}{\partial R_{i}\partial d_{k}}\, b_{mk}=\delta_{im}\,, \quad 
\sum_{j}\dfrac{\partial^{2}\chi^{2}}{\partial R_{i}\partial R_{j}}\,
b_{jr}+\sum_{k}\dfrac{\partial^{2}\chi^{2}}{\partial R_{i}\partial d_{k}}\,
a_{kr}=0\,,
\label{eq:general_block_1}\\
&&\!\!\!\!\!\!\!\!\!\!\!\!\!\!\!\!\!\!\!\!\!\!\!\!
\sum_{j}\dfrac{\partial^{2}\chi^{2}}{\partial d_{l}\partial R_{j}}\,
c_{jm}+\sum_{k}\dfrac{\partial^{2}\chi^{2}}{\partial d_{l}\partial d_{k}}\,
b_{mk}=0\,, 
\qquad\;\, \sum_{j}\dfrac{\partial^{2}\chi^{2}}{\partial 
d_{l}\partial R_{j}}\, b_{jr}+\sum_{k}\dfrac{\partial^{2}\chi^{2}}{\partial d_{l}\partial d_{k}} a_{kr}=\delta_{lr}\,.\label{eq:general_block_2}
\end{eqnarray}
Combining Eqs.~(\ref{eq:general_block_1}a) and (\ref{eq:general_block_2}a)
we find the inverse of the upper left \mbox{$R$-block},
\begin{equation}
(c^{-1})_{ij}=2(V^{-1})_{ij}=\dfrac{\partial^{2}\chi^{2}}{\partial R_{i}\partial R_{j}}-\sum_{kl}\dfrac{\partial^{2}\chi^{2}}{\partial d_{l}\partial R_{i}}\left[\dfrac{\partial^{2}\chi^{2}}{\partial d_{l}\partial 
d_{k}}\right]^{-1}\dfrac{\partial^{2}\chi^{2}}{\partial d_{k}\partial R_{j}}\,,
\label{eq:noEOM_Rj_bk}
\end{equation}
where $\left[\dfrac{\partial^{2}\chi^{2}}{\partial d_{l}\partial
    d_{k}}\right]^{-1}$ stands for the $(l,k)$ element of the inverse matrix of
$\dfrac{\partial^{2}\chi^{2}}{\partial d_{l}\partial d_{k}}$ (and not the
inverse of the element). 
Combining Eqs.~(\ref{eq:general_block_1}b) and (\ref{eq:general_block_2}b)
one can obtain relations for the $a$ and $b$ submatrices:
\begin{eqnarray}
b_{jr} & = & -\sum_{i,k}\left[\dfrac{\partial^{2}\chi^{2}}{\partial R_{j}\partial R_{i}}\right]^{-1}\dfrac{\partial^{2}\chi^{2}}{\partial d_{k}\partial R_{i}}\, a_{kr}\,,\\
\delta_{lr}&=&\sum_{k}\left(\dfrac{\partial^{2}\chi^{2}}{\partial d_{l}\partial d_{k}}-\sum_{i,j}\dfrac{\partial^{2}\chi^{2}}{\partial d_{l}\partial R_{j}}\left[\dfrac{\partial^{2}\chi^{2}}{\partial R_{j}\partial 
R_{i}}\right]^{-1}\dfrac{\partial^{2}\chi^{2}}{\partial d_{i}\partial R_{k}}\right)a_{kr}\,,\\
\delta_{im} & = & \sum_{j}\dfrac{\partial^{2}\chi^{2}}{\partial R_{i}\partial R_{j}}\, c_{jm}-\sum_{k,l,s}\dfrac{\partial^{2}\chi^{2}}{\partial d_{k}\partial R_{i}}\,\left[\dfrac{\partial^{2}\chi^{2}}{\partial R_{m}\partial 
R_{s}}\right]^{-1}\dfrac{\partial^{2}\chi^{2}}{\partial d_{l}\partial R_{s}}\, a_{lk}\,,
\end{eqnarray}
where again $\left[\dfrac{\partial^{2}\chi^{2}}{\partial R_{j}\partial R_{i}}\right]^{-1}$
stands for the $(j,i)$ element of the inverse matrix of $\dfrac{\partial^{2}\chi^{2}}{\partial R_{j}\partial R_{i}}$.
We ``integrate out" the auxiliary parameters $d_k$ by substituting their
minimum conditions $d^{(0)}_k=\tilde d_k(R^{(0)}_i)$
(which is analogous to using the equation of motion when integrating out heavy
particles): 
\begin{equation} 
\left.\dfrac{\partial\chi^{2}}{\partial d_{j}}\right|_{d_k=\tilde d_k(R_i)}=0\,,\qquad \tilde \chi^2(R_i)=\chi^2(R_i,\tilde d_k(R_i))\,.\label{EOM}
\end{equation}
Their first derivatives with respect to $R_{j}$ read
\begin{eqnarray}
\dfrac{\partial}{\partial R_{j}}\left.\dfrac{\partial\chi^{2}}{\partial d_{i}}\right|_{\tilde d_m(R_{k})} & = & \left.\dfrac{\partial^{2}\chi^{2}}{\partial d_{i}\partial R_{j}}\right|_{\tilde 
d_m(R_{k})}+\sum_{l}\left.\dfrac{\partial^{2}\chi^{2}}{\partial d_{i}\partial d_{l}}\right|_{{\tilde d}_{m}(R_{k})}\dfrac{\partial {\tilde d}_{l}(R_{n})}{\partial R_{j}}=0\,,\nonumber \\
\dfrac{\partial \tilde d_i(R_{n})}{\partial R_{j}} & = & -\sum_l\left[\dfrac{\partial^{2}\chi^{2}}{\partial d_i\partial d_l}\right]_{\tilde d_m(R_{k})}^{-1}\,\left.\dfrac{\partial^{2}\chi^{2}}{\partial d_{l}\partial 
R_{j}}\right|_{\tilde d_m(R_{k})}
\,.
\end{eqnarray}
The minimum of $\tilde \chi^2(R_i)$ is indeed located at $R^{(0)}_{i}$ because
\begin{equation}
\dfrac{\partial\tilde\chi^2}{\partial R_{i}}=\sum_k\left.\dfrac{\partial\chi^{2}}{\partial d_k}\right|_{d_k=\tilde d_k(R_i)}\dfrac{{\rm d} \tilde d_k}{{\rm d} R_i}+
\left.\dfrac{\partial\chi^{2}}{\partial R_i}\right|_{d_k=\tilde
  d_k(R_i)}=\left.\dfrac{\partial\chi^{2}}{\partial R_i}\right|_{d_k=\tilde
  d_k(R_i)}
\,,\label{first_der}
\end{equation}
and because the first term vanishes by the condition in Eq.~(\ref{EOM}). When
evaluating Eq.~(\ref{first_der}) for $R_i=R_i^{(0)}$ it vanishes because of
Eq.~(\ref{Mincond}).  
Finally, let us calculate the inverse correlation matrix:
\begin{eqnarray}
&&\left.\dfrac{\partial^{2}\tilde{\chi}^{2}}{\partial R_{i}\partial R_{j}}\right|_{\tilde{R}_{k}} = 
\left.\dfrac{\partial^{2}\chi^{2}}{\partial R_{i}\partial R_{j}}\right|_{d^{(0)}_{m},R^{(0)}_{k}}+
\sum_{k}\left.\dfrac{\partial^2\chi^{2}}{\partial d_k\partial R_j}\right|_{d^{(0)}_m,R^{(0)}_k}
\left.\dfrac{\partial \tilde d_{k}(R_{i})}{\partial R_{i}}\right|_{\tilde{R}_{k}}\nonumber \\
 & = & \left.\dfrac{\partial^{2}\chi^{2}}{\partial R_{i}\partial R_{j}}\right|_{d^{(0)},R^{(0)}_{k}}-\sum_{k,l}\left[\dfrac{\partial^{2}\chi^{2}}{\partial d_{k}\partial 
d_{l}}\right]_{d^{(0)}_{m},R^{(0)}_{k}}^{-1}\,\left.\dfrac{\partial^{2}\chi^{2}}{\partial d_{l}\partial R_{i}}\right|_{d^{(0)}_{m},R^{(0)}_{k}}\left.\dfrac{\partial^{2}\chi^{2}}{\partial d_{k}\partial 
R_{j}}\right|_{d^{(0)}_{m},R^{(0)}_{k}},\label{eq:EOM_Rj_bk}\end{eqnarray}
which agrees with Eq.~(\ref{eq:noEOM_Rj_bk}).

We can now apply the previous results to Eq.~(\ref{additive}):
\begin{eqnarray}
\chi^{2} & = & \sum_{k}\left[d_k^2+\sum_{i,m}\left(\dfrac{R_i^{k,m}+d_k\,
\Delta_i^{k,m}-R_m}{\sigma_i^{k,m}}\right)^2\right]\label{eq:with_bk_Rj}\\
 & = & \sum_k\left\{ d_k^2\left[1+
\dfrac{\left\langle \Delta^2\right\rangle_k}{\left\langle \sigma\right\rangle _k^2}\right]+2\, d_k\sum_{i,m}\dfrac{(R_i^{k,m}-R_m)\Delta_i^k}{\sigma_{i}^{k,m\,2}}+\sum_{i,m}
\left(\dfrac{R_i^{k,m}-R_{m}}{\sigma_i^{k,m}}\right)^2\right\} 
\,.\nonumber
\end{eqnarray}
The equation of motion for $d_{k}$ reads
\begin{equation}
d_k(R)=-\,\dfrac{1}{1+\dfrac{\left\langle \Delta^2\right\rangle_k}{\left\langle \sigma\right\rangle_k^2}}
\sum_{i,m}\dfrac{(R_i^{k,m}-R_m)\Delta_i^{k,m}}{\sigma_i^{k,m\,2}}
\,.
\end{equation}
This renders for $\tilde\chi^2$ the form
\begin{eqnarray}
\tilde\chi^2 & = & \sum_k\left[d_k(R)\sum_{i,m}\dfrac{(R_i^{k,m}-R_m)\Delta_i^{k,m}}{\sigma_i^{k,m\,2}}+
\sum_{i,m}\left(\dfrac{R_i^{k,m}-R_m}{\sigma_i^{k,m}}\right)^2\right]\\
 & = & \sum_k\left[\sum_{i,m}\left(\dfrac{R_i^{k,m}-R_m}{\sigma_i^{k,m}}\right)^2-
\dfrac{1}{1+\dfrac{\left\langle \Delta^2\right\rangle_k}{\left\langle \sigma\right\rangle_k^2}}\sum_{i,j,m,n}\dfrac{(R_i^{k,m}-R_m)\Delta_i^{k,m}}
{\sigma_i^{k,m\,2}}\dfrac{(R_j^{k,n}-R_n)\Delta_j^{k,n}}{\sigma_j^{k,n\,2}}\right]\,,\nonumber 
\end{eqnarray}
which reproduces Eq.~(\ref{traditional}), as we wanted to demonstrate. 

We conclude this appendix presenting an alternative way to write
Eq.~(\ref{chi2total}) after using the equations of motion
for $d_k$. We again concentrate on the simpler case without subtraction of the
non-charm contribution:
\begin{eqnarray}
\chi^2 & = & \sum_k\left\{ d_k^{2}+\sum_{i,m}\left[\dfrac{R_i^{k,m}-(1+d_k\Delta f_k^{i,m})\, R_m}{\sigma_k^{i,m}}\right]^{2}\right\} \nonumber \\
 & = & \sum_k\left\{ \left[1+\sum_{i,m}
\left(\dfrac{\Delta f_k^{i,m}\, R_m}{\sigma_k^{i,m}}\right)^2\right]d_k^2-2d_k\sum_{i,m}\dfrac{\left(R_{k}^{i,m}-R_m\right)
\Delta f_k^{i,m}\, R_m}{(\sigma_k^{i,m})^2}\right.\nonumber \\
&+&\left.\sum_{i,m}\left[\dfrac{R_k^{i,m}-R_m}
{\sigma_k^{i,m}}\right]^2\right\} .\label{eq:chi2_auxiliary}
\end{eqnarray}
The EOM for $d_k$ now reads
\begin{equation}
\left[1+\sum_{i,m}\left(\dfrac{\Delta f_k^{i,m}\, R_m}{\sigma_k^{i,m}}\right)^2\right]
\tilde d_k(R_n)-\sum_{i,m}\dfrac{\left(R_k^{i,m}-R_m\right)\Delta f_k^{i,m}\,
  R_m}{(\sigma_k^{i,m})^2}=0
\,,
\end{equation}
which upon insertion into Eq.~(\ref{eq:chi2_auxiliary}) renders
\begin{eqnarray}
\widetilde{\chi}^2 & = & \sum_k\left\{
  \sum_{i,m}\left[\dfrac{R_k^{i,m}-R_m}{\sigma_k^{i,m}}\right]^2-\tilde
  d_k(R_n)\sum_{i,m}\dfrac{\left(R_k^{i,m}-R_m\right)\Delta f_k^{i,m}\, R_m} 
{(\sigma_k^{i,m})^2}\right\} \nonumber\\
 & = & \sum_k\Biggr\{ \sum_{i,m}\left[\dfrac{R_k^{i,m}-R_m}{\sigma_k^{i,m}}\right]^2\nonumber\\
&-&
\dfrac{1}{\left[1+\sum\left(\dfrac{\Delta f_k^{i,m}\,
 R_m}{\sigma_k^{i,m}}\right)^2\right]}
\left[\sum_{i,m}\dfrac{\left(R_k^{i,m}-R_m\right)\Delta_k^{i,m}\, R_m}
{(\sigma_k^{i,m})^2}\right]^2\Biggr\}
\,.\label{eq:chi2_final} 
\end{eqnarray}
One can rewrite Eq.~(\ref{eq:chi2_final}) in the matrix form
\begin{eqnarray}
\widetilde{\chi}^2 & = & \sum_k\left\{ \sum_{i,j,m,n}\left[R_{k}^{im}-R_m\right]\left[V_{k}^{-1}\right]^{mn}_{ij}\left[R_{k}^{jn}-R_n\right]\right\} \,,\label{eq:Matrix-Form} \\
\left[V_{k}^{-1}\right]^{mn}_{ij} & = & \dfrac{\delta_{ij}\delta^{mn}}{(\sigma_{k}^{i,m})^{2}}-
\dfrac{\left(R_k^{i,m}-R_m\right)\Delta f_k^{i,m}\, R_m
\left(R_k^{jn}-R_n\right)\Delta f_k^{jn}\, R_n}
{\left[1+\sum\left(\dfrac{\Delta_k^{i,m}\, R_m}{\sigma_k^{i,m}}\right)^2\right]}\,.\nonumber
\end{eqnarray}
In Eq.~(\ref{eq:Matrix-Form}) one can interpret the second term of
the inverse correlation matrix,  
as a non-linear correlation among the measurements, where the correlation matrix
itself depends on the value of the fit parameters. 
The total inverse correlation matrix is block diagonal, and the blocks correspond to $V_k^{-1}$.

\section*{\boldmath Appendix C: Dependence on $\alpha_s$ for Higher Moment Analyses}
\addcontentsline{toc}{section}{Appendix C: Dependence on $\alpha_s$ for Higher Moment Analyses}
\newcounter{beta}
\renewcommand{\thesection}{\Alph{beta}}
\renewcommand{\theequation}{\Alph{beta}.\arabic{equation}}
\renewcommand{\thetable}{\Alph{beta}}
\setcounter{beta}{3} \setcounter{equation}{0} \setcounter{table}{0}

In this appendix we display the dependence of $\overline m_c(\overline m_c)$ and for 
$\overline m_c(3~\mbox{GeV})$ on the value of the strong coupling constant at the $Z$ pole,
when fitted from the second, third and fourth moments. These results are the equivalent of
Eq.~(\ref{mcfinalalphas}). The results for the canonical value $\alpha_s(m_Z) = 0.1184 \pm 0.0021$
are shown in Tab.~\ref{tab:high-n}.

\begin{itemize}
 \item $n = 2$
 \begin{align}
\overline m_c(\overline m_c) = \,&(1.276 + 2.90\,[\alpha_s(m_Z) - 0.1184]) 
\, \pm \, (0.004)_{\rm stat}
\, \pm \, (0.004)_{\rm syst}\\ &
\, \pm \, (0.017 + 1.14\,[\alpha_s(m_Z) - 0.1184])_{\rm pert} 
\, \pm \, (0.004)_{\langle GG\rangle}\, {\rm GeV}
\,,\nonumber
\\[2mm]
\overline m_c(3~\mbox{GeV}) = \,&(0.988 - 5.32\,[\alpha_s(m_Z) - 0.1184]) 
\, \pm \, (0.002)_{\rm stat} 
\, \pm \, (0.002)_{\rm syst}\\ &
\, \pm \, (0.019 + 1.31\,[\alpha_s(m_Z) - 0.1184])_{\rm pert} 
\, \pm \, (0.004)_{\langle GG\rangle}\, {\rm GeV}\,.\nonumber
\end{align}
\item $n = 3$
\begin{align}
\overline m_c(\overline m_c) = \,&(1.277 + 2.12\,[\alpha_s(m_Z) - 0.1184]) 
\, \pm \, (0.003)_{\rm stat} 
\, \pm \, (0.003)_{\rm syst}\\ &
\, \pm \, (0.016 + 1.08\,[\alpha_s(m_Z) - 0.1184])_{\rm pert} 
\, \pm \, (0.004)_{\langle GG\rangle}\, {\rm GeV}
\,,\nonumber
\\[2mm]
\overline m_c(3~\mbox{GeV}) = \,&(0.989 - 6.14\,[\alpha_s(m_Z) - 0.1184]) 
\, \pm \, (0.003)_{\rm stat} 
\, \pm \, (0.003)_{\rm syst}\\ &
\, \pm \, (0.018 + 1.25\,[\alpha_s(m_Z) - 0.1184])_{\rm pert} 
\, \pm \, (0.005)_{\langle GG\rangle}\, {\rm GeV}\,.\nonumber
\end{align}
\item $n = 4$
\begin{align}
\overline m_c(\overline m_c) = \,&(1.279 + 1.55\,[\alpha_s(m_Z) - 0.1184]) 
\, \pm \, (0.002)_{\rm stat}
\, \pm \, (0.002)_{\rm syst}\\ &
\, \pm \, (0.016 + 1.05\,[\alpha_s(m_Z) - 0.1184])_{\rm pert} 
\, \pm \, (0.005)_{\langle GG\rangle}\, {\rm GeV}
\,,\nonumber
\\[2mm]
\overline m_c(3~\mbox{GeV}) =\, &(0.992 - 6.75\,[\alpha_s(m_Z) - 0.1184]) 
\, \pm \, (0.003)_{\rm stat}
\, \pm \, (0.003)_{\rm syst}\\ &
\, \pm \, (0.017 + 1.21\,[\alpha_s(m_Z) - 0.1184])_{\rm pert} 
\, \pm \, (0.006)_{\langle GG\rangle}\, {\rm GeV}\,.\nonumber
\end{align}
\end{itemize}

\bibliography{charm2}
\bibliographystyle{JHEP}

\end{document}